\documentclass[11pt]{article}
\usepackage[a4paper,margin=25mm]{geometry}
\usepackage[T1]{fontenc}
\usepackage{times}
\usepackage{newtxmath}
\usepackage{textcomp}
\usepackage{amsmath}
\usepackage{graphicx}
\usepackage{booktabs}
\usepackage{longtable}
\usepackage{array}
\usepackage{needspace}
\usepackage{xcolor}
\usepackage{tikz}
\usetikzlibrary{arrows.meta,angles,quotes}
\usepackage{authblk}
\usepackage[authoryear]{natbib}
\usepackage{url}
\usepackage[hidelinks]{hyperref}
\usepackage{orcidlink}
\usepackage[capitalise]{cleveref}
\DeclareUnicodeCharacter{0394}{\ensuremath{\Delta}}
\DeclareUnicodeCharacter{03A3}{\ensuremath{\Sigma}}
\DeclareUnicodeCharacter{2265}{\ensuremath{\geq}}
\DeclareUnicodeCharacter{2212}{\ensuremath{-}}
\graphicspath{{figures/}}

\title{Response-balanced sampling relocates rather than improves accuracy in slope-stability surrogates under a fixed simulation budget}
\author[1]{Iancu-Bogdan Teodoru\thanks{Corresponding author: \href{mailto:iancu-bogdan.teodoru@academic.tuiasi.ro}{iancu-bogdan.teodoru@academic.tuiasi.ro}} \orcidlink{0000-0001-8298-9374}}
\author[2]{Alessia Amelio \orcidlink{0000-0002-3568-636X}}
\author[3]{Mircea Aniculăesi \orcidlink{0000-0002-1900-5819}}
\author[3]{Irina Lungu \orcidlink{0000-0003-0963-7027}}
\affil[1]{Department of Land Improvement and Environmental Protection, Technical University Gheorghe Asachi of Iasi, Romania}
\affil[2]{HPC Laboratory, Department InGeo, University ``G. d'Annunzio'' Chieti-Pescara, Italy}
\affil[3]{Department of Transportation Infrastructure and Foundations, Technical University Gheorghe Asachi of Iasi, Romania}
\date{}
\begin{document}
\maketitle

\begin{abstract}
Machine-learning surrogates of slope stability learn factors of safety (FOS) that only physics-based analyses produce, so analyses are the budget to allocate. Because Latin hypercube sampling (LHS) of the parameters covers the FOS unevenly, a natural remedy is to choose part of the training points to even out the FOS distribution. We test this response-balanced sampling at a fixed budget. A random-forest screener trained on a 300-point LHS seed predicts the FOS of 200\,000 candidates, and 300 further limit-equilibrium analyses are allocated to even out the FOS over the selection range, $[0.5, 6]$. The design is compared with 300 further LHS analyses over five realizations of the added points, on two test sets, under two independently tuned tree-ensemble families, with ablations throughout; an oracle drawing from 22\,000 solved analyses meets the intended density exactly. The primary criterion is the log-scale root-mean-square error (log-RMSE) on the decision range, FOS $\in [0.5, 2)$, against continued LHS. Response-balanced sampling does not systematically improve accuracy; it relocates accuracy along the FOS axis, consistently across realizations, from $[1, 2)$ to $[0.5, 1)$, following the training density actually achieved, even when the intended density is met exactly. On the decision range no screener-selected design shows a demonstrated gain, and with known bin membership that verdict stands, in this design. Within the selection range, LHS gives $[0.5, 1)$ 11--12\% of the points against a log-uniform target's 27.9\%, but $[1, 2)$ already 35--37\%: the region that is rare need not be where decisions are made.

\end{abstract}

\section{Introduction}\label{sec:introduction}

Machine learning (ML) surrogates are increasingly used to replace computationally expensive physics-based analyses across the geosciences \citep{Lu2019, Guardiani2021, Aminpour2022, Furtney2022, Degen2023, Li2023}, and slope stability has become one of their canonical applications \citep{Lin2021, Aminpour2023, Xu2023, Gao2024, Gao2025, Cheng2026}. A trained surrogate predicts the factor of safety (FOS) of a slope in milliseconds, which makes practical what direct limit-equilibrium or finite-element analysis makes slow: regional screening, probabilistic analysis with large Monte Carlo samples, parametric studies and interactive design tools \citep{Kang2016, Ji2017, Zhu2020, Guardiani2021, Xu2023a, Aminpour2023}. As model classes have matured, attention in the wider ML community has moved from models to data, a shift often labelled data-centric machine learning \citep{Zha2023, Jakubik2024, Malerba2024}: for a fixed model class, the composition of the training set often matters more than further refinement of the model \citep{Jain2020, Gupta2021, Ursu2025}.

Surrogates of physics-based analyses hold a particular place in this landscape. A factor of safety cannot be measured: a case history or an experiment tells whether a slope remained stable or failed, but not its margin of safety, so FOS values exist only as the output of an analysis. The modeller therefore generates the training data by running a solver on parameter combinations of their choosing. Data are not scarce; solver runs are. The question is one of allocation: given a fixed number of analyses, which parameter combinations should be solved? In current practice it is answered implicitly, by a design of experiments in the input space, most often Latin hypercube sampling (LHS), valued for its space-filling behaviour and its one-dimensional uniformity \citep{Stein1987, McKay2000, Husslage2008, Viana2016, Lin2022}.

An input-space design is blind to the quantity the surrogate has to learn. Because the map from strength, unit weight and geometry to the FOS is strongly nonlinear \citep{Lin2021, Li2019}, a design that covers the parameters evenly covers the response unevenly. The intuition that follows, that a response covered unevenly yields poor predictions where it is sparse, is well established in imbalanced regression \citep{Branco2016, Branco2017, Steininger2021}, and designs that fill the output space rather than the input space have been proposed where the outputs must be covered evenly, as in inverse design and calibration \citep{Wang2024, LuAndersonCook2021}. In slope stability, \citet{Shakya2024} have shown that the sampling strategy markedly affects the accuracy of ML predictions, and \citet{Mazloom2026BARL} have applied batch active learning to a continuous FOS target, gating candidates on a predicted FOS and reporting a more even FOS distribution as their main gain. That gain, however, is not measured at an equal number of analyses. What has not been measured is the effect of shaping the response distribution when the number of analyses is held fixed, and where along the FOS axis that effect falls.

This paper measures it. We call the strategy under study \emph{response-balanced sampling}: a preliminary model, the screener, trained on an initial LHS seed, predicts the FOS of a large pool of candidate parameter combinations, and the remaining budget is spent on candidates chosen so that the training set as a whole approaches a target distribution of the FOS over a declared selection range. ``Balanced'' names the distribution aimed at, not a judgement of its quality. The term is distinct from balanced sampling in finite-population survey sampling, where a sample is balanced when its Horvitz--Thompson estimates reproduce known population totals of auxiliary variables \citep{DevilleTille2004}.

The main variant, C-log, fills logarithmic bins in the FOS against the seed's deficit in a single batch; its sequential version, C-seq, spends the same budget in three batches and retrains the screener on everything solved before each batch. An earlier linear-bin variant, C, is read in the single realization only. Every variant is compared against continued LHS at the same total number of analyses, and ablations that train on the added points alone are reported with every arm. With a real screener, the training density achieved departs from the one intended. Two further arms read what that departure does. Both draw from a reference pool of 22\,000 solved analyses, with C-log's allocation. The oracle, C-oracle, bins candidates on their solved FOS, and so meets the intended density exactly; C-logP bins them on the screener's prediction. Their paired contrast measures the effect of the departure on accuracy and on relocation, and its readings were declared before the pool was solved. The oracle is a diagnostic reference, not a sampling strategy. Accuracy is read along the FOS axis, zone by zone, on two test sets that stand for two deployment scenarios. Two independently tuned tree-ensemble families, random forests and gradient boosting, serve as the measuring instrument, and the results are stated as consistent across these two families, and no further. The primary criterion is the log-RMSE, the root-mean-square error of the logarithm of the FOS, on the decision range, FOS $\in [0.5, 2)$, against continued LHS; the range spans unstable slopes awaiting remediation and the zone where design decisions are made.

The slope is deliberately simple: homogeneous, drained and with no water table, its FOS governed by the slope angle, the friction angle and the stability ratio $\gamma H/c$. It serves as a controlled testbed. Its ground truth is cheap, so every arm can be solved in full, the points each strategy adds can be drawn again in five realizations, and the oracle can draw from a solved pool of 22\,000 analyses, which on a layered finite-element model would be far more costly. Its response structure is known, so the quantities the reading rests on, the training density achieved against the one intended and where along the FOS axis LHS leaves the response sparse, are not mixed with the effects of stratigraphy or pore pressure. Cases of practical complexity motivate the question; what this case is meant to carry over to them is a protocol and a diagnostic, not effect sizes (\cref{sec:disc-interpretation}).

The result is not the one the intuition predicts. Under LHS in this parameter domain, only the zone FOS $\in [0.5, 1)$ is rare: within the selection range it holds 11--12\% of the points, against the 27.9\% a log-uniform target would give it, while $[1, 2)$ already holds 35--37\%, more than the target would give it. Evening out the response over the selection range therefore adds training points below FOS~1 and withdraws them from the zone where design decisions are made. Response-balanced sampling does not systematically improve accuracy over continued LHS. It relocates accuracy along the FOS axis, from $[1, 2)$ to $[0.5, 1)$, following the training density actually achieved. Over five new realizations of the added points, the relocation has the same direction for C-log and C-seq, in every combination of learner, test set and half of the decision range. On the decision range, no design that selects with the screener shows a demonstrated gain over continued LHS, and in one of the four cells of learner and test set C-log is worse, with an interval that excludes zero. Relocation is present even when the intended density is met exactly: the oracle relocates accuracy in three of the four cells. With known bin membership, ``no demonstrated gain'' on the decision range stands, in this design. The paired contrast of C-oracle with C-logP is inconclusive: it neither assigns a share of the effect to the screener's error nor excludes one.

The study contributes in method and in evidence. As method, it gives a budget-fair protocol for testing a sampling strategy for a surrogate: arms that spend the same number of analyses, ablations that separate the added points from the seed, accuracy read zone by zone on a log scale, replicated realizations of the added points, and a solved reference pool with declared readings, against which the effect of the departure from the intended density is measured. It also gives a diagnostic that needs only the solved seed: compare the response distribution that LHS already gives with the target, and with the range in which the surrogate will be used. As evidence, it shows for a slope-stability surrogate that response-balanced sampling relocates accuracy rather than improving it, even when the intended density is met, and that it brings no demonstrated gain on the decision range. It also reports a declared prediction that failed: the advantage of sequential retraining seen in the first realization did not replicate.

The remainder of the paper is organized as follows. \Cref{sec:related-work} reviews slope-stability surrogates and their training data, sampling studies for them, and reliability-oriented active learning, and then the machine-learning literature on surrogate modelling, design of experiments, active learning, imbalanced regression and conformal prediction. \Cref{sec:methods} describes the data generation, the sampling strategies, the reference pool and the oracle, the replications, the tuning protocol and the evaluation. \Cref{sec:results} presents the results, \cref{sec:discussion} discusses them and their limitations, and \cref{sec:conclusions} concludes.

\section{Related work}\label{sec:related-work}

\subsection{Slope-stability surrogates and their training data}\label{sec:rw-surrogates}

The training data of a slope-stability surrogate are produced by a solver, on parameter combinations fixed by a design in the input space. \Citet{Kang2016} draw the training points of a support vector regression by LHS, so that they cover the design space uniformly, with fifteen points per random variable, and solve them by Bishop's simplified or Spencer's method. \Citet{Guardiani2021} draw 2\,000 LHS samples of the strength and hydraulic parameters of a reservoir slope under rapid drawdown and solve each by limit equilibrium. \Citet{Li2023} build 880 homogeneous slopes from a full factorial design and solve them with FLAC 3D, a finite-difference code. \Citet{Gao2025} make the design itself the object of study: comparing orthogonal and full factorial training sets for a finite-element strength-reduction model, they find that the choice changes the accuracy reached. Their comparison, too, is between designs in the input space.

In these studies the distribution of the FOS in the training set is an outcome of the design, not a target. Where it is shown, it is described after the fact. \Citet{Li2023} give the frequency distribution of their FOS values, state that the dataset was built without considering data balance, attribute their lowest class accuracy, in the marginally stable class, to that imbalance, and list the imbalance among their limitations. \Citet{Zhong2025} train slope-reliability surrogates as failure/safe classifiers on LHS samples and balance the two classes with synthetic failure samples generated by adaptive synthetic sampling; the balancing acts on the labelled set, not on which analyses are run. Where the distribution is controlled, it is on the test side. \Citet{Azmoon2021} draw slope geometries and soil properties uniformly at random and label them by limit equilibrium. To build a test set that covers all ranges of the FOS, they generate more than 30\,000 further cases and draw 1\,000 at random from each of nine FS classes. They describe no such selection for their training set.

Several slope-stability surrogate studies use a homogeneous slope. \Citet{Guardiani2021} consider one with a simplified geometry, and \citet{Mazloom2026BARL} model a single-layer homogeneous soil; \citet{Li2023} describe theirs as a deliberate choice, made to understand the behaviour of the model under simplified conditions. A homogeneous slope is also the first example of the finite-element analyses of \citet{GriffithsLane1999}. The present study uses such a slope in the same spirit, as a controlled testbed (\cref{sec:data}).

\subsection{Sampling studies for slope-stability surrogates}\label{sec:rw-sampling}

Two studies address sampling for slope-stability surrogates directly. \Citet{Shakya2024} train a model of the soil's stress--strain response for strain-dependent slope stability analysis under six sampling methods, one of them introduced to represent the low stress range better. They report a decisive effect of the sampling method on the resulting FOS. Their surrogate stands in for the constitutive response, and the region they target lies in its input space; the distribution of the FOS is not the object.

The closest work is BARL, batch active regression learning \citep{Mazloom2026BARL}, a batch active-learning scheme for a continuous FOS target. A committee of five gradient-boosting models scores the candidates of an LHS pool by the spread of its predictions. The score is gated by the committee's mean predicted FOS, which penalizes candidates whose prediction falls outside the target range. A cosine-similarity threshold keeps the members of a batch apart. The central element of the present design is thus already in BARL: selection gated on a predicted FOS, before the analysis is paid for. BARL's initial training set is an LHS design filtered to FOS $\in [0.3, 6]$. The authors note that most practical cases lie between 0.9 and 2, which is consistent with the upper end of the selection range and of the decision range used here. In their domain, the initial set is skewed towards low FOS (lower quartile 1.29, median 2.09), so the under-represented zone is the upper one, 4--6. In ours, the rare zone under LHS is $[0.5, 1)$ (\cref{sec:disc-interpretation}). Which zone is rare therefore depends on the parameter domain. BARL presents the more even FOS distribution after active learning as its greatest improvement, and reports test errors well below those of the initial model. That comparison is between a committee trained on the final set, 5\,410 analyses, and one trained on the initial 300. It shows the value of a larger, adaptively chosen training set, which is not in question. It does not compare BARL with another selection rule at the same number of analyses.

The present study makes that comparison for response-balanced sampling. It holds the number of analyses fixed and compares the strategy with continued LHS, replicated over independent realizations of the added points. It is a budget-fair test of shaping the response distribution of the training set; it does not introduce that idea.

\subsection{Reliability-oriented active learning}\label{sec:rw-reliability}

In reliability analysis, slope reliability included, surrogates can be refined adaptively, with a different aim. AK-MCS \citep{Echard2011}, an active-learning method combining Kriging and Monte Carlo simulation, refines a Kriging surrogate in a loop over Monte Carlo samples, to estimate a probability of failure with few calls to the model. In slope reliability, \citet{Zhu2020} start a Gaussian-process surrogate from random training points and add points near the critical state until a target fit is reached; BARL's group has developed active-learning schemes for structural reliability and Bayesian updating \citep{ZhangShafieezadeh2022, SongShafieezadehXiao2022, SongWang2022}. These methods direct their analyses towards the limit state, because it determines the probability of failure. The present study asks a different question: how accuracy over a range of FOS values is distributed when a fixed number of analyses is allocated by a target distribution of the response. A gain of active learning over uniform designs is not assured either: across 16 benchmark problems of distribution estimation, \citet{Su2025} find no systematic improvement of active learning over a uniform space-filling design.

\subsection{Surrogate modelling}\label{sec:rw-surrogate-modelling}

Surrogate models approximate the input--output map of a computationally expensive simulator using a finite set of evaluated input configurations \citep{Forrester2009}. Once trained, the surrogate can produce predictions at substantially lower computational cost than a new solver evaluation, making it useful for repeated prediction, sensitivity analysis, and design exploration. The accuracy of this approximation depends not only on the regression algorithm and its tuning but also on where the expensive training evaluations are placed. In global metamodelling, adaptive sampling has therefore been investigated as an alternative to designs fixed entirely before the simulator is run. The review by \citet{Liu2018} discusses how adaptive allocation can improve metamodel accuracy with fewer evaluations, whereas the comparison of Kriging sampling schemes by \citet{Fuhg2021} emphasizes that the relative merits of a selection rule depend on the problem and the objective. In the present study, the solver budget is fixed and the final regressors are used to evaluate alternative allocations of that budget, rather than to introduce a new surrogate architecture. This isolates the empirical effect of shaping the training response distribution on forward prediction, with results assessed using two independently tuned tree-ensemble families.

\subsection{Design of experiments and output-space-filling designs}\label{sec:rw-doe}

Design of experiments determines the configurations at which a computer model is evaluated. Input-space-filling methods seek broad coverage of the parameter domain without requiring prior knowledge of the response. LHS, in particular, stratifies each input marginal, but it does not impose any corresponding distribution on the resulting outputs. Consequently, even an input design with broad coverage can produce many evaluations in some response intervals and relatively few in others. An alternative is to account explicitly for output-space coverage. \citet{LuAndersonCook2021} formulate input--response space-filling as a multi-objective design problem, using Pareto fronts to trade off spacing in the inputs and in the response; they motivate this construction with applications including calibration, sensitivity studies, and inverse problems. \citet{Wang2024} develop sequential designs to fill an output or feature space, particularly for inverse design and feature-based modelling. Their work also distinguishes output coverage from accurate estimation of the underlying input--output relationship: achieving the former does not guarantee the latter. These objectives should therefore not be conflated. In our setting, response-balanced sampling specifies a target distribution across the FOS selection range $[0.5,6]$, but the evaluation criterion is the accuracy of a forward surrogate under an equal simulation budget, especially over the decision range $[0.5,2)$. Continued LHS provides the budget-matched input-space baseline against which that question can be tested.

\subsection{Active learning}\label{sec:rw-active-learning}

Active learning selects additional labelled examples using information from a model trained on observations acquired so far \citep{Cohn1996, Settles2009}. In regression with costly labels, uncertainty-based acquisition directs evaluations towards locations where predictions are least certain \citep{Settles2009}, while committee-based schemes use disagreement among predictors as a selection signal \citep{Seung1992, Settles2009}. Bayesian experimental design similarly ranks prospective observations through a utility, such as expected information gain \citep{Chaloner1995}; other acquisition functions seek diversity among the outputs. In the greedy sampling framework of \citet{Wu2019}, the strategy of greedy sampling on the output (GSy) selects new samples by considering diversity in the predicted output space, using a regression model trained on the available labelled observations. As the authors acknowledge, estimates derived from a predictor fitted to a small initial training set may be inaccurate. This provides a methodological precedent for prediction-dependent selection in output space, without implying that the same limitations necessarily characterize the screener adopted in the present study. More generally, adaptive sampling performance depends on the aim and on the properties of the response surface (\cref{sec:rw-surrogate-modelling}); reliability-oriented approaches have already been distinguished in \cref{sec:rw-reliability}. Our design uses a preliminary predictor to allocate solver evaluations towards a prescribed training-response distribution, rather than to maximize an uncertainty or information-gain criterion. The experiment tests what that particular allocation does to accuracy at fixed cost; it does not evaluate whether uncertainty-based or committee-based active learning would perform better or worse.

\subsection{Imbalanced regression and reweighting}\label{sec:rw-imbalanced}

Imbalanced regression concerns a continuous target whose observed values are unevenly distributed, especially when predictive performance matters in sparsely represented response regions \citep{Branco2016}. Unlike class imbalance, the rarity of a regression target is tied to its local density and to how the application assigns relevance across the response range. Several methods address the imbalance after a labelled dataset has been obtained. SMOGN (synthetic minority over-sampling for regression with Gaussian noise) combines resampling and synthetic-example generation for under-represented, relevant target regions \citep{Branco2017}. DenseWeight assigns sample weights as a function of target density, and DenseLoss uses them to change the contribution of existing observations to the training loss \citep{Steininger2021}. Deep imbalanced regression introduces label- and feature-distribution smoothing to support learning from uneven continuous targets \citep{Yang2021}. An empirical comparison of resampling strategies across learners and datasets finds that none is best everywhere, the choice depending on the learner, the dataset and the metric \citep{Avelino2024}; the same study illustrates that a global mean squared error cannot distinguish a model accurate on frequent targets from one accurate on rare ones. Such approaches modify how an existing set of labels is used, rather than which expensive simulator evaluations are performed in the first place. In the present setting, an additional observation requires a new limit-equilibrium analysis: weighting or synthesizing observations cannot supply an uncomputed solver response. Conversely, response-balanced sampling changes the empirical distribution of available solver labels but does not, by itself, prescribe how a regression learner should weight them. These are complementary interventions. Training a density-reweighted surrogate on the continued-LHS dataset would be a cheaper and scientifically distinct comparison, but it is not tested here; our conclusions concern the allocation of new solver runs, not the effectiveness of reweighting.

\subsection{Conformal prediction}\label{sec:rw-conformal}

Prediction intervals supplement point predictions with a representation of uncertainty in an individual response. Conformal methods construct such intervals by calibrating predictive errors and, under suitable exchangeability assumptions and a specified construction, can provide finite-sample marginal coverage statements \citep{Angelopoulos2023}. For regression, the jackknife+ and its cross-validation extension, CV+, incorporate predictions from leave-one-out or fold-specific fitted models when constructing intervals; their coverage analysis is given by \citet{Barber2021}. This distinction matters because simply pooling out-of-fold residuals and centring their quantiles on a separately refitted model is not automatically equivalent to CV+. The intervals of the present study are built in that pooled way (\cref{sec:evaluation}); we therefore describe them as \emph{cross-conformal-style}, not as CV+. Furthermore, response-balanced sampling may introduce differences between the response distributions represented in the training and calibration data and those encountered at prediction time, potentially affecting the applicability of standard exchangeability-based coverage guarantees. A response-stratified test set introduces a separate consideration, as empirical coverage is evaluated under a deliberately modified response distribution rather than necessarily under the target population distribution. These issues are relevant to conformal inference beyond exchangeability \citep{Barber2023}. Accordingly, we report empirical interval coverage on the specified evaluation sets and interpret it descriptively, without claiming a formal distribution-free coverage guarantee.

\section{Methods}\label{sec:methods}

\subsection{Slope model and data generation}\label{sec:data}

\begin{figure}
    \centering
    \begin{tikzpicture}[scale=0.65]
    \small
  \coordinate (A) at (0.00, 0.00);
  \coordinate (B) at (0.00, 3.00);
  \coordinate (C) at (3.00, 3.00);
  \coordinate (D) at (9.00, 6.00);
  \coordinate (E) at (13.50, 6.00);
  \coordinate (F) at (13.50, 0.00);
  \coordinate (G) at (5.00, 4);
  \coordinate (H) at (7.00, 5);
  \coordinate (L) at (3.00, 13.00);
  \coordinate (S) at (5.62, 8.23);
  \coordinate (AD) at (6.25, 5.50);
    \coordinate (Cx)   at (4.40, 3.00); %
\coordinate (c1) at (0, 7.5);
\coordinate (c2) at (3, 7.5);
\coordinate (c3) at (9, 7.5);
\coordinate (c4) at (13.5, 7.5);
\coordinate (c5) at (-1, 3);
\coordinate (c6) at (-1, 6);
\coordinate (c7) at (-1, 0);
  \filldraw[very thick, fill=gray!35,opacity=0.5] (A) -- (B) -- (C) -- (D) -- (E) -- (F) -- cycle;
\draw[thick, teal] (1.35, 3.00) arc (-129.2:-19.3:6.75);
  \draw[thick, dashed, purple] (3.66, -0.00) arc (-115.2:-64.8:6.08);
  \draw[purple, thick] (8.84, -0.00) arc (-64.8:4.7:6.08);
  \draw[purple, thick] (0.71, 3.00) arc (-155.7:-115.2:6.08);
  \draw[thick, purple] (-0.00, 3.00) arc (-121.2:-33.7:10.00);
  \draw[<->, >=Stealth, thin] (c1) -- node[above, text=black, midway] {$\max(H;0.5L)$} (c2);  
  \draw[thin] ([yshift=-0.25cm] c1) -- ([yshift=0.25cm] c1);
  \draw[thin] ([yshift=-0.25cm] c2) -- ([yshift=0.25cm] c2);
  \draw[thin] ([yshift=-0.25cm] c3) -- ([yshift=0.25cm] c3);
  \draw[thin] ([yshift=-0.25cm] c4) -- ([yshift=0.25cm] c4);
    \draw[<->, >=Stealth, thin] ([yshift=-1cm] c2) -- node[above, text=black, midway] {$L=H\cot\beta$} ([yshift=-1cm] c3);
    \draw[<->, >=Stealth, thin] (c3) -- node[above, text=black, midway] {$\max(1.5H;0.5L)$} (c4);
    \draw[<->, >=Stealth, thin] (c5) -- node[above, text=black, midway, rotate=90] {$H$} (c6);
     \draw[<->, >=Stealth, thin] ([yshift=5.5cm] B) -- node[above, text=black, midway] {trial surfaces exit} ([yshift=4.5cm] G);
      \draw[<->, >=Stealth, thin] ([yshift=3.5cm] H) -- node[above, text=black, midway] {trial surfaces entry} ([yshift=2.5cm] E);
        \draw[<->, >=Stealth, thin] (c7) -- node[above, text=black, midway, rotate=90] {$\max(H; 0.5L)$} (c5);
    \draw[<->, >=Stealth, thin] ([yshift=4.5cm] C) -- node[above, text=black, midway] {$L/3$} ([yshift=3.5cm] G);
       \draw[<->, >=Stealth, thin] ([yshift=2.5cm] H) -- node[above, text=black, midway] {$L/3$} ([yshift=1.5cm] D);
    \draw[thin] ([xshift=-0.25cm] c6) -- ([xshift=0.25cm] c6);
    \draw[thin] ([xshift=-0.25cm] c5) -- ([xshift=0.25cm] c5);
    \draw[thin] ([xshift=-0.25cm] c7) -- ([xshift=0.25cm] c7);
    \draw[thin] ([yshift=5.25cm] B) -- ([yshift=5.75cm] B);
    \draw[thin] ([yshift=3.25cm] C) -- ([yshift=3.75cm] C);
    \draw[thin] ([yshift=0.25cm] D) -- ([yshift=.75cm] D);
    
    \draw[thin] ([yshift=4.25cm] G) -- ([yshift=4.75cm] G);
     \draw[thin] ([yshift=3.25cm] G) -- ([yshift=3.75cm] G);
    \draw[thin] ([yshift=3.25cm] H) -- ([yshift=3.75cm] H);
    \draw[thin] ([yshift=2.25cm] H) -- ([yshift=2.75cm] H);
    \draw[thin] ([yshift=1.25cm] E) -- ([yshift=1.75cm] E);
    \draw[thin] ([yshift=2.25cm] E) -- ([yshift=2.75cm] E);
      \draw[ultra thick] (B) -- (C) -- (G);
   
   \draw[ultra thick] (H) -- (D) -- (E);
  
\foreach \p in {A,B,C,D,E,F,G,H} \fill[black] (\p) circle (3pt);
\node at (12, 1) {$c,\ \phi,\ \gamma$};
 \draw[thin] ([xshift=0.75cm] C) -- ([xshift=0.5cm] Cx);
   \pic[draw, <->, "$\beta$", angle radius=11mm, angle eccentricity=1.25]
        {angle = Cx--C--D};
\draw[->] (C) -- ++(2,0) node[right] {\small $x$};
\draw[->] (C) -- ++(0,1.5) node[above] {\small $y$};
\draw[fill=purple!5, draw=purple!30, thick] (11.25,2.041) circle (0.3cm);
\node[purple] at (11.25,2.041) {$\times$};
\draw[fill=purple!5, draw=purple!30, thick] (13,5.317) circle (0.3cm);
\node[purple] at (13,5.317) {$\times$};
\draw[fill=teal!5, draw=teal!30, thick] (11.25,4.506) circle (0.3cm);
\node[teal] at (11.25,4.506) {\checkmark};
\end{tikzpicture}

    \caption{Model geometry and adaptive analysis domain, with $L = H\cot\beta$. Solid curves are trial circular slip surfaces of the Bishop analysis. The surfaces marked $\times$ reach the lateral or basal edge of the model, the condition flagged by the boundary check of \cref{sec:data}; the surface marked $\checkmark$ clears it and is accepted as the critical surface. The toe is the coordinate origin.}
    \label{fig:geometry}
\end{figure}
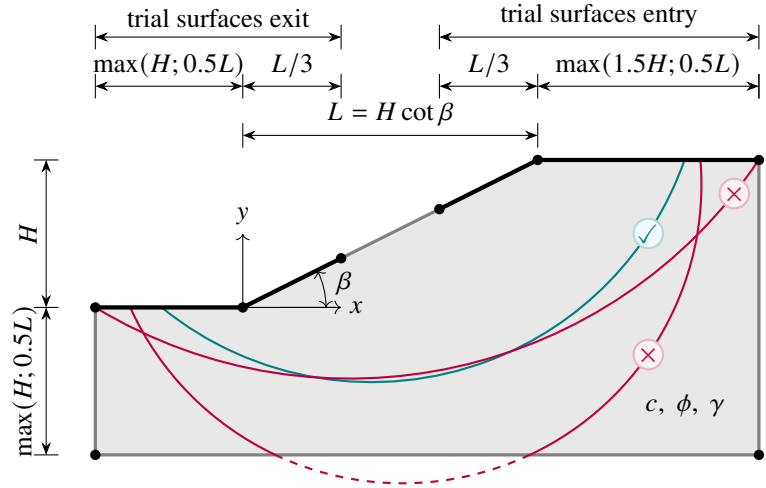

The study concerns a homogeneous soil slope of height $H$ and inclination $\beta$, under drained conditions and with no water table (\cref{fig:geometry}). The soil is a single Mohr--Coulomb material in effective stresses, with cohesion $c$, friction angle $\phi$ and unit weight $\gamma$. A configuration is therefore a point of the five-dimensional domain of \cref{tab:domain}. The quantity the surrogates learn is the factor of safety: the factor by which the shear strength must be reduced to bring the slope to limiting equilibrium, so that $\mathrm{FOS} = 1$ marks incipient failure \citep{Duncan1996}.

The case is deliberately simple. Its ground truth is cheap, and its response structure is known: the FOS is governed by the slope angle, $\phi$ and the stability ratio $\gamma H/c$, the reciprocal of Taylor's stability number \citep{Taylor1937}. This is what allows every arm to be solved in full, replicated five times and read against an oracle built on a solved reference pool (\cref{sec:oracle}).

\begin{table}[htbp]
\centering
\caption{Parameter domain. Every design in the study, including the candidate pools and the reference pool, is drawn from these ranges. Strengths are effective-stress (drained) values.}
\label{tab:domain}
\begin{tabular}{llrl}
\toprule
Parameter & Symbol & Range & Unit \\
\midrule
Slope angle    & $\beta$  & 10--65 & $^\circ$ \\
Slope height   & $H$      & 1--25  & m \\
Cohesion       & $c$      & 1--75  & kPa \\
Friction angle & $\phi$   & 5--40  & $^\circ$ \\
Unit weight    & $\gamma$ & 14--22 & kN/m$^3$ \\
\bottomrule
\end{tabular}
\end{table}

\paragraph{Generator.} Every FOS in the study, for training, testing and the reference pool, is computed with the simplified Bishop method \citep{Bishop1955}, as implemented in xslope 0.5.2 \citep{JonesWilliams2026, xslope}, with 30 slices and a search over circular slip surfaces. The search starts from a coarse grid sweep and then refines a grid of circle centres. It has two tolerances. It converges when two successive refinements each lower the FOS by less than $5 \times 10^{-4}$, once the centre spacing is below 3\% of the model height. A centre spacing below 0.01\,m also ends it, as a geometric backstop. A search that meets neither tolerance within 50 iterations stops there and is flagged as not converged. Bishop's method satisfies moment equilibrium but not horizontal force equilibrium, and it is restricted to circular surfaces. It is among the most widely used methods in practice and in the automated generation of machine-learning datasets \citep{Azmoon2021}. The comparison made here is internal to this one method: every arm's labels come from the same solver with the same settings. A bias of the method relative to more complete formulations, such as Spencer's \citep{Spencer1967, Fredlund1977}, therefore bears on every arm alike. The conclusions concern which configurations are analysed, not how they are analysed.

\paragraph{Adaptive geometry.} Across the domain, $H$ varies by a factor of 25 and $\beta$ from 10$^\circ$ to 65$^\circ$, so no fixed model extent contains the governing mechanism at every scale. A domain truncated too close to the slope biases the FOS upward, because it excludes deeper or more distant surfaces that may govern. The extent is therefore scaled with the geometry. With $L = H\cot\beta$, the horizontal run of the face, the model extends $\max(H, 0.5L)$ below the toe and beyond it, and $\max(1.5H, 0.5L)$ beyond the crest. Trial circles exit on the toe side, between the left edge and $L/3$ from the toe. They enter on the crest side, between $2L/3$ and the right edge (\cref{fig:geometry}).

\paragraph{Boundary check and repair.} A critical surface that reaches the edge of the model signals a mechanism that the model is still too small to contain, and its FOS may reflect the imposed boundary. Every analysis is therefore checked. A surface is flagged when its ends come within $0.01H$ of the left or right edge, or when the lowest point of the arc comes within $0.01H$ of the base. A flagged analysis is re-run on enlarged models, in a fixed order, until the surface clears the boundary:
\begin{enumerate}
  \item the crest-side extension doubled;
  \item the depth below the toe, which also sets the toe-side extent, doubled;
  \item both multiplied by 1.5;
  \item both doubled;
  \item both tripled.
\end{enumerate}
The original extent is not retried, since the search is deterministic and would return the same surface. The study solved 30\,250 analyses:
\begin{itemize}
  \item 2\,250 in the ten fixed training and test sets;
  \item 6\,000 in the replications (\cref{sec:replications});
  \item 22\,000 in the reference pool.
\end{itemize}
Of these, 255 were flagged: 21, 34 and 200 respectively. Each was freed by one of the five enlargements. No analysis was excluded, and none failed to solve.

\paragraph{Non-converged searches.} In 22 of the 30\,250 analyses, 2 in the fixed sets, 6 in the replications and 14 in the pool, the search reached its 50-iteration limit before meeting its tolerances. These rows are kept and flagged, not replaced. They were re-solved outside the pipeline with up to 400 iterations, and all 22 converged within 200.
\begin{itemize}
  \item The FOS never increased: it decreased in 19 cases and was unchanged in 3.
  \item The largest change was $6.8 \times 10^{-4}$ relative ($\approx 10^{-3}$ absolute), of the order of the search's own FOS tolerance.
  \item No case changed its selection bin or its reporting zone.
\end{itemize}
Replacing only these rows would mix two solver configurations, and the rows to replace would be chosen by a property of the result. Keeping them gives every row the same solver configuration and keeps the data exactly reproducible from the released configuration.

\paragraph{Independent check.} The methods of xslope are verified by its authors against the verification manuals of commercial codes and against closed-form solutions \citep{JonesWilliams2026}. What this study adds is how each model is built from its inputs and how the search is restricted: the geometry, the entry and exit windows and the search tolerances. To check these, a sample of 15 analyses, three per reporting zone, was re-computed with Slide2 9.044 (Rocscience), Bishop simplified, with 50 slices and its own circular search, and without access to the xslope results. xslope gave the lower FOS in 13 of the 15 cases and was higher by at most 0.6\% in the other two, so Slide2's own search finds no lower critical circle than xslope's restricted one. The FOS differed by less than 0.6\% in 14 of the 15 cases, and by 0.017 (1.8\%, at FOS 0.94) in the last. Every arm is trained and tested on xslope's FOS, so a difference between programs does not enter the comparisons. In about one critical circle in eight, the circle enters the crest almost vertically, and there the FOS depends on how the last slice is discretized; dropping these points from the test sets changes no verdict on the decision range and no relocation gate.

\subsection{Sampling strategies and budget}\label{sec:arms}

The comparison follows the practitioner's position: an initial budget of 300 analyses has been spent on a Latin hypercube design, and the question is how to spend the next 300. Every budget arm contains the same 300-point seed. The arms differ in how the next 300 analyses are allocated, and arm A adds none (\cref{tab:budget}).

\paragraph{Seed and continued LHS.} The seed is a Latin hypercube sample of 300 points over the domain of \cref{tab:domain} \citep{McKay2000, Stein1987}. Arm A trains on the seed alone. Arm B adds a second, independently drawn LHS of 300 points. LHS is the reference because it is the standard space-filling design for surrogate construction; it is blind to the response. Every set is checked to lie farther than $10^{-3}$ from every other set it is used with, in the domain normalized to the unit cube.

\paragraph{Response-balanced sampling.} The guided arms implement the strategy defined in \cref{sec:introduction}. Operationally it has four ingredients:
\begin{enumerate}
  \item \emph{a screener}, a preliminary model trained on the solved seed;
  \item \emph{a candidate pool} of 200\,000 parameter combinations, drawn uniformly at random over the domain, whose FOS the screener predicts at no solver cost;
  \item \emph{an allocation} of the 300 analyses to 15 bins of the predicted FOS over the selection range, FOS $\in [0.5, 6]$;
  \item \emph{a within-bin rule} that picks the candidates of each bin.
\end{enumerate}
The selection range is the interval the selection targets, and nothing more. Configurations with FOS below 0.5 are gross instabilities, and those above 6 lie far beyond any design threshold. The primary accuracy criterion is set on a narrower decision range (\cref{sec:evaluation}). The screener is a random forest of 500 trees \citep{Breiman2001} trained on $\log(\mathrm{FOS})$, with the five parameters and $\gamma H/c$ as features. The selection acts on the predicted FOS only. Every selected point is then analysed, and it is kept whatever FOS the analysis returns, including values outside the selection range.

\paragraph{The guided variants.} Three variants of the rule are compared, together with a sensitivity arm for the screener.
\begin{itemize}
  \item \textbf{C: linear bins, max--min within bins.} The selection range is cut into 15 bins of equal width, and each bin receives 20 points. Within a bin, a greedy max--min selection \citep{KennardStone1969} draws 30 candidates in the standardized parameter space from a random start, and 20 of them are kept at random. The guided points alone are spread evenly over the predicted FOS; the seed is not taken into account.
  \item \textbf{C-log: logarithmic bins, filled against the seed, random within bins.} The selection range is cut into 15 geometric bins, which are equal in $\log(\mathrm{FOS})$, the trained response. The 300 points are allocated by water-filling against the seed's solved FOS: each point in turn goes to the bin where seed and selection together hold the fewest points, with ties going to the lower FOS. The aim is a training set, seed and selection together, that is even over the bins, not a guided set that is even on its own. The resulting allocation puts 38, 35 and 31 points in the three lowest bins, where the seed is sparsest, and 1 point in $(1.14, 1.35]$, where the seed already holds 37. Within a bin the points are drawn at random. Max--min is not used there: it selects the mutually most distant candidates, which in a box are its corners. In C, 54\% of the guided points lie within 5\% of a bound in two or more parameters, against 9\% for LHS. A random draw instead keeps the distribution of the parameters given the FOS.
  \item \textbf{C-seq: C-log in three batches.} The same budget is spent in three batches of 100. Before each batch, the screener is refitted and the deficit recomputed on the solved FOS of the seed and of every earlier batch, and a new pool is drawn. Each batch thus corrects where the earlier ones actually landed. Only the retraining distinguishes C-seq from C-log.
  \item \textbf{C-logG: C-log with another screener.} The random forest is replaced by a gradient-boosting screener, of the same family as the gradient-boosting learner. It uses the grid point that cross-validation on the seed selects for the gradient-boosting learner (\cref{sec:tuning}). The candidate pool, its random stream and the intended allocation are exactly those of C-log, so only the screener differs, and with it the candidates available to the within-bin draw. C-logG is a sensitivity check on the screener, not a fourth variant of the rule. Gradient boosting was chosen post hoc, after two diagnostics. The first compared the two screeners on the points the random-forest screener had selected for C-log, and is therefore conditional on that selection. The second compared them on 1\,800 solved LHS points that neither screener had selected: the natural test set (\cref{sec:evaluation}) and the five replication LHS sets. The natural test set was thus read once outside its role, to compare screeners; no surrogate was fitted, tuned or selected on that reading.
\end{itemize}
C is the original design. C-log and C-seq were designed after C's results, and were declared in the configuration before any of their points was solved. The bins, the allocation rule and the within-bin rule are stated once, with the reasons above, and were not tuned.

\paragraph{Budget.} Every budget arm spends the same number of solver analyses, 600 (\cref{tab:budget}). Only arm A, the seed alone, spends 300. The screener is trained on the seed, which every budget arm already contains. Predicting the FOS of the candidate pool costs no analyses. Response-balanced sampling therefore spends no analyses beyond those counted in the table.

\begin{table}[htbp]
\centering
\small
\setlength{\tabcolsep}{4pt}
\caption{The arms. The seed is the initial Latin hypercube of 300 analyses. Screened points are selected among 200\,000 candidates by the screener's predicted FOS, in 15 bins on the selection range [0.5, 6.0]; the screener is a random forest trained on the seed (C-seq: refitted on the seed and its earlier batches; C-logG: gradient boosting). The budget arms B, C, C-log, C-seq and C-logG spend the same 600 analyses; A is the seed alone. The pool arms C-oracle and C-logP draw C-log's allocation from one sub-pool of the solved reference pool per realization: their points were solved before they were selected, so their count (†) is not a budget result. They are a diagnostic reference, not a sampling strategy. C$_0$, C$_0$-log, C$_0$-seq, C$_0$-logG, C$_0$-oracle and C$_0$-logP are ablations: they train on the added points without the seed, carry no budget and are not like-for-like comparisons with the budget arms. r0 is the existing sets and is exploratory; r1–r5 are new realizations of the added points on the fixed seed, so A, the seed alone, has none.}
\label{tab:budget}
\begin{tabular}{llllr}
\toprule
arm & training data & added points chosen by & realizations & analyses \\
\midrule
\multicolumn{5}{l}{\textit{Budget arms}} \\
A & seed only & \parbox[t]{4.6cm}{\raggedright --\strut} & r0 & 300 \\
B & seed + 300 LHS & \parbox[t]{4.6cm}{\raggedright Latin hypercube\strut} & r0, r1–r5 & 600 \\
C & seed + 300 screened & \parbox[t]{4.6cm}{\raggedright equal counts in linear FOS bins, max-min within bin\strut} & r0 & 600 \\
C-log & seed + 300 screened & \parbox[t]{4.6cm}{\raggedright log bins filled against the seed's deficit, random within bin\strut} & r0, r1–r5 & 600 \\
C-seq & seed + 300 screened & \parbox[t]{4.6cm}{\raggedright as C-log, in 3 batches of 100, screener refitted before each\strut} & r0, r1–r5 & 600 \\
C-logG & seed + 300 screened & \parbox[t]{4.6cm}{\raggedright as C-log, gradient-boosting screener\strut} & r0, r1–r5 & 600 \\
\addlinespace
\multicolumn{5}{l}{\textit{Pool arms}} \\
C-oracle & seed + 300 from the pool & \parbox[t]{4.6cm}{\raggedright C-log's allocation from a solved sub-pool, binned on solved FOS\strut} & r1–r5 & 600† \\
C-logP & seed + 300 from the pool & \parbox[t]{4.6cm}{\raggedright as C-oracle, binned on the screener's predicted FOS\strut} & r1–r5 & 600† \\
\addlinespace
\multicolumn{5}{l}{\textit{Ablations}} \\
C$_0$ & 300 screened (no seed) & \parbox[t]{4.6cm}{\raggedright as C\strut} & r0 & -- \\
C$_0$-log & 300 screened (no seed) & \parbox[t]{4.6cm}{\raggedright as C-log\strut} & r0, r1–r5 & -- \\
C$_0$-seq & 300 screened (no seed) & \parbox[t]{4.6cm}{\raggedright as C-seq\strut} & r0, r1–r5 & -- \\
C$_0$-logG & 300 screened (no seed) & \parbox[t]{4.6cm}{\raggedright as C-logG\strut} & r0, r1–r5 & -- \\
C$_0$-oracle & 300 from the pool (no seed) & \parbox[t]{4.6cm}{\raggedright as C-oracle\strut} & r1–r5 & -- \\
C$_0$-logP & 300 from the pool (no seed) & \parbox[t]{4.6cm}{\raggedright as C-logP\strut} & r1–r5 & -- \\
\bottomrule
\end{tabular}
\end{table}

\paragraph{Ablations.} Each guided arm has an ablation, marked C$_0$, that trains on the added points alone, without the seed: C$_0$, C$_0$-log, C$_0$-seq and C$_0$-logG. The pool arms' ablations, C$_0$-oracle and C$_0$-logP, are introduced in \cref{sec:oracle}. The ablations separate the contribution of the guided points from that of the seed. They are not budget arms. The guided points cannot be selected without the screener, and hence without the seed, so an ablation's training set costs 600 analyses while containing 300. Every ablation is reported in every analysis.

\subsection{The reference pool and the oracle}\label{sec:oracle}

With a real screener, the training density a guided arm achieves departs from the density it intended. A candidate whose solved FOS falls outside the bin it was drawn for is \emph{misbinned}; the oracle is a diagnostic reference with no misbinning. It uses the same allocation and the same within-bin rule as C-log, but it bins candidates on their solved FOS. It needs a pool of candidates that are already solved, and that pool costs far more analyses than the budget. The oracle is therefore not a sampling strategy, and nothing about its cost is a budget result.

\paragraph{The reference pool.} The pool is a design of up to 50\,000 points, drawn once, independently and uniformly over the domain. Points within $10^{-3}$ of the seed or of either test set are removed; none was. Row $i$ of the design belongs to sub-pool $(i \bmod 5) + 1$, independently of its FOS. The result is five disjoint sub-pools, one for each replicated realization (\cref{sec:replications}). The pool is solved in chunks of 1\,000 rows (200 per sub-pool), in design order, with the generator, checks and repairs of \cref{sec:data}.

\paragraph{Stopping rule.} The rule, its cap and the allocation it reads were declared and frozen before the first chunk was solved (tag \texttt{oracle-declared} in the code repository). The allocation is C-log's, $a$: the seed's solved FOS water-filled to 300 points over the 15 geometric bins (\cref{sec:arms}). The seed is fixed, so $a$ is the same in every realization. After each chunk, the rule is met when, in every sub-pool and every bin $b$:
\begin{itemize}
  \item at least $a_b$ rows have their solved FOS in $b$, so the oracle can draw its points;
  \item at least $a_b$ rows have their FOS, as predicted by the seed-trained random-forest screener, in $b$, so the screened contrast arm below can draw its points.
\end{itemize}
Solving stops for good at the first chunk that meets the rule. Reaching the cap without meeting it would have been an error, not a reason to relax the rule. The rule reads only counts of pool FOS. It reads no test set, no fit and no error, so it cannot favour an arm.

\paragraph{Where the pool stopped.} The rule was met at chunk 22: 22\,000 analyses, 4\,400 per sub-pool, none excluded. The solved side of the rule had been met at chunk 15. The predicted side decided the stop, in the lowest bin, FOS $\in [0.5, 0.59]$, where the seed-trained random forest predicted a FOS in that bin less often than one occurred there. This is contrary to the expectation declared with the rule, that the predicted side would not bind. The first 15 chunks are what the oracle needs. The seven chunks after them, 7\,000 analyses, are the price of drawing the screened contrast arm from the same solved pool, which is what makes the paired contrast below possible.

\paragraph{The pool arms.} In each realization $r$, two arms draw 300 points from the rows of sub-pool $r$, with C-log's allocation $a$ and C-log's random draw within bins. They differ only in the value used to place a row in a bin:
\begin{itemize}
  \item \textbf{C-oracle} bins on the solved FOS. It meets the intended allocation exactly by construction: every point lies in the bin it was drawn for.
  \item \textbf{C-logP} bins on the FOS predicted by the seed-trained random-forest screener, the one C-log uses. It is C-log drawn from the solved pool instead of from 200\,000 unsolved candidates.
\end{itemize}
Each arm trains on the seed and its 300 points, and is paired with an ablation that trains on the 300 points alone, C$_0$-oracle and C$_0$-logP. Since C-oracle meets the intended allocation, C-logP's departure from it is misbinning, by definition. What the pool arms measure is the effect of that departure on accuracy and on relocation, through the paired contrast C-oracle versus C-logP. The two arms share the sub-pool, its eligible rows, the seed, the allocation and the within-bin rule. Against C-log, both also differ in the candidate pool: a sub-pool of 4\,400 solved rows instead of 200\,000 unsolved candidates, from which a point that fails to solve can never be drawn. This difference is checked with a declared sanity check (\cref{sec:statistics}).

\paragraph{Declared readings.} The predictions for the pool arms, and the readings that turn their intervals into statements, were declared and committed with the stopping rule, before any pool analysis was solved. They are given with the statistics in \cref{sec:statistics}. Every statement about the pool arms holds \emph{in this design}. That qualifier fixes:
\begin{itemize}
  \item the seed-trained random-forest screener;
  \item C-log's allocation and its random draw within bins;
  \item this reference pool, solved on one platform;
  \item the seed, the test sets and the common hyper-parameters of the replications.
\end{itemize}

\subsection{Replicated realizations}\label{sec:replications}

A single draw of the added points could make an arm look better or worse by chance. Each strategy's added points are therefore drawn again in five new realizations, $r_1$ to $r_5$. The design was declared before any replication was solved. The arms added later, C-logG and the pool arms, were declared before any of their points was generated.

\paragraph{What a realization varies.} Only the points each strategy adds to the seed. Each realization has its own random stream, keyed on the strategy, the realization and the batch.
\begin{itemize}
  \item \textbf{B:} a new LHS of 300 points.
  \item \textbf{C-log:} a new pool of 200\,000 candidates and new draws within bins.
  \item \textbf{C-seq:} the same, batch by batch. The screeners of the second and third batches also differ between realizations, because they learn from that realization's earlier batches.
  \item \textbf{C-logG:} the candidate pool and the draw stream of the same realization of C-log, with the gradient-boosting screener.
  \item \textbf{C-oracle and C-logP:} sub-pool $r$ of the reference pool (\cref{sec:oracle}), each with its own draw stream.
\end{itemize}
Arm A adds no points and is the same in every realization. C, the linear-bin design, is not replicated, and neither is its ablation C$_0$. Every other ablation is replicated with its arm: C$_0$-log, C$_0$-seq, C$_0$-logG, C$_0$-oracle and C$_0$-logP.

\paragraph{What is fixed.} Realizations vary the added points only; the seed, and with it the screener, the allocation and the hyper-parameters, is common to all of them. In detail, every realization shares:
\begin{itemize}
  \item the seed's 300 points, in every arm that contains the seed;
  \item the seed-trained screeners: C-log's random forest is the same fitted function in all five realizations, and so are C-seq's first-batch screener and C-logG's gradient-boosting screener;
  \item the allocation, computed from the seed's deficit and identical in every realization;
  \item the hyper-parameters, common to every arm and chosen by cross-validation on the seed alone (\cref{sec:tuning});
  \item both test sets.
\end{itemize}
Every conclusion drawn from the replications is conditional on the seed, the test sets and the common hyper-parameters. The intervals computed over the realizations cover the variability between designs under this seed, not the variability between seeds.

\paragraph{The existing sets, $r_0$.} The sets used by the main, tuned fits are refitted in the same way, with the common hyper-parameters, as realization $r_0$. They shaped the guided variants and the declared predictions. $r_0$ is therefore exploratory: it is reported apart and enters no interval. The pool arms have no $r_0$.

\subsection{Learners and tuning protocol}\label{sec:tuning}

\paragraph{Learners.} Two tree-ensemble families serve as surrogates: a random forest (RF) of 500 trees \citep{Breiman2001} and a histogram-based gradient boosting machine (GBM) \citep{Friedman2001, Ke2017}, both as implemented in scikit-learn \citep{Pedregosa2011}. Each tree of the random forest is grown on a bootstrap sample of the training set, and the forest averages their predictions. For trees with variance $\sigma^2$ and pairwise correlation $\rho$, the variance of the average of $B$ trees is $\rho\sigma^2 + (1-\rho)\sigma^2/B$, so averaging reduces variance mainly as far as the trees are decorrelated \citep[ch.~15]{Hastie2009}. The fraction of features considered at each split is tuned (\cref{tab:grids}); the grid selects all features on every arm, so the trees are decorrelated by bootstrap resampling alone. Both learn $\log(\mathrm{FOS})$ from the features the screener uses, the five parameters and $\gamma H/c$. Predictions are exponentiated back to the FOS only where a metric is computed on the FOS scale. The log target matches the multiplicative character of the FOS: a given absolute error matters more near $\mathrm{FOS} = 1$ than in the stable tail. The learner is the study's measuring instrument, not its object. Every comparison is made between arms under the same learner, never between learners on the same arm. Two families are used so that each pattern can be read with two independently tuned instruments; a pattern found under both is consistent across two tree-ensemble families, and nothing more general is claimed.

\paragraph{Tuning protocol.} Each family has one fixed grid (\cref{tab:grids}), identical for every arm. The hyper-parameters of an arm are selected by five-fold cross-validation, with shuffled folds, on that arm's training data alone; no test set is read during selection. The selection criterion is the RMSE on the log target, consistent with the trained response and with the multiplicative intervals of \cref{sec:evaluation}. The selected configuration is then refitted on the arm's whole training set. As a sensitivity check, every arm is also tuned and fitted under the RMSE on the FOS scale. The selected hyper-parameters of every arm, ablations included, are reported under both criteria (\cref{tab:scorings:params}).

\begin{table}[htbp]
\centering
\caption{Tuning grids, one per family, identical for every arm. Settings not listed are fixed: 500 trees for the random forest, early stopping disabled for gradient boosting (so the number of iterations is set by the grid), and scikit-learn defaults otherwise.}
\label{tab:grids}
\begin{tabular}{lll}
\toprule
Family & Hyper-parameter & Values \\
\midrule
RF  & share of features per split (\texttt{max\_features}) & 0.3, 0.5, 1.0 \\
    & minimum samples per leaf (\texttt{min\_samples\_leaf}) & 1, 2, 5 \\
\addlinespace
GBM & learning rate (\texttt{learning\_rate}) & 0.03, 0.06, 0.10 \\
    & leaves per tree (\texttt{max\_leaf\_nodes}) & 2, 3, 4, 7 \\
    & minimum samples per leaf (\texttt{min\_samples\_leaf}) & 1, 2, 3, 5 \\
    & iterations (\texttt{max\_iter}) & 2\,000, 4\,000, 8\,000 \\
\bottomrule
\end{tabular}
\end{table}

\paragraph{Why the grid is not the library's defaults.} Library defaults are not neutral here, because the arms differ in size. The scikit-learn default for gradient boosting, 100 iterations with at least 20 samples per leaf, constrains a model trained on 300 points, as arm A and the ablations are, far more than one trained on 600. The GBM grid was extended before C-log was declared, using training-fold cross-validation on the arms that existed then (A, B, C and C$_0$), until the selected values lay inside it or on a plateau (\texttt{max\_iter}). One exception was declared: on C$_0$ under the FOS-scale criterion, a minimum leaf size of 10, outside the grid, scored about 5\% better, but in only two of the five folds, less than the spread between folds; the grid was not extended for it. The grid was then kept fixed, so that every arm is tuned on the same grid. For some later arms the selected learning rate or minimum leaf size lies on an edge of the grid (\cref{tab:scorings:params}). For the random forest, the grid selects all features and leaves of one sample on every arm under the log criterion, which are the library's defaults. These tuned fits enter only the single-realization comparisons; the replicated results, on which the conclusions rest, use common hyper-parameters fixed by cross-validation on the seed.

\paragraph{Common hyper-parameters.} The replications are not tuned per realization (\cref{sec:replications}). For RF the common configuration is 500 trees with the library's defaults, which is what the grid selects on every arm under the log criterion. For GBM it is the grid point that cross-validation selects on the seed alone: a learning rate of 0.06, 8\,000 iterations, 3 leaves per tree and at least 3 samples per leaf. The configuration is selected on the seed alone, before any point is added, so it is not selected on the points of any guided or LHS extension. The seed is part of every budget arm's training set, and is arm A's whole training set.

A learning curve for LHS and C (the linear-bin design), with fixed hyper-parameters (library defaults for RF, the seed-selected grid point for GBM), is given in the Supplementary material (\cref{sec:s-learning-curve}). It does not cover C-log or the pool arms, and it is not used to extrapolate the relocation to other budgets.

\subsection{Evaluation}\label{sec:evaluation}

\paragraph{Two test sets, two deployment scenarios.} Every arm is evaluated on the same two test sets, which stand for two ways a surrogate may be used, not for two tests of the same thing.
\begin{itemize}
  \item \textbf{Test-natural}, 300 points from an independent LHS over the domain. It weights the FOS as uniform use of the parameter domain does, and answers the question of average accuracy over the domain.
  \item \textbf{Test-stratified}, 150 points spread evenly over the predicted FOS. An independent pool of 200\,000 candidates, drawn uniformly over the domain, is screened by the seed-trained random forest; the selection range is cut into 15 bins of equal width, and 10 candidates are drawn at random in each. It weights every safety level of the selection range alike. Its points are placed by the seed-trained random forest that also screens C, C-log and the first batch of C-seq; their FOS, like every FOS in the study, comes from the solver, and may fall outside the selection range. A test set stratified on the FOS has a precedent in \citet{Azmoon2021}, there on the solved FOS rather than a predicted one.
\end{itemize}
Both test sets are disjoint from every training set and from each other (\cref{sec:arms}), and the check is repeated before every fit and every evaluation. Cross-validation scores serve only for tuning: each is computed on its own arm's training distribution, so they are not comparable across arms.

\paragraph{Zones.} Errors are read along the FOS axis in five reporting zones, half-open: FOS $< 0.5$, $[0.5, 1)$, $[1, 2)$, $[2, 6)$ and $\geq 6$. The edges 0.5 and 6 are the ends of the selection range, 1 is limiting equilibrium, and 2 mirrors 0.5 around 1 in log space. Below 1 the slope is unstable, between 1 and 2 design decisions are taken, and above 2 the reserve is comfortable. No normative threshold, such as 1.3 or 1.5, is an edge, so no zone reads as a code requirement. The zones were fixed after the C-log results and before C-seq was run, together with the decision range below, replacing an earlier set with edges at 1, 1.3, 3 and 6; this is a reporting choice, and no design or fit depends on it.

\paragraph{The decision range.} The primary criterion is the log-RMSE on the decision range, FOS $\in [0.5, 2)$, against arm B. The range joins the two zones around equilibrium and is symmetric around 1 in log space. It was defined after the C-log results and before C-seq was run, from the lower end of the selection range and its mirror around 1 in log space. Both of its halves, $[0.5, 1)$ and $[1, 2)$, are reported, and so are the whole test set and the selection range.

\paragraph{Metrics.} The log scale is primary and the FOS scale secondary. On the log scale an error is $\ln\widehat{\mathrm{FOS}} - \ln\mathrm{FOS}$, a relative error, consistent with the trained response; the log-RMSE is the RMSE of these errors. Each arm is scored, on each test set, by the RMSE and log-RMSE over the whole test set, the selection range and the decision range, and within each zone; by the worst-zone (minimax) RMSE on each scale; and by $R^2$ on the FOS scale. No classification metric is used. This is a regression study: an engineer treats a FOS of 0.9 and one of 1.1 as effectively the same, and a count of predictions on the wrong side of $\mathrm{FOS} = 1$ would penalize an error of 0.05 like one of 0.5. The zone errors, the worst zone and the prediction intervals in the low zones carry the engineering reading instead.

\paragraph{Prediction intervals.} Each tuned fit carries a cross-conformal-style prediction interval \citep{Vovk2015}. The conformity score is the absolute residual on the log scale, $|\ln\mathrm{FOS} - \ln\widehat{\mathrm{FOS}}|$. The scores are out-of-fold residuals from a nested cross-validation on the arm's training data: each held-out fold is predicted by a model whose hyper-parameters were selected by an inner search that did not see that fold. With $n$ scores and $\alpha = 0.10$, the quantile $q$ is the $k$-th smallest score, $k = \min(\lceil (n+1)(1-\alpha) \rceil, n)$. The interval around the prediction of the refitted model is multiplicative,
\[
  \bigl[\,\widehat{\mathrm{FOS}}\,e^{-q},\ \widehat{\mathrm{FOS}}\,e^{q}\,\bigr],
\]
so its width is a fixed proportion of the prediction. This suits the FOS, which spans two orders of magnitude: an additive interval would be implausibly wide near $\mathrm{FOS} = 1$ and implausibly narrow in the stable tail. The procedure carries no formal coverage guarantee. The scores come from several fold models, while the test predictions come from one refitted model; a CV+ construction would earn a guarantee \citep{Barber2021}. Nor are the training and test points exchangeable for the guided arms, whose training points are not drawn from the test distribution, or on the stratified test set; exchangeability is the premise of every such guarantee \citep{Tibshirani2019, Barber2023}. Coverage is therefore reported empirically, over each test set and by zone, together with the mean width. The intervals are computed for the single-realization fits only.

\subsection{Statistical comparisons}\label{sec:statistics}

Two kinds of comparison are made. On the single realization of the tuned fits, arms are compared point by point on the test sets. Across the replicated realizations, they are compared through intervals on per-realization metrics. The conclusions rest on the replications.

\paragraph{Single-realization comparisons.} All arms are evaluated on the same test points, so every comparison is paired by test point and made under the same learner and selection criterion. For a pair of arms, the difference $d_i = |e_{1,i}| - |e_{2,i}|$ of absolute errors on test point $i$ is tested with the Wilcoxon signed-rank test \citep{Wilcoxon1945, Demsar2006}. The median of $d$, the share of points with $d < 0$ and the $p$-value are reported. $d < 0$ means that the first arm has the smaller absolute error on that point; the test is read as a statement about absolute errors, not about RMSE. Only pairs declared in the configuration are tested, each with the alternative declared there. The hypothesized entries, C and C-log against A and against B, are one-sided, with the alternative that the guided arm has smaller absolute errors. The one-sided direction was fixed with the evaluation code, before any test result was computed, for the FOS scale on the whole test set and on the selection range. The C-log entries were declared with the same direction before the C-log set was solved, after the results of C were known. The log scale and the decision range were added after the C-log results; their one-sided $p$-values are post hoc and are reported as such. Every other entry is two-sided: those involving an ablation, C-log against C, and every entry of C-seq and C-logG. Each comparison is run on both scales and on three subsets: the whole test set, the selection range and the decision range. Beside each test, the difference in RMSE, $\Delta\mathrm{RMSE} = \mathrm{RMSE}_1 - \mathrm{RMSE}_2$, is given with a 95\% percentile interval from 10\,000 paired bootstrap resamples of the test points, drawn jointly for both arms \citep{EfronTibshirani1993}. This interval covers test-sampling variability only, not the variability of the selected points or of the seed. No family-wise correction is applied: no claim rests on a single $p$-value, and the claims rest on the declared rules over the replications below. The pool arms have no single realization, and no test.

\paragraph{Replication intervals.} In each realization $r_1$ to $r_5$, each arm's log-RMSE is computed on each test set, subset and zone. Two arms are compared through the difference of their means over the realizations, with a 95\% $t$-interval.
\begin{itemize}
  \item Arms that share no draw within a realization get a Welch interval \citep{Welch1947}: every arm against B, C-seq against C-log, and C-logP against C-log.
  \item Arms that share a realization's added points get a paired interval, with four degrees of freedom: each guided or pool arm against its ablation, C-logG against C-log, which share the candidate pool and its draw stream, and C-oracle against C-logP, which share the sub-pool.
\end{itemize}
Each difference is computed on the log-RMSE and, as a relative scale, on its logarithm. Beside each interval, $\hat{P}(\text{first} < \text{second})$ gives the share of realization pairs in which the first arm has the lower value, ties counting one half: over the 25 pairs of realizations for a Welch comparison, over the five realizations for a paired one. It is descriptive only, since the pairs are not independent. Against B, the direction of a guided arm is consistent when its mean difference has the same sign in all four cells; only an interval entirely below 0 in all four cells would read as a gain. An interval containing 0 is inconclusive; it is never read as a tie or as evidence of equivalence. The intervals cover the variability between designs under this seed. The test sets are the same in every realization, so their sampling error is common to all realizations: it neither widens the spread between them nor averages out over them. It is covered separately by the paired bootstrap above, and the two are never combined. $r_0$ enters no interval.

\paragraph{Learner interaction.} Whether the effect of a sampling strategy depends on the learner is read from one quantity. For each arm and realization, $\ln \mathrm{logRMSE}_{\mathrm{RF}} - \ln \mathrm{logRMSE}_{\mathrm{GBM}}$ measures the gap between the two families; the interval of its difference between two arms is the interaction interval. It is a Welch interval for C-log, C-seq and C-logG against B, and a paired one for C-logG against C-log and for C-oracle against C-logP. A learner interaction is claimed only where this interval excludes 0, never from comparing the separate intervals of two arms. It is read on the decision range, the primary criterion; the interaction intervals on every subset and zone are given in the Supplementary material (\cref{tab:rep:interactions:test-natural,tab:rep:interactions:test-stratified}).

\paragraph{Relocation.} A cell is a learner and a test set, four cells in all. For an arm $X$, with $E$ the log-RMSE on a zone,
\[
  L(X) = \ln\frac{E_{X,[0.5,1)}}{E_{B,[0.5,1)}}, \qquad H(X) = \ln\frac{E_{X,[1,2)}}{E_{B,[1,2)}}, \qquad R(X) = H(X) - L(X).
\]
Each is read as its median over $r_1$ to $r_5$ within a cell. The relocation gate for $X$ passes when the median $L$ is below 0 and the median $H$ above 0 in at least three of the four cells: $X$ gains in $[0.5, 1)$ and loses in $[1, 2)$ against B. The margin is $m = \ln 1.10$. The same gate and margin serve the readings declared for C-logG and for the pool arms.

\paragraph{Declared readings of the pool arms.} The predictions and readings below were declared with the stopping rule, before any pool analysis was solved (\cref{sec:oracle}), and cover $r_1$ to $r_5$ only.

\emph{Stage 1, on training data only, before any fit.}
\begin{itemize}
  \item A construction check on C-oracle: its achieved allocation equals the intended one, and it has no point below 0.5 or above 6.
  \item A sanity check on the size of the candidate pool: on $\sum_b |\text{achieved}_b - \text{intended}_b|$, the Welch 95\% interval of C-logP $-$ C-log contains 0, and the mean difference is at most 15 points in absolute value.
  \item The density ratio $\rho_Z$, the median over realizations of the number of training points, seed and selection together, of C-oracle over that of C-logP in zone $Z$, for $L = [0.5, 1)$ and $H = [1, 2)$. It fixes the direction of the stage-2 density prediction before any fit.
\end{itemize}

\emph{Stage 2, on the test sets.} The contrast is $dR_O = R(\text{C-oracle}) - R(\text{C-logP})$, in which B cancels exactly, and, for each zone, $dZ_O = \ln(E_{\text{C-oracle},Z} / E_{\text{C-logP},Z})$. Each is read as its median per cell, beside a paired 95\% interval per cell.
\begin{itemize}
  \item \emph{Step P.} C-logP's gate is reported first. If it fails, no reading of $dR_O$ is described as a change in relocation.
  \item \emph{Density prediction.} $\rho_Z \geq 1.10$ predicts $dZ_O < 0$, the oracle being more accurate where it has more data; $\rho_Z \leq 1/1.10$ predicts $dZ_O > 0$; otherwise no direction is predicted for $Z$. The direction is met when the predicted sign holds in at least three cells, and the magnitude when, in addition, $|{\rm median}\ dZ_O| \geq m$ in at least three cells.
  \item \emph{Levels against B}, on the decision-range log-RMSE, with Welch intervals: level 1, the mean difference is below 0 in at least one cell; level 2, the interval lies entirely below 0 in at least one cell; level 3, it lies entirely below 0 in all four cells.
\end{itemize}
The same levels were declared for C-logG, before it was run, and are read for it in the same way.
The readings are fixed in \cref{tab:readings}. Groups A and B depend on C-oracle's gate, and within each the first matching row applies; group C is read independently, also in order. Five realizations were kept on purpose, for the resolution of every other replicated arm. It was declared with the readings that, at this $n$, the equivalence readings A3 and B2 are practically unreachable, and that ``no difference detected'' or a mixed reading is the expected outcome of a null or small effect.

\begin{table}[htbp]
\centering
\small
\caption{Readings of the pool arms, declared before any pool analysis was solved. Median and interval are per cell, over $r_1$ to $r_5$; ``in $\geq 3/4$'' means in at least three of the four cells; $m = \ln 1.10$. ``In this design'' fixes the seed-trained random-forest screener, C-log's allocation and its random draw within bins, this reference pool, and the seed, test sets and common hyper-parameters. The rows are condensed; the full declared text is in the declaration, reproduced in the Supplementary material (\cref{sec:s-oracle}).}
\label{tab:readings}
\begin{tabular}{@{}lp{0.36\linewidth}p{0.5\linewidth}@{}}
\toprule
& Condition & Reading \\
\midrule
\multicolumn{3}{@{}l}{\emph{A: C-oracle's gate passes}} \\
A1 & median $dR_O \geq m$ in $\geq 3/4$ & relocation is larger with known membership, as density-following predicts when $\rho_L \geq 1.10$ \\
A2 & median $dR_O \leq -m$ in $\geq 3/4$ & relocation is smaller with known membership, against density-following for this contrast \\
A3 & paired $dR_O$ interval inside $(-m, m)$ in $\geq 3/4$ & relocation is practically unchanged with known membership; the only equivalence reading \\
A4 & $|\text{median } dR_O| < m$ in $\geq 3/4$, A3 not met & no difference detected \\
A5 & otherwise & mixed / inconclusive on $dR_O$ \\
\addlinespace
\multicolumn{3}{@{}l}{\emph{B: C-oracle's gate fails}} \\
B1 & C-logP's gate fails too & relocation not detected on this pool for either arm \\
B2 & paired interval of $R(\text{C-oracle})$ inside $(-m, m)$ in $\geq 3/4$ & relocation vanishes with known membership \\
B3 & median $R(\text{C-oracle}) \leq -m$ in $\geq 3/4$ & relocation reverses with known membership \\
B4 & $|\text{median } R(\text{C-oracle})| < m$ in $\geq 3/4$, B2 not met & relocation not detected with known membership \\
B5 & otherwise & mixed / inconclusive \\
\addlinespace
\multicolumn{3}{@{}l}{\emph{C: the decision range}} \\
C1 & C-oracle at level 3, the paired C-oracle $-$ C-logP interval entirely below 0 in all four cells, and C-logP below level 3 & C-oracle has lower decision-range log-RMSE than B in all four cells, in this design; the difference in decision-range accuracy between C-oracle and C-logP is attributable to the screener's error, in this design \\
C2 & C-oracle at level 3, C1 not met & C-oracle has lower decision-range log-RMSE than B in all four cells, in this design; not a budget result \\
C3 & C-oracle below level 3 & ``No demonstrated gain'' stands, now also with known bin membership, in this design \\
\bottomrule
\end{tabular}
\end{table}

\paragraph{Wording about the screener.} One attribution is permitted, and only for the paired contrast of C-oracle against C-logP: ``attributable to the screener's error, in this design''. It is written only where the paired interval excludes 0 in at least three of the four cells, on the side of the reading: entirely above 0 for A1, entirely below 0 for A2, B2 and B3. Where the interval contains 0, neither ``attributable'' nor ``not attributable'' is written. C-logP's departure from the intended allocation is misbinning by construction (\cref{sec:oracle}), and is not presented as an empirical attribution. C-oracle against B is not read as a statement about the screener, except through C1.

\paragraph{Achieved against intended density.} A descriptive analysis sets, bin by bin over the 15 bins of the selection range and over $r_1$ to $r_5$, the change in each guided arm's training density against B beside its change in log-RMSE, and, for the screened arms, the achieved allocation against the intended one. The difference between achieved and intended counts is the net distortion of the requested distribution, not a prediction error of the screener. The procedure was committed before any code or output of the analysis was committed, on replication results already computed; it is not a preregistration. It fixes the bins, a rule that marks, and never drops, bins with fewer than 10 test points, and a single summary: the number of unmarked bins in which the two changes have opposite signs. It introduces no test. Its pattern is read only as consistent with density-driven relocation, never causally, and it is never cited as evidence about the screener. The procedure and the limit of its blinding are given in the Supplementary material (\cref{sec:s-structural-procedure}), and the ablation rows in \cref{sec:s-structural-ablation}.

\section{Results}\label{sec:results}

Where four values are listed for the four cells of learner and test set without naming them, they are in the order GBM on the natural test set, GBM on the stratified one, RF on the natural one, RF on the stratified one. A difference against B is the arm's log-RMSE minus B's, with its 95\% interval; a negative difference is a lower log-RMSE than B's.

\subsection{The decision range against B}\label{sec:res-decision}

\begin{figure}[htbp]
\centering
\includegraphics[width=\linewidth]{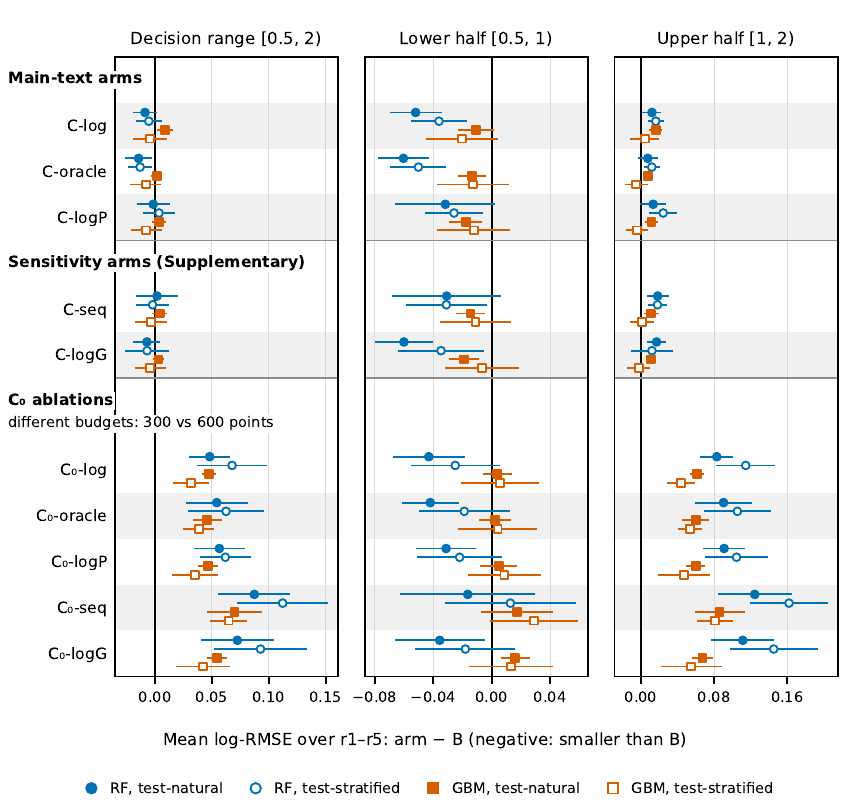}
\caption[The primary criterion against B]{Primary criterion: log-RMSE against B on the decision range, FOS in [0.5, 2), and on its two halves, [0.5, 1) and [1, 2). Each point is the mean over the confirmatory realizations r1-r5 of the arm's log-RMSE minus B's, with its 95\% Welch interval; a negative difference means a smaller log-RMSE than B. Each row carries the four learner-by-test cells: circles RF, squares GBM; filled markers test-natural, hollow markers test-stratified. The rows share one x axis within each column; the columns have their own. The intervals cover variability between designs, conditional on the fixed seed, the fixed test sets and the common hyper-parameters fixed from the seed; they constitute no test. An interval containing 0 is inconclusive, never a tie or evidence of equivalence. C-seq and C-logG are sensitivity arms, reported in Supplementary. The C$_0$ rows are ablations: they train on the 300 guided points without the seed, against the 600 of B, so they join different budgets and are not a like-for-like comparison. A is absent: it has no replications, only its existing set (r0), and therefore no interval; C and the original C$_0$ are absent for the same reason, as they were never replicated. r0 is exploratory and enters no interval, so it is not drawn; C-oracle, C-logP and their C$_0$ ablations have no r0. Zones are half-open, [lo, hi), as everywhere the replication metrics are read.}
\label{fig:decision-forest}
\end{figure}

\Cref{fig:decision-forest} gives the primary criterion, the log-RMSE on the decision range, FOS $\in [0.5, 2)$, of every arm against B, over the realizations $r_1$ to $r_5$, in the four cells of learner and test set. B's mean decision-range log-RMSE is 0.049 and 0.053 for GBM on the natural and the stratified test set, and 0.104 and 0.101 for RF.

No arm that selects with the screener shows a demonstrated gain on this criterion. C-log's mean difference from B is below 0 in three cells, between $-0.009$ and $-0.004$, and each of these intervals contains 0. In the fourth cell, GBM on the natural test set, C-log's log-RMSE is higher than B's, by 0.009 [0.002, 0.016]. No other interval of a screened arm against B excludes 0. C-logP, which applies C-log's screener and allocation to a sub-pool of the solved reference pool, has its mean below 0 in two cells and every interval containing 0: level~1. C-seq and C-logG, reported in \cref{sec:res-sensitivity}, likewise have every interval containing 0.

The only intervals entirely below 0 are C-oracle's in the two RF cells: $-0.015$ [$-0.027$, $-0.003$] on the natural test set and $-0.013$ [$-0.024$, $-0.003$] on the stratified one. For GBM its intervals contain 0, so C-oracle reaches level~2, not level~3. C-oracle bins on the solved FOS and needs the solved pool, so this is not a budget result; its reading is given in \cref{sec:res-oracle}.

Every ablation has a higher decision-range log-RMSE than B in every cell, with each interval entirely above 0, from 0.031 (C$_0$-log, GBM, stratified) to 0.112 (C$_0$-seq, RF, stratified). The ablations train on 300 points against B's 600, so this is not a comparison at equal budget; the contribution of the seed is read in \cref{sec:res-ablations}.

For C-log, the decision-range difference is the net of two changes of opposite sign in its halves, a gain on $[0.5, 1)$ and a loss on $[1, 2)$, in all four cells. The halves of the other arms are taken up in \cref{sec:res-relocation}.

The contrast between C-log and B depends on the learner on the natural test set. The interaction is read on the ratio of the two families' decision-range log-RMSE, RF over GBM. There this ratio is 0.78 [0.70, 0.86] times as large for C-log as for B, about 22\% smaller, so the contrast is more favourable to C-log under RF than under GBM. On the stratified test set the interval contains 1.

\subsection{Relocation along the FOS axis}\label{sec:res-relocation}

The two halves of the decision range are the second and third columns of \cref{fig:decision-forest}. Against B, C-log's log-RMSE is lower on $[0.5, 1)$ and higher on $[1, 2)$ in each of the four cells, so the direction is consistent in all eight combinations of cell and half. On $[0.5, 1)$ the mean difference is $-0.052$ [$-0.070$, $-0.034$] and $-0.036$ [$-0.055$, $-0.017$] for RF on the natural and the stratified test set, where B's log-RMSE is 0.150 and 0.138. For GBM it is $-0.011$ and $-0.020$, against B's 0.062 and 0.060, and both intervals contain 0. On $[1, 2)$ the difference is $+0.016$ [$+0.009$, $+0.023$] and $+0.004$ [$-0.011$, $+0.020$] for GBM, against B's 0.044 and 0.048, and $+0.012$ [$+0.002$, $+0.022$] and $+0.016$ [$+0.007$, $+0.025$] for RF, against B's 0.082 and 0.076. Five of the eight intervals exclude 0. Under RF the gain on $[0.5, 1)$ is resolved on both test sets; under GBM neither of its intervals excludes 0.

The sequential variant C-seq, reported in \cref{sec:res-sensitivity}, has the same direction in all eight of its combinations. C-log and C-seq together have the same direction in 16 of 16 combinations, and 10 of the 16 intervals exclude 0.

The relocation gate passes for both pool arms, C-oracle and C-logP, in three of the four cells. Both fail it in the same cell, GBM on the stratified test set, where their median $L$ and $H$, the log ratios of their zone log-RMSE to B's (\cref{sec:statistics}), are both below 0: $-0.206$ and $-0.274$ for C-oracle, $-0.010$ and $-0.156$ for C-logP. What the two pool arms show together is read in \cref{sec:res-oracle}.

The ablations lose on $[1, 2)$ in every cell, each of the 20 intervals entirely above 0. On $[0.5, 1)$ the two learners part. Under RF, nine of the ten ablations have a lower log-RMSE than B, although they train on 300 points against B's 600; four of these intervals, all on the natural test set, lie entirely below 0. Under GBM all ten are above B, and one interval, that of C$_0$-logG on the natural test set, lies entirely above 0. The ablations are taken up in \cref{sec:res-ablations}.

\subsection{Training density and error per bin}\label{sec:res-bins}

\begin{figure}[htbp]
\centering
\includegraphics[width=\linewidth]{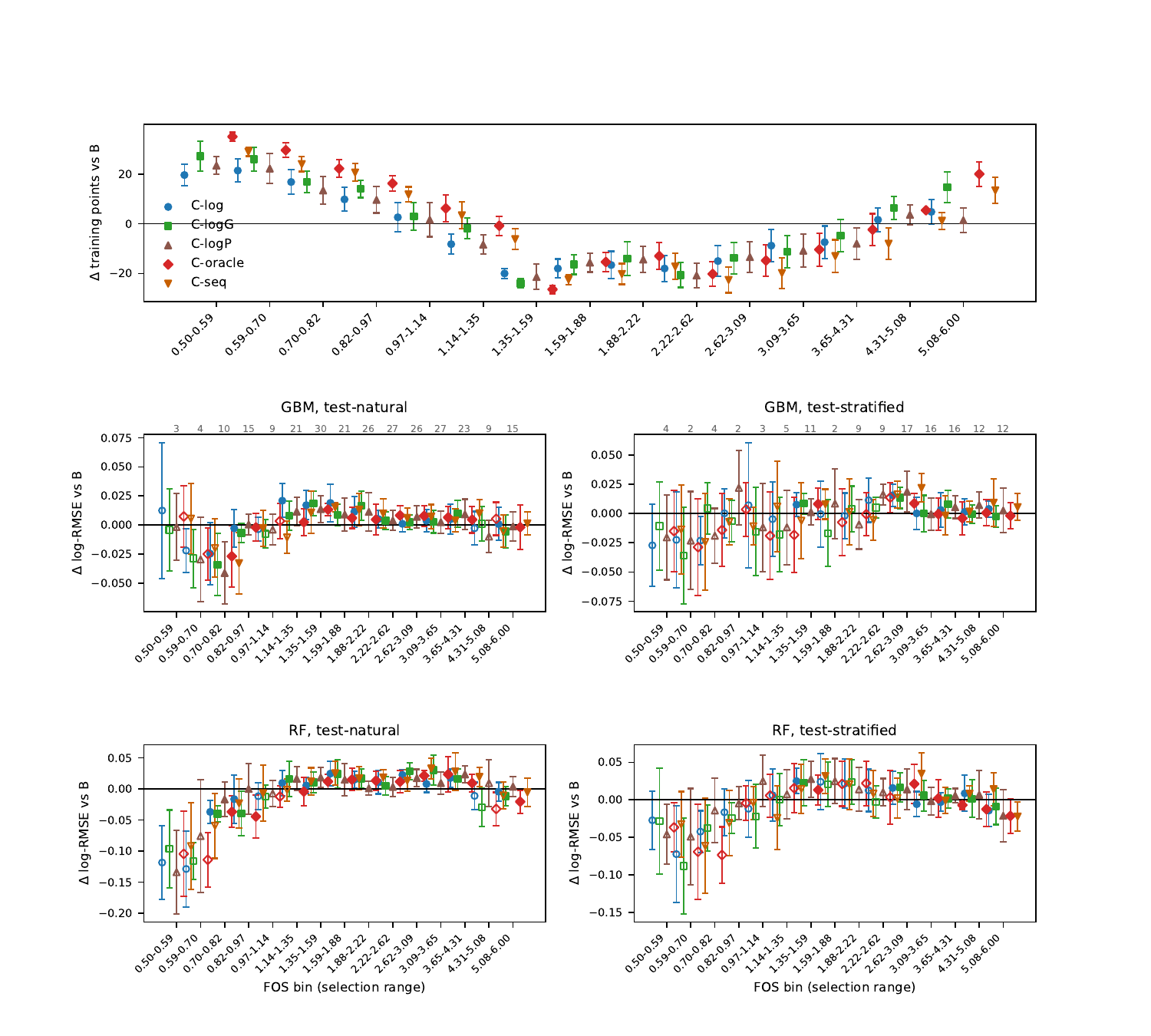}
\caption[Training density and log-RMSE per bin against B]{Per-bin training density (raw counts; B, C-log, C-seq, C-logG, C-oracle and C-logP all train on 600 points) and per-bin log-RMSE, each as the difference between an arm's mean over the five confirmatory realizations and B's, with 95\% Welch intervals. A negative error difference means a smaller log-RMSE than B. Every interval, on the density panel and on the error panels alike, visualizes variability between designs and constitutes no test; read together they are a per-bin diagnostic, not a set of tests. The pattern is consistent with density-driven relocation; it is not evidence that density caused it. The small numbers above the upper error panels are the test points per bin. Hollow markers, on the error panels, are bins with fewer than 10 of them; they are shown but carry no statement in the text. Both the densities and the errors in this figure are read from the replication metrics and use the half-open reporting zones; the allocation and displacement tables use the generator's binning instead, which differs only on an exact bin edge. C-oracle, C-logP and their C$_0$ ablations exist only in r1-r5; they have no r0.}
\label{fig:bins}
\end{figure}

\Cref{fig:bins} sets, bin by bin over the 15 bins of the selection range and over $r_1$ to $r_5$, the change in each guided arm's training density against B beside its change in log-RMSE. The analysis is descriptive, under the procedure given in \cref{sec:statistics}; it introduces no test, and its intervals visualize the variability between designs. Every arm in the figure trains on 600 points, so points gained in some bins are lost elsewhere, outside the selection range included. C-log has more training points than B in the four bins below FOS 0.97, between 9.8 and 21.4 more per bin on average, and fewer in the eight bins from 1.14 to 4.31, between 7.4 and 20.0 fewer. The upper edge of the decision range, FOS = 2, falls inside the bin 1.88--2.22.

The procedure fixes one summary: the number of unmarked bins, those with at least 10 test points, in which the change in density and the change in log-RMSE have opposite signs. Over the unmarked bins, 11 of the natural test set and 6 of the stratified one, the counts are 10 of 11, 4 of 6, 11 of 11 and 3 of 6 for C-log; 11 of 11, 3 of 6, 10 of 11 and 5 of 6 for C-oracle; and 10 of 11, 3 of 6, 9 of 11 and 4 of 6 for C-logP. Over all 15 bins, marked ones included, they are 13, 8, 15 and 12 of 15 for C-log, 12, 8, 14 and 13 for C-oracle, and 14, 9, 12 and 12 for C-logP. On the natural test set the pattern is consistent with density-driven relocation: where an arm has more training points than B, its log-RMSE is mostly lower, and where it has fewer, mostly higher. On the stratified test set 9 of the 15 bins are marked, so its counts rest on six bins. The two lowest bins, 0.50--0.70, where the guided arms add most points, are marked on both test sets, so no statement rests on them here; the gain on $[0.5, 1)$ is read on the zone in \cref{sec:res-relocation}. Under GBM the per-bin differences in log-RMSE are small, at most 0.027 in absolute value over the unmarked bins of the natural test set for C-log and C-oracle, and their intervals are wide.

C-oracle meets the intended allocation by construction, and C-logP, drawn from the same sub-pool, departs from it. Against B, C-oracle has 35.0 more training points in the lowest bin and 20.0 more in the highest, 5.08--6.00; C-logP has 23.4 and 1.4 more. What this departure does to accuracy is measured by the paired contrast of the two arms in \cref{sec:res-oracle}, not by this figure. C-seq and C-logG appear in the figure and are reported in \cref{sec:res-sensitivity}; the allocation tables, the predicted-against-true bin matrices and the ablation rows are in the Supplementary material (\cref{sec:s-structural-allocation,fig:s:matrix,sec:s-structural-ablation}).

\subsection{The oracle}\label{sec:res-oracle}

The pool arms are read as declared before the pool was solved (\cref{sec:statistics}, \cref{tab:readings}), over $r_1$ to $r_5$. \Cref{tab:rep:oracle} gives the medians and intervals per cell.

\begin{table}[htbp]
\centering
\small
\setlength{\tabcolsep}{4pt}
\caption{The oracle, read as declared before the pool was solved: C-oracle against C-logP on the same sub-pool, r1-r5, medians per learner $\times$ test cell and the paired 95\% interval of $dR_O = R(\mbox{C-oracle}) - R(\mbox{C-logP})$. `Attributable to the screener's error, in this design' only where that interval excludes 0 in at least three of four cells; an interval containing 0 is inconclusive. C$_0$-oracle and C$_0$-logP are reported in the replication tables.}
\label{tab:rep:oracle}
\begin{tabular}{lrrrr}
\toprule
 & \multicolumn{2}{c}{C-oracle} & \multicolumn{2}{c}{C-logP} \\
\cmidrule(lr){2-3} \cmidrule(lr){4-5}
cell & median $L$ & median $H$ & median $L$ & median $H$ \\
\midrule
GBM, natural & -0.256 & 0.155 & -0.417 & 0.281 \\
GBM, stratified & -0.206 & -0.274 & -0.010 & -0.156 \\
RF, natural & -0.487 & 0.096 & -0.242 & 0.078 \\
RF, stratified & -0.431 & 0.181 & -0.218 & 0.215 \\
reading & gate passes & -- & gate passes & -- \\
\bottomrule
\end{tabular}
\par\medskip
\begin{tabular}{lrrrrrrr}
\toprule
 & \multicolumn{3}{c}{$dR_O$} & \multicolumn{2}{c}{$dL_O$} & \multicolumn{2}{c}{$dH_O$} \\
\cmidrule(lr){2-4} \cmidrule(lr){5-6} \cmidrule(lr){7-8}
cell & median & CI low & CI high & median & CI & median & CI \\
\midrule
GBM, natural & -0.242 & -0.439 & 0.072 & 0.047 & [-0.098, +0.315] & -0.008 & [-0.239, +0.089] \\
GBM, stratified & -0.063 & -0.570 & 0.531 & -0.116 & [-0.424, +0.417] & -0.028 & [-0.238, +0.192] \\
RF, natural & 0.027 & -0.221 & 0.619 & -0.262 & [-0.644, +0.123] & -0.046 & [-0.247, +0.126] \\
RF, stratified & 0.108 & -0.094 & 0.330 & -0.304 & [-0.509, +0.015] & -0.161 & [-0.336, +0.079] \\
reading & A5 & \shortstack[r]{no\\attribution} & C3 & \shortstack[r]{$\rho_L$\\1.227} & -- & \shortstack[r]{$\rho_H$\\1.088} & -- \\
\bottomrule
\end{tabular}
\end{table}

\begin{figure}[htbp]
\centering
\includegraphics[width=\linewidth]{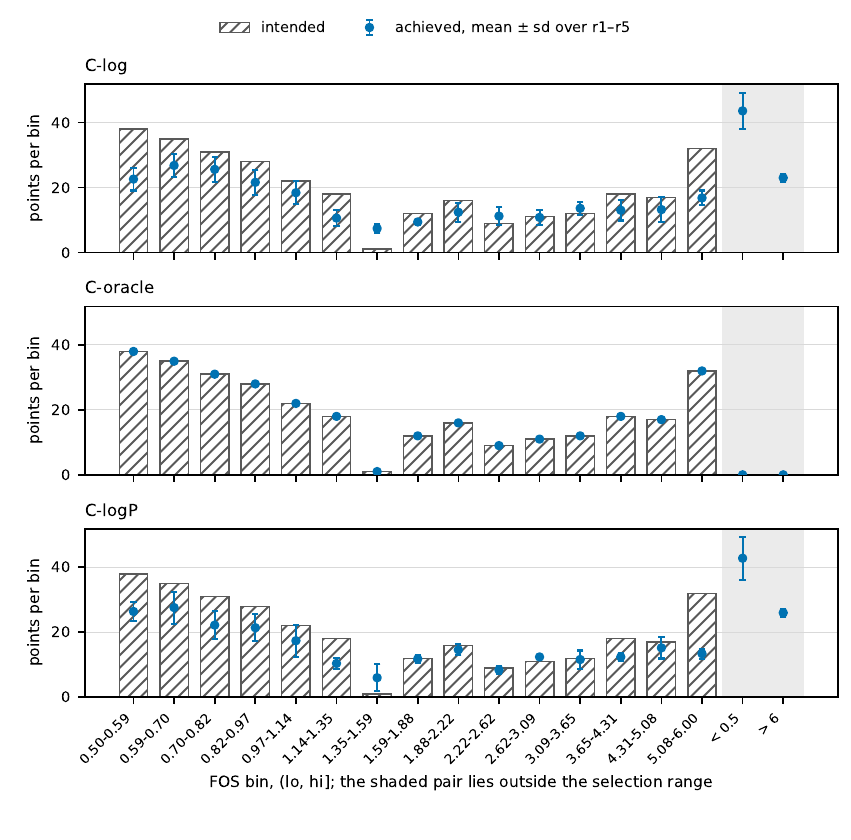}
\caption[Intended against achieved allocation]{Intended against achieved allocation of the 300 added points for C-log, C-oracle and C-logP: per bin, the intended count (hatched) and the achieved count as the mean ± sd over the confirmatory realizations r1-r5; r0 is not drawn, and C-oracle and C-logP have none. The intended allocation is the same in every realization. Descriptive only: achieved minus intended is the net distortion of the requested bin-count distribution, not a prediction error, and two points that swap bins leave it unchanged. Bins follow the generator, (lo, hi] with the first bin closed at 0.5, as the allocation tables do; the two shaded bins, FOS < 0.5 and FOS > 6, lie outside the selection range, where nothing was intended. The counts are those of the allocation tables. C-oracle bins on the solved FOS and meets the intended allocation by construction; its panel is a check, not a finding. What known bin membership changes in test accuracy is read only from the paired contrast C-oracle against C-logP (\cref{tab:rep:oracle}), in this design. C$_0$-log, C$_0$-oracle and C$_0$-logP train on the same guided points as C-log, C-oracle and C-logP in every realization, so the allocation of each is that of the arm it ablates, shown once rather than duplicated.}
\label{fig:allocation}
\end{figure}

\paragraph{Stage 1.} On training data alone, before any fit, the construction check held: in every realization C-oracle's achieved allocation equals the intended one, and it has no point outside the selection range. C-logP's achieved allocation departs from the intended one by 74 to 99 points per realization, summed in absolute value over the bins; this is the net distortion of the requested distribution, not a prediction error. The sanity check on the candidate pool, recorded before any fit, was met: on that sum, C-logP minus C-log is $-3.0$ points on average, with a Welch interval of [$-16.2$, $10.2$]. The density ratios of C-oracle to C-logP are $\rho_L = 1.227$ and $\rho_H = 1.088$, so a direction was predicted for $[0.5, 1)$ only: there C-oracle, with more training points, would have the smaller log-RMSE. \Cref{fig:allocation} shows the construction check beside C-logP's departure, with C-log for reference: C-oracle lands on the intended allocation by construction, and C-logP does not. What the departure does to accuracy is read only from the paired contrast below.

\paragraph{Relocation with known bin membership.} The relocation gate passes for C-logP in three of the four cells (step P), and for C-oracle in three of the four; both fail in the second cell, GBM on the stratified test set. Relocation is therefore present even when the intended density is met exactly. The paired contrast of the two arms, $dR_O = R(\text{C-oracle}) - R(\text{C-logP})$, is a difference of log ratios (\cref{sec:statistics}), read against the margin $m = \ln 1.10 = 0.095$. Since $R = H - L$ measures the size of the relocation, $dR_O > 0$ means that C-oracle relocates more than C-logP. Its medians are $-0.242$, $-0.063$, $+0.027$ and $+0.108$: at least $m$ in one cell, at most $-m$ in one, and smaller than $m$ in absolute value in two. Every paired interval contains 0, the narrowest being [$-0.094$, $0.330$] in the fourth cell. The reading is row A5: mixed or inconclusive on $dR_O$. Since no interval excludes 0, no attribution to the screener is made, in either direction. At five realizations, a mixed reading was declared to be the expected outcome of a null or small effect (\cref{sec:statistics}).

\paragraph{The density prediction.} For $[0.5, 1)$, the medians of $dL_O$, the log ratio of C-oracle's log-RMSE to C-logP's, so that $dL_O < 0$ means the smaller log-RMSE for C-oracle, are $+0.047$, $-0.116$, $-0.262$ and $-0.304$. The predicted sign holds in three cells, and in the same three the median reaches $m$ in absolute value, so the direction and the magnitude declared for this prediction are both met. Every paired interval contains 0, from [$-0.098$, $0.315$] to [$-0.509$, $0.015$], so the result is consistent with density-following, and no more. For $[1, 2)$ no direction was predicted; the medians of $dH_O$ are all below 0, from $-0.008$ to $-0.161$, and every interval contains 0.

\paragraph{The decision range.} C-oracle reaches level~2 against B, in the two RF cells (\cref{sec:res-decision}), and C-logP level~1. On the decision range, the paired intervals of C-oracle against C-logP all contain 0, from [$-0.010$, $0.005$] in the first cell to [$-0.036$, $0.003$] in the fourth. The reading is row C3: ``no demonstrated gain'' stands, now also with known bin membership, in this design. C-oracle's level against B is not a budget result, since its points come from the solved pool, and it is not read as a statement about the screener. The pool arms' ablations are taken up in \cref{sec:res-ablations}.

\subsection{The ablations}\label{sec:res-ablations}

Each ablation trains on the 300 points its arm adds to the seed, without the seed. Its arm and B train on 600, so no comparison in this subsection is made at equal budget.

\paragraph{Against their own arms.} Paired within each realization, every arm has a lower log-RMSE than its ablation on the decision range, in every cell and for all five pairs (C-log, C-seq, C-logG, C-oracle and C-logP): the difference, arm minus ablation, lies between $-0.114$ and $-0.036$, and each of the 20 intervals lies entirely below 0. The same holds on $[1, 2)$, between $-0.144$ and $-0.040$, and on the whole test set, between $-0.138$ and $-0.032$, again in 20 of 20 intervals. On $[0.5, 1)$ the differences are smaller, between $-0.044$ and $-0.001$; 14 of the 20 intervals lie entirely below 0, and none above. No guided or pool design replaces the seed.

\paragraph{Against B.} On the decision range every ablation has a higher log-RMSE than B, in every cell (\cref{sec:res-decision}). On $[0.5, 1)$, under RF, nine of the ten ablations have a lower log-RMSE than B, four of them with an interval entirely below 0 (\cref{sec:res-relocation}); under GBM all ten are higher. The ablations concentrate their training points there: in the four bins from 0.50 to 0.97, between 32\% and 44\% of an ablation's training set, against 8.5\% of B's. That an ablation trained on 300 points has, under RF, a lower log-RMSE than B on $[0.5, 1)$ is consistent with density-driven relocation. In the bins from 0.97 to 1.88 the ablations hold between 15\% and 18\% of their training set, against 31\% for B.

\paragraph{Outside the selection range.} At FOS $\geq 6$, on the natural test set, every ablation has a higher log-RMSE than B, by 0.37 to 0.49, with each of the ten intervals entirely above 0. On the stratified test set the differences are smaller, from $-0.043$ to $+0.131$, and 6 of the 10 intervals lie entirely above 0. C$_0$-oracle has no training point outside the selection range, since its points are drawn by their solved FOS inside it; the other ablations hold between 2.5\% and 8.7\% of their training set at FOS $\geq 6$, against 7.2\% for B. The pattern at FOS $\geq 6$ does not follow the training fraction: C$_0$-log, C$_0$-seq and C$_0$-logP hold as large a share there as B or larger, between 7.6\% and 8.7\%, and their log-RMSE on the natural test set is still about 0.4 higher.

The original ablation C$_0$, of the linear-bin design C, exists in the single realization only, and is reported with it in \cref{sec:res-sensitivity}. The bin-level rows of every replicated ablation, against B and against its own arm, are given in the Supplementary material (\cref{fig:s:ablation,tab:structural:ablation:fractions}).

\subsection{Single realization and sensitivity}\label{sec:res-sensitivity}

\paragraph{The single realization.} The tuned fits of the single realization $r_0$ are exploratory: $r_0$ shaped the later variants and their predictions, enters no interval, and is reported apart. The pool arms and their ablations have no $r_0$. The tables of $r_0$ are in the Supplementary material (\cref{sec:s-r0}). The hypothesized entries, C and C-log against A and against B, were declared one-sided on the FOS scale, for the whole test set and the selection range: 16 entries against each reference, over two arms, two learners, two test sets and two subsets. Against A, which trains on the seed alone, all 16 have $p < 0.05$, the guided arm having the smaller absolute errors. Against B, 2 of the 16 do, both for C under RF on the stratified test set: on the whole test set ($p = 0.016$, median $d = -0.032$, $d < 0$ on 58\% of the points) and on the selection range ($p = 0.033$, median $d = -0.022$, 57\%). No correction for multiplicity is applied, and C, the linear-bin design, was not replicated. For C-log against B no declared entry has $p < 0.05$. On the log scale and on the decision range, where the one-sided $p$-values are post hoc, no entry against B has $p < 0.05$. The original ablation C$_0$ has a higher decision-range log-RMSE than C in all four cells, from 0.098 against 0.057 (GBM, natural) to 0.177 against 0.117 (RF, stratified). In $r_0$, C-seq had a lower decision-range log-RMSE than C-log under RF on the natural test set, 0.095 against 0.119; this is the observation on which the prediction for C-seq below was built, and it did not replicate.

\paragraph{C-seq.} Over $r_1$ to $r_5$, C-seq has every decision-range interval against B containing 0 (\cref{sec:res-decision}), and the direction of relocation in all eight of its combinations (\cref{sec:res-relocation}). Its declared prediction against C-log was that under RF the mean difference would be negative on both test sets. It failed: on the decision range, where reading it is post hoc because the declaration names no measure, the difference is $+0.011$ [$-0.007$, $+0.028$] on the natural test set and $+0.003$ [$-0.011$, $+0.018$] on the stratified one. Both intervals contain 0, so nothing is established in either direction; the sign is the reverse of the prediction. Under GBM, as predicted, both intervals contain 0. Of the predictions declared with C-seq and read on $r_0$, the one on training data, that its three batches would put at most 10 points below FOS 0.5, failed, with 32; those on the test sets held in full or in most cells (Supplementary material, \cref{tab:s:cseq-train,tab:s:cseq-test}).

\paragraph{C-logG.} With the gradient-boosting screener, relocation is present by the gate, in three of the four cells; it fails in GBM on the stratified test set. Its size relative to C-log is not classified: $dR = R(\text{C-logG}) - R(\text{C-log})$, a difference of log ratios, is at least $\ln 1.10$ in two cells and smaller than $\ln 1.10$ in absolute value in one, so no label reaches three of four, and the reading is mixed or inconclusive under the declared rule. On the decision range, C-logG's mean difference from B is below 0 in three cells and no interval lies entirely below 0: level~1. On the natural test set, the ratio of the two families' decision-range log-RMSE, RF over GBM, is 0.88 [0.79, 0.97] times as large for C-logG as for B; on the stratified test set the interval contains 1.

\paragraph{Selection criterion on the FOS scale.} Tuning on the FOS-scale RMSE instead of the log-RMSE changes the selected hyper-parameters, under RF, only for A, and under GBM for every arm but C-logG. The hypothesized entries against B are unchanged: the same two, for C under RF on the stratified test set, have $p < 0.05$, and no other. Against A, 14 of the 16 keep $p < 0.05$; the two that do not are C-log under GBM on the stratified test set ($p = 0.094$ on the whole test set and $0.137$ on the selection range). The tables are in the Supplementary material (\cref{tab:test-natural:scorings,tab:test-stratified:scorings,tab:scorings:params}).

\paragraph{Prediction intervals.} At a nominal 90\%, the cross-conformal-style intervals of the single realization cover, on the natural test set, between 86.7\% and 93.3\% of the points for A, B and the guided arms, and between 78.0\% and 89.7\% for the ablations. On the stratified test set the ranges are 90.0\% to 98.0\% and 84.0\% to 95.3\%. Coverage is not uniform along the FOS axis. Below FOS 0.5, on the natural test set's 10 points, it ranges from 0\% to 70\% across arms; on $[0.5, 1)$, over 33 points, from 73\% to 97\%. The intervals are multiplicative, with a half-width factor $e^{q}$ between 1.20 and 1.33 for RF and between 1.09 and 1.20 for GBM. Coverage by zone, and mean width, are given in the Supplementary material (\cref{tab:test-natural__log-rmse:conformal,tab:test-natural__log-rmse:conformal:width,tab:test-stratified__log-rmse:conformal,tab:test-stratified__log-rmse:conformal:width}).

\section{Discussion}\label{sec:discussion}

\subsection{Interpretation}\label{sec:disc-interpretation}

The results support three statements, which differ in strength. First, response-balanced sampling relocates accuracy along the FOS axis. Against continued LHS, C-log and C-seq have a lower log-RMSE on $[0.5, 1)$ and a higher one on $[1, 2)$ in all 16 combinations of variant, learner, test set and half of the decision range, and 10 of the 16 intervals exclude 0 (\cref{sec:res-relocation}). Second, on the decision range, the primary criterion, no arm that selects with the screener shows a demonstrated gain over B: none of their intervals lies entirely below 0, and in one cell, C-log under GBM on the natural test set, the interval lies entirely above 0, $+0.009$ [$0.002$, $0.016$] (\cref{sec:res-decision}). This is a statement about the absence of evidence for a gain, not evidence that the arms are equivalent. Third, relocation is present even when the intended density is met. C-oracle, which bins on the solved FOS and meets the intended allocation by construction, passes the relocation gate in three of the four cells, and on the decision range it reaches level~2 against B, not level~3 (\cref{sec:res-oracle}). What the results do not support is an attribution of the departure's effect to the screener: the paired contrast of C-oracle and C-logP is mixed or inconclusive, with every interval containing 0.

Relocation may look self-evident: a model is more accurate where it has more training data, and at a fixed budget the points added in one zone are withheld from another. The expectation under test was a different one. Imbalanced regression holds that models perform worse on rare target values, and that resampling or reweighting towards them helps there \citep{Branco2016, Branco2017, Steininger2021}. The premise of response-balanced sampling was that the rare zone of the FOS is also the zone where accuracy matters, so that the gain there would be net. That intuition does not fix which outcome follows. The training density is set in the response, but the learner fits a function of the parameters: a FOS bin is a level-set region of the five-dimensional parameter space, and more points with a FOS in that bin need not lie where the model errs. A global gain was possible, had the rare zone dominated the error; so were no effect, and a loss in every zone, had the added points been spent where they taught the model little. Nor did any source promise the gain under test. The gains reported for BARL compare 5\,410 analyses with 300, not two selection rules at the same budget \citep{Mazloom2026BARL}, and \citet{Wang2024} caution that filling the output space ``does not guarantee a precise estimation of the input-output relationship''. The finding is which of these outcomes occurred, where along the FOS axis it fell, and that it replicated: a gain on $[0.5, 1)$ and a cost on $[1, 2)$, with the same direction in 16 of 16 combinations.

Where the cost falls depends on the parameter domain. Within the selection range, a log-uniform target gives each of the zones $[0.5, 1)$ and $[1, 2)$ 27.9\% of the points. LHS in this domain gives $[0.5, 1)$ between 11\% and 12\% of the points that fall within the selection range, in the seed, in B's training set and in the natural test set, and $[1, 2)$ between 35\% and 37\%. The zone $[0.5, 1)$ is rare under LHS; $[1, 2)$ is not, and it already holds more than the target would give it. Evening out the response over the selection range therefore adds points below FOS~1 and withdraws them from the zone where design decisions are made: against B, C-log has between 7.4 and 20.0 fewer training points in each of the eight bins from 1.14 to 4.31 (\cref{sec:res-bins}). The gain falls where the response is rare, and the cost where it is relevant. The region that is rare need not be the region where decisions are made, and for a surrogate meant to support those decisions it is the second that sets the criterion.

Three arms vary what selects the points, and each is read as its own result. C-seq retrains the screener on everything solved before each of its three batches. Against C-log on the decision range under RF, read post hoc because the declared prediction names no measure, its intervals contain 0 on both test sets, $+0.011$ [$-0.007$, $0.028$] and $+0.003$ [$-0.011$, $0.018$]: the comparison is inconclusive, and the advantage seen in the single realization did not replicate (\cref{sec:res-sensitivity}). C-logG, with a gradient-boosting screener chosen post hoc, shows no demonstrated gain against B on the decision range, and relocation is present by the gate in three of the four cells. C-oracle bins on the solved FOS and meets the intended allocation exactly. It relocates accuracy in three of the four cells, and on the decision range its intervals lie entirely below 0 only in the two RF cells, $-0.015$ [$-0.027$, $-0.003$] and $-0.013$ [$-0.024$, $-0.003$]: level~2, not level~3. The verdict on the decision range is thus unchanged with known bin membership, in this design. Two readings that the oracle invites are not made. Its level against B is not a budget result: its points come from a solved pool of 22\,000 analyses, and it is a diagnostic reference, not a sampling strategy. Nor is a share of the effect assigned to misbinning. In the paired contrast of C-oracle against C-logP, which share the sub-pool and the allocation, every interval of the relocation measure contains 0, and the predicted direction on $[0.5, 1)$, met in three cells, is consistent with density-following and no more. These data neither assign a share of the relocation to the screener's error nor exclude one.

The ablations train on the 300 points their arm adds, without the seed; they are not budget arms, and no comparison involving them is made at equal budget. Two readings follow. No guided or pool design replaces the seed: every arm has a lower decision-range log-RMSE than its own ablation, with each of the 20 intervals entirely below 0 (\cref{sec:res-ablations}). Under RF, nine of the ten ablations have a lower log-RMSE than B on $[0.5, 1)$ with 300 training points against B's 600, four of them with an interval entirely below 0, and they hold between 32\% and 44\% of their training points in the bins from 0.50 to 0.97, against 8.5\% for B. This is consistent with density-driven relocation. It does not make a 300-point design preferable, since the same ablations have a higher log-RMSE than B on $[1, 2)$ and on the decision range in every cell. Under GBM, all ten ablations have a higher log-RMSE than B on $[0.5, 1)$. At FOS $\geq 6$, where the ablations' error on the natural test set does not follow their training fraction, the two test sets also differ in what they hold: 24 natural points between FOS 6.49 and 20.53, with a median of 8.15, and 23 stratified points between 6.04 and 9.52, with a median of 7.04.

Every comparison above is vertical: one learner, different sampling arms. The learner interaction is read on the decision range, the primary criterion, as the ratio of the two families' decision-range log-RMSE, RF over GBM. For C-log against B on the natural test set, this ratio is 0.78 [0.70, 0.86] times as large for C-log as for B, and the same holds on each half of the decision range, 0.79 [0.65, 0.97] on $[0.5, 1)$ and 0.84 [0.75, 0.94] on $[1, 2)$; on the stratified test set every interval contains 1. On $[0.5, 1)$, C-log's difference from B under RF is $-0.052$ [$-0.070$, $-0.034$] on the natural test set and $-0.036$ [$-0.055$, $-0.017$] on the stratified one; under GBM it is $-0.011$ [$-0.023$, $0.002$] and $-0.020$ [$-0.045$, $0.004$], both intervals containing 0. Each of these is a statement about one family, and no comparison of their sizes is drawn from them. That the contrast could depend on how closely a learner's local accuracy follows its local training density is a hypothesis that this study does not test.

The object of the study is the allocation of a fixed budget to a target distribution over the response, not adaptive sampling in general. Uncertainty- and committee-based active learning, Bayesian experimental design and output-diversity sampling \citep{Wu2019} choose points by other criteria (\cref{sec:rw-active-learning}), and nothing here says how they would fare at the same budget. C-seq, the one sequential variant tested, retrains the screener between batches but keeps the same target distribution. A negative comparison against a uniform design is not without precedent: across 16 benchmark problems of distribution estimation, \citet{Su2025} find no systematic improvement of active learning over a uniform space-filling design. Reweighting acts on a different lever. It changes how an existing set of labels is used, while the question here is which analyses to run; a density-reweighted surrogate fitted on B's training set would be a cheaper, complementary comparison, which this study does not make (\cref{sec:rw-imbalanced}).

Which zone is rare depends on the parameter domain. In the domain of BARL the initial LHS set is skewed the other way, and the under-represented zone is 4--6 \citep{Mazloom2026BARL}. The effect sizes are likewise properties of this domain, this budget and these learners, and are not claimed to transfer. Two things are meant to. One is a diagnostic that needs only the solved seed: before reshaping the training response, compare the FOS distribution that LHS already gives within the selection range with the target, and with the decision range. Here it would have shown that the target withdraws points from $[1, 2)$. The other is the protocol: arms compared at an equal number of analyses, accuracy read zone by zone on a log scale rather than as one global score, replicated realizations of the added points, and ablations that separate the added points from the seed. The deliberately simple case is what made this protocol affordable. Every arm could be solved in full and replicated, and the oracle could be read against a solved pool of 22\,000 analyses, which on a layered finite-element model would be far more costly. In aggregate the effects are of the order of 0.01 in decision-range log-RMSE, against B's 0.05 to 0.10; they are reported by zone because a single score would hide where they fall.

\subsection{Limitations}\label{sec:limitations}

Every conclusion is conditional on the design of the replications. Realizations vary the added points only; the seed, and with it the screener, the allocation and the hyper-parameters, is common to all of them. The 16 of 16 combinations therefore show that relocation is not an accident of one candidate draw under this seed. They do not show robustness to another pattern of misbinning: for C-log, the screener is the same fitted function in all five realizations, sampled only through different candidates. Two arms cover part of that dimension, still under the same seed: C-logG, with another screener on the same candidate pool and draw stream, and C-oracle, with no misbinning at all. All realizations share one LHS seed. The allocation is computed from that seed's deficit, so another seed would add points in other bins. The mechanism --- accuracy following the achieved training density, present even when the intended density is met exactly --- is not expected to depend on the seed, but its size and profile along the FOS axis may. The expectation rests on the domain: under LHS, $[1, 2)$ holds 35--37\% of the points within the selection range, above the log-uniform target's 27.9\%, which is not a property of the seed; so the direction of relocation is expected to hold under another seed, and its size and the departure are not.

The intervals cover the variability between designs under this seed, and nothing else. The two test sets are fixed, so their sampling error is common to every realization and does not average out over them. The stratified test set is placed by the same seed-trained random forest that screens the guided arms; the natural test set is not. The replications use common hyper-parameters fixed from the seed, so they do not replicate the tuned comparison of the single realization, which is exploratory. Five realizations give wide intervals: for the oracle's relocation contrast, the paired half-widths expected when the readings were declared were 0.14 to 0.42, against a margin of 0.095, and a mixed reading was declared to be the expected outcome of a null or small effect. An interval that contains 0 is therefore read as inconclusive, and ``no demonstrated gain'' is a statement about the absence of evidence. No correction for multiplicity is applied: the claims rest on the declared rules over several cells, not on single $p$-values. The decision range was defined after the C-log results and before C-seq was run, and the results on the whole test set and on the selection range are reported beside it.

The scope is bounded in five further ways. The surrogates belong to two tree-ensemble families, and the results are consistent across these two independently tuned families, and no further; whether a smooth learner, such as a Gaussian process or a neural network, relocates accuracy in the same way is not tested. The budget is one, 600 analyses per arm, and the learning curve, which covers LHS and C only, with fixed hyper-parameters, is not used to extrapolate relocation to other budgets. The generator is one, xslope 0.5.2 with the Bishop simplified method, applied to a homogeneous, drained slope with no water table, chosen as a controlled testbed (\cref{sec:data}); what is meant to transfer from it, and what is not, is set out in \cref{sec:disc-interpretation}. The screener is platform-dependent: differences of floating-point rounding between processor architectures change the fitted random forest and, through it, the guided designs, so the reference is the solved sets, recorded with their platform, not the selection code alone. Finally, the prediction intervals are cross-conformal-style and carry no formal coverage guarantee. Their coverage is reported empirically and is not uniform along the FOS axis: on the ten points of the natural test set below FOS 0.5, it ranges from 0\% to 70\% across arms.

\section{Conclusions}\label{sec:conclusions}

This study asked how a fixed budget of slope-stability analyses should be allocated along the response, and tested one answer, response-balanced sampling, against continued LHS at the same number of analyses. At that budget, evening out the training distribution of the FOS does not systematically improve accuracy. It relocates accuracy along the FOS axis, from $[1, 2)$ to $[0.5, 1)$, following the training density actually achieved. The direction of relocation replicated over five new realizations of the added points, and relocation is present even when the intended density is met exactly. On the decision range, FOS $\in [0.5, 2)$, no design that selects with the screener shows a demonstrated gain over continued LHS, and with known bin membership the verdict is unchanged, in this design.

The region that is rare under LHS need not be the region where decisions are made. In this domain the rare zone lies below FOS~1, while $[1, 2)$, where design decisions are taken, is already denser under LHS than the target would make it, so evening out the response withdraws points from it. Before the training response of a surrogate is reshaped, the FOS distribution that LHS already gives should be compared with the target and with the range in which the surrogate will be used, and accuracy should be judged zone by zone rather than by one global score.

Three directions follow from the limits of the study. (i) A smooth learner, such as a Gaussian process, fitted on the same training sets would test whether relocation extends beyond tree ensembles; it needs no new analyses. (ii) Replications over several LHS seeds would vary the screener and the allocation with the seed, and would measure how the size and profile of relocation depend on it. (iii) The solved data allow further comparisons at no cost in analyses: a density-reweighted surrogate fitted on the continued-LHS data, the complementary lever of reweighting labels already computed; and, drawing from the solved reference pool, other selection rules, such as uncertainty-based active learning, at the same budget.

\section*{Abbreviations}
AI, artificial intelligence; AK-MCS, active learning with Kriging and Monte Carlo simulation; BARL, batch active regression learning; CV+, the cross-validation extension of the jackknife+; FOS, factor of safety; GBM, gradient boosting machine; GSy, greedy sampling on the output; LHS, Latin hypercube sampling; ML, machine learning; RF, random forest; RMSE, root-mean-square error (log-RMSE, the RMSE on the log scale); SMOGN, synthetic minority over-sampling for regression with Gaussian noise.

\section*{Acknowledgments}
Claude Code (Anthropic), with Claude Opus and Sonnet models, was used throughout the study to assist in implementing and testing the analysis code and in drafting and revising the manuscript. The study design, the analyses and the interpretation of the results are the authors', and every reported number comes from the released pipeline. The authors reviewed all code and text and take full responsibility for the content.

\section*{Funding Statement}
There are no funders to report for this submission.

\section*{Competing Interests}
The authors declare none.

\section*{Data Availability Statement}
The data that support the findings of this study are openly available on Zenodo at \url{https://doi.org/10.5281/zenodo.23618253}: every solved analysis of the fixed sets, the replications and the reference pool, with the designs they were drawn from and their provenance record. The code that generates the data, fits and evaluates the surrogates and reports the results is hosted at \url{https://github.com/bteodoru/surrogate-guided-slope-stability}. The release used for this article is archived on Zenodo (\href{https://doi.org/10.5281/zenodo.23727659}{doi:10.5281/zenodo.23727659}), and the repository, with its full history, on Software Heritage (\texttt{swh:1:rev:3e9855324d121f53c0bc5cd0c7f30d25aeb74817}). The guided designs depend on the computing platform: the screener's predictions, and with them the points it selects, differ between x86 and arm64 processors, so a rerun on another platform gives different guided sets. The solved sets in the deposit, generated on macOS (arm64), are the reference.

\section*{Author Contributions}
Conceptualization: I.-B.T., M.A. Methodology: I.-B.T., A.A. Software: I.-B.T. Validation: I.-B.T., M.A., I.L. Formal analysis: I.-B.T. Investigation: I.-B.T. Data curation: I.-B.T. Visualization: I.-B.T. Project administration: I.-B.T. Writing -- original draft: I.-B.T., A.A. Writing -- review and editing: I.-B.T., A.A., M.A., I.L. All authors approved the final submitted draft.

\section*{Supplementary Material}
The Supplementary material, in the appendix, contains: a chronology of the study's declarations, designs and results, with the commits that record them (\cref{sec:s-chronology}); the declared texts, namely the oracle declaration as committed, the C-logG reading rule, the procedure of the structural analysis with its amendments and the limit of its blinding, and the C-seq predictions read on the single realization with their outcomes (\cref{sec:s-declared}); the hyper-parameters selected for every arm (\cref{sec:s-params}); the learning curve (\cref{sec:s-learning-curve}); the per-realization values, replicated comparisons, learner interactions and bin diagnostics of the replications (\cref{sec:s-replications}); the tables and figures of the structural analysis, the ablations included (\cref{sec:s-structural}); and the tables of the single realization (\cref{sec:s-r0}).

\bibliographystyle{apalike}
\bibliography{refs}

\clearpage
\section*{Supplementary material}
\setcounter{section}{0}\setcounter{table}{0}\setcounter{figure}{0}
\renewcommand{\thesection}{S\arabic{section}}
\renewcommand{\thetable}{S\arabic{table}}
\renewcommand{\thefigure}{S\arabic{figure}}
\renewcommand{\theHsection}{S\arabic{section}}
\renewcommand{\theHtable}{S\arabic{table}}
\renewcommand{\theHfigure}{S\arabic{figure}}

This appendix holds the supplementary material of the article. References to sections, tables and figures without an S belong to the main text. Every table and figure here is written by the study's pipeline (\texttt{sgs report}) from its stored metrics, with its generated caption. Every C$_0$ ablation is reported wherever its arm is, unless a caption states otherwise. As in the article, an interval containing 0 is inconclusive, never a tie or evidence of equivalence, and the intervals of the replications are conditional on the fixed seed, the fixed test sets and the common hyper-parameters fixed from the seed.

\section{Chronology}\label{sec:s-chronology}

The article states order relations only where they change how a result is read (\cref{sec:arms,sec:oracle,sec:tuning,sec:evaluation,sec:statistics}). \Cref{tab:s:chronology} gives the dates and commits behind them, and the other milestones of the study. Times are the committer times of the code repository, in local time (UTC+03:00), except where another source is named. The repository, with its full history, is identified in the article's Data Availability Statement. A commit time shows when a change was recorded, and the order of the commits; it shows nothing about when the work was done or first run, unless the row says so.

{\small
\begin{longtable}{@{}r@{\hspace{6pt}}>{\raggedright\arraybackslash}p{0.13\linewidth}>{\raggedright\arraybackslash}p{0.12\linewidth}>{\raggedright\arraybackslash}p{0.62\linewidth}@{}}
\caption{Chronology of declarations, designs and results. ``Committed'' means recorded in the code repository; the commit shows that a file existed in that state at that time, not when it was computed.}\label{tab:s:chronology}\\
\toprule
\# & time (+03:00) & commit & event \\
\midrule
\endfirsthead
\toprule
\# & time (+03:00) & commit & event \\
\midrule
\endhead
\bottomrule
\endfoot
1 & 2026-09-18 17:26:27 & fa9a323 & Configuration committed, with the hypothesized comparisons and their one-sided direction (C against B and against A). \\
2 & 2026-09-18 19:20:41 & 5c7c272 & Evaluation code committed, with the first test metrics (arms A, B, C and C$_0$). \\
3 & 2026-09-19 02:44:42 & 4c5724e & State before the GBM grid extension (the earlier grid), tagged \texttt{gbm-grid-v1}. \\
4 & 2026-09-19 12:19:58 & 6fb04d9 & GBM grid extended, by training-fold cross-validation on A, B, C and C$_0$. \\
5 & 2026-09-19 12:43:17 & f8ef46d & C-log and C$_0$-log declared, with their comparisons and the one-sided direction of the C-log entries. \\
6 & 2026-09-19 12:50:04 & aa95f60 & C-log set committed. \\
7 & 2026-09-19 14:24:46 & d74bf66 & C-log and C$_0$-log evaluated. \\
8 & 2026-09-19 20:15:32 & 0c4cd2f & Reporting zones with edges 0.5, 1, 2 and 6, the log scale and the decision range [0.5, 2). \\
9 & 2026-09-19 20:40:05 & ca86071 & C-seq and C$_0$-seq declared, with their predictions (\cref{sec:s-cseq}). \\
10 & 2026-09-19 20:54:06 & 84a78ab & C-seq batches committed. \\
11 & 2026-09-19 22:40:44 & e8eafe3 & C-seq and C$_0$-seq fitted and evaluated; the test outcomes of \cref{tab:s:cseq-test}. \\
12 & 2026-09-20 12:08:09 & 32c3d51 & Replications declared: five new realizations, common hyper-parameters, intervals, predictions. \\
13 & 2026-09-20 19:32:25 & da46ddb & Replication sets committed. \\
14 & 2026-09-20 19:44:27 & 075edb7 & Replications evaluated; their bin-level results and figures committed. \\
15 & 2026-09-21 10:57:13 & c1681ce & Structural procedure committed (\cref{sec:s-structural-procedure}); the commit changes only the procedure's document and the repository's instructions. \\
16 & 2026-09-21 10:59:10 to 11:02:07 & d5023b3 to 919fd95 & Structural code committed; the artifact in 919fd95. \\
17 & 2026-09-21 11:24:27 & fe77b9a & Amendment 1 to the structural procedure (the C$_0$ ablations); the commit changes only the procedure's document. \\
18 & 2026-09-21 11:26:45 & 9ece938 & The C$_0$ rows added to the structural code and artifact. \\
19 & 2026-09-21 12:05:46 and 12:36:11 & 1be46d6 and 3ea03f3 & The record of the structural procedure corrected: fixed on computed data, not preregistered; the limit of the blinding stated as reproduced in \cref{sec:s-structural-procedure}. \\
20 & 2026-09-23 14:02:37 & 975fe22 & Relocation reading committed in code: the gate, the labels and the constants of \cref{sec:s-clogg}. \\
21 & 2026-09-23 14:52:49 & 6b6b822 & C-logG predictions and thresholds declared (\cref{sec:s-clogg}). \\
22 & 2026-09-23 15:09:52 & e1e57d5 & In the code of the relocation reading, dR paired against C-log, R(C-logG) $-$ R(C-log); before, it was C-logG's own R against B (\cref{sec:s-clogg}). \\
23 & 2026-09-23 15:50:47 & 40de8fc & C-logG set of the single realization committed. \\
24 & 2026-09-23 17:02:22 & caf7a04 & C-logG replication sets r1--r5 committed. \\
25 & 2026-09-23 17:21:51 & d770ab9 & C-logG stage 1 (training data) recorded, before any C-logG test evaluation. \\
26 & 2026-09-23 18:42:39 & 1cf61ae & C-logG and C$_0$-logG fitted and evaluated (stage 2). \\
27 & 2026-09-24 01:00:20 & 4795cf0 & Reference pool and oracle declared in the configuration (a summary of the declaration). \\
28 & 2026-09-24 11:41:11 & ab43902 & Oracle declaration committed, tagged \texttt{oracle-declared} (\cref{sec:s-oracle}). \\
29 & 2026-09-24 11:59:08 & --- & Solver log of the first pool chunk, c001, created. Source: the authors' run log, kept outside the repository; this is the only evidence for the start time. A chunk takes about 45 minutes. \\
30 & 2026-09-24 12:44:43 & --- & Chunk c001 finished: \texttt{run\_metadata.json} records 09:44:43 UTC; the record was committed in 0f4dfc2 (row~31). \\
31 & 2026-09-24 13:33:48 & 0f4dfc2 & Pool chunk c001 committed. \\
32 & 2026-09-25 11:10:09 & 8209389 & Pool chunks up to c015 committed; the solved side of the stopping rule is met at c015. \\
33 & 2026-09-25 17:08:45 & cbaa4e0 & Pool chunks up to c022 committed; the stopping rule is met and solving stops. \\
34 & 2026-09-25 17:45:03 & 409c18a & Oracle stage 1 (training data) recorded, before any fit of C-oracle or C-logP. \\
35 & 2026-09-25 17:47:58 & bbca97d & C-oracle, C-logP and their C$_0$ ablations fitted and evaluated (stage 2). \\
36 & 2026-09-25 18:20:15 & f280243 & Amendment 2 to the structural procedure (the pool arms). \\
37 & 2026-09-28 20:19:32 & fccc8d3 & Check of the non-converged searches completed (22 cases). \\
38 & 2026-09-28 21:16:22 & 184255c & Screener diagnostic on 1\,800 independent LHS points committed (the second diagnostic behind C-logG, \cref{sec:arms}). \\
\end{longtable}}

\section{Declared texts}\label{sec:s-declared}

\subsection{The oracle declaration}\label{sec:s-oracle}

The text below is reproduced from the version of \texttt{docs/oracle-declaration.md} committed at tag \texttt{oracle-declared} (\cref{tab:s:chronology}, row~28); the file has not changed since. Only the markup is converted to typeset form; the wording, including its dates, is unchanged. Its section numbers are those of the design document it was copied from, and its cross-references (§1, §12) point there. \Cref{tab:readings} of the article condenses §9.3. The outcome is reported in \cref{sec:res-oracle}.

\paragraph{Preamble of the declared file.}
Copied verbatim from the design spec of 2026-09-23 (local, not tracked): §4.4 (the stopping
rule), §9 (predictions and readings) and §10 (intervals). Committed with the tag
\texttt{oracle-declared}, before any pool chunk was solved, as the git history shows. Frozen from
that commit; the stage-2 reading quotes this file. \texttt{config.py} carries a summary of it.

\subsubsection*{4.4 Stopping rule (declared)}

Allocation \texttt{a} is C-log's: the seed's solved FOS water-filled to 300 points over the 15 log bins of the selection range. The seed is fixed, so \texttt{a} is the same in every realization: \texttt{[38, 35, 31, 28, 22, 18, 1, 12, 16, 9, 11, 12, 18, 17, 32]}.

After each completed chunk k, over the eligible rows of chunks 1..k, the rule is met when, for every sub-pool r and every bin b:
\begin{quote}\footnotesize\begin{verbatim}
solved_count[r, b]    >= a[b]     (C-oracle can draw)
predicted_count[r, b] >= a[b]     (C-logP can draw; RF screener trained on the seed)
\end{verbatim}\end{quote}

Binning is the generator's: \texttt{(lo, hi]}, the first bin closed (\texttt{design.bin\_counts}).

\begin{itemize}
\item At the first chunk that meets the rule, solving stops for good. The pool is chunks 1..k.
\item If the rule is not met at the cap, that is an error. The rule is not relaxed, and a new decision is needed.
\item The rule reads only pool FOS counts. It reads no test set, no fit and no error, so it cannot favour an arm.
\item \textbf{Expected stop.}
  \begin{itemize}
  \item Bin 0 has p = 0.96\% under uniform sampling: 23 of 2400 LHS points, with a 95\% Clopper--Pearson interval of 0.61--1.43\%. No LHS row is excluded.
  \item About 5.8k rows per sub-pool give a \textbf{per-sub-pool} probability of $\approx$ 99.5\% of reaching 38 points. The rule needs all five sub-pools, so the \textbf{joint} probability at that size is $\approx$ 97.5\%.
  \item Simulated on the empirical bin rates, with 200 rows per sub-pool per chunk, the stop falls at a median of 24 chunks (p95 29). At the CP lower bound for bin 0 it falls at a median of 37 (p95 44), and P(cap of 50 reached) $\approx$ 0.35\%. ``About 30'' is therefore a cautious planning figure.
  \end{itemize}
\item \textbf{The predicted side does not bind.} On a uniform pool of 200k, the RF screener puts 1.16--1.17\% of candidates in bin 0, against a true rate of 0.96\%.
  \begin{itemize}
  \item The most demanding predicted bin is bin 0: 38 / 1.17\% $\approx$ 3.2k rows per sub-pool in expectation. Next comes bin 1: 35 / 1.52\% $\approx$ 2.3k.
  \item Both are below the solved side's bin 0, 38 / 0.96\% $\approx$ 4.0k.
  \item The predicted side is kept in the rule anyway, so that a C-logP draw can never fail.
  \end{itemize}
\item \textbf{At the stop,} the binding sub-pool's bin 0 may hold exactly 38 points, and C-oracle then takes all of them. The rule stops on the \emph{joint} counts (true bin $\times$ predicted bin). Conditional on those counts at the stop, the rows are i.i.d.\ within each (true bin, predicted bin) cell. This is stated, not corrected; its practical effect is negligible.
\end{itemize}

\subsubsection*{9. Declared predictions and readings}

These are declared \textbf{before the first chunk is solved}, in \texttt{config.py} beside the C-logG block. They cover r1..r5 only.

\textbf{``In this design''} means:
\begin{itemize}
\item the RF screener trained on the seed;
\item C-log's allocation \texttt{a};
\item a random draw within each bin;
\item this reference pool, solved on darwin arm64;
\item the fixed seed, test sets and \texttt{LC\_PARAMS}.
\end{itemize}

Every attribution below carries that qualifier.

\textbf{Terms.}
\begin{itemize}
\item A \emph{cell} is learner $\times$ test (four cells).
\item A \emph{median} is over r1..r5 within a cell.
\item \emph{m} = ln 1.10.
\item \emph{Paired interval} is \texttt{evaluate.paired\_interval}: a 95\% t-interval over the five per-realization values, n $-$ 1 degrees of freedom.
\item ``In $\geq$ 3/4'' always means in at least 3 of the 4 cells.
\end{itemize}

\subsubsection*{9.1 Stage 1 --- training data only, before any fit}

{\small\noindent
\begin{tabular}{p{0.40\linewidth}p{0.30\linewidth}p{0.20\linewidth}}
\toprule
quantity & expectation & type \\
\midrule
C-oracle: $\Sigma|$achieved $-$ intended$|$, points $<$ 0.5, points $>$ 6, share staying in bin & 0, 0, 0, 100\% & construction check (asserted) \\
C-logP $-$ C-log: $\Sigma|$achieved $-$ intended$|$ & Welch 95\% interval contains 0 \textbf{and} $|$mean $\Delta| \leq$ 15 points (C-log: 76--100) & sanity check on pool size \\
\texttt{$\rho$\_Z} = median over r of \texttt{n\_Z(C-oracle) / n\_Z(C-logP)}, training set = seed + selection, Z $\in$ \{L = [0.5, 1), H = [1, 2)\} & expected $\rho_L \approx$ 1.27, $\rho_H \approx$ 1.06 (below) & \textbf{fixes the direction of §9.2 before any fit} \\
C-oracle $\cap$ C-logP rows, per realization & reported & descriptive \\
\bottomrule
\end{tabular}}

\begin{itemize}
\item \textbf{Binning.} The first two rows bin as the generator does, \texttt{(lo, hi]}. \texttt{$\rho$} uses the evaluation zones \texttt{[lo, hi)}, because §9.2 compares it with zone errors.
\item \textbf{Expected $\rho$, declared now.} It was estimated on existing training data only, with no test set read:
  \begin{itemize}
  \item C-log r1..r5 stands in for C-logP;
  \item for C-oracle, a random within-true-bin draw from the solved LHS rows;
  \item per realization, $\rho_L$ = 1.29, 1.22, 1.32, 1.25, 1.30 and $\rho_H$ = 1.07, 1.05, 1.08, 1.09, 1.02.
  \end{itemize}
  The oracle recovers in range the \textasciitilde 45 points C-log loses below 0.5, mostly into L. The binding values of $\rho$ are the stage-1 ones, not these.
\item \textbf{If the sanity check fails,} the paired contrast C-oracle vs C-logP is unaffected, since the two share the pool. Only statements comparing with C-log carry ``differs also in the candidate pool''.
\end{itemize}

\subsubsection*{9.2 Stage 2 --- test: quantities and predictions}

\textbf{Quantities}, as in \texttt{evaluate.relocation\_reading}:
\begin{itemize}
\item \texttt{L(X) = ln(E\_X,L / E\_B,L)} and \texttt{H(X) = ln(E\_X,H / E\_B,H)}, where E is the log-RMSE on the zone;
\item \texttt{R(X) = H(X) $-$ L(X)};
\item \texttt{dR\_O(r) = R(C-oracle, r) $-$ R(C-logP, r)}, in which B cancels exactly;
\item \texttt{dZ\_O(r) = ln(E\_oracle,Z / E\_logP,Z)} for Z $\in$ \{L, H\}.
\end{itemize}

Each is read as its median per cell. The per-realization values also go into a paired interval per cell.

\textbf{Relocation gate, applied to C-oracle and to C-logP separately:} median L $<$ 0 \textbf{and} median H $>$ 0, in $\geq$ 3/4.

\textbf{Density prediction} (the testable form of ``follows the achieved density''), for each zone Z:
\begin{itemize}
\item \texttt{$\rho$\_Z $\geq$ 1.10} predicts \texttt{dZ\_O $<$ 0}: the oracle is more accurate where it has more data;
\item \texttt{$\rho$\_Z $\leq$ 1/1.10} predicts \texttt{dZ\_O $>$ 0};
\item otherwise no directional prediction is made for Z; the values are reported.
\end{itemize}

It is read at two levels:
\begin{itemize}
\item \textbf{direction met:} the predicted sign in $\geq$ 3/4;
\item \textbf{magnitude met:} additionally, \texttt{|median dZ\_O| $\geq$ m} in $\geq$ 3/4.
\end{itemize}

Both levels are reported with the four medians. Direction alone is weak: under a coin flip, P = 5/16, and the cells share training sets, so they are not independent.

\textbf{Implied prediction for dR\_O.} With the expected $\rho$ (L above the margin, H inside it), density-following predicts \texttt{dR\_O $>$ 0}. This is the direction declared for §9.3.

\textbf{Levels, C-oracle against B, on the decision range.} The quantity is the per-realization difference in decision-range log-RMSE, C-oracle $-$ B, under a Welch interval (the arms do not share realizations):
\begin{enumerate}
\item \textbf{point improvement:} the mean difference is below 0 in one cell;
\item \textbf{sustained improvement:} the Welch 95\% interval is entirely below 0 in one cell;
\item \textbf{systematic gain:} the interval is entirely below 0 in all four cells.
\end{enumerate}

The same levels are computed for C-logP.

\textbf{Background prediction:} C-oracle does \textbf{not} reach level 3. The reason: halving the distortion with the GBM screener ($\Sigma|$achieved $-$ intended$|$ 76--100 $\rightarrow$ 43--55) produced no gain against B.

\subsubsection*{9.3 Readings fixed in advance}

\texttt{$\rho$} is the stage-1 value. Within groups A and B, the first matching row applies, in the order listed. Group C is read independently of A and B, and is also ordered.

\textbf{Step P, preliminary: the gate for C-logP.} It is reported first, as passes or fails, with its four median pairs (L, H). Every A and B reading below names C-logP's gate state. If the gate fails for C-logP, no reading of dR\_O is described as a change \emph{in} relocation: it is a difference between an arm that relocates and one that does not.

\needspace{0.45\textheight}
\noindent\textbf{Group A: the gate passes for C-oracle.}

{\small\noindent
\begin{tabular}{p{0.05\linewidth}p{0.24\linewidth}p{0.60\linewidth}}
\toprule
\# & condition & reading \\
\midrule
A1 & median dR\_O $\geq$ m in $\geq$ 3/4 & With $\rho_L \geq$ 1.10, as density-following predicts: relocation is larger with known membership, where the density in L is higher. With both $\rho$ inside the band, relocation is larger \textbf{without} a density difference; this is against density-following for this contrast, and is stated. Attribution follows §9.4: ``C-logP's smaller relocation is attributable to the screener's error, in this design'' only if the paired dR\_O interval excludes 0 in $\geq$ 3/4. Otherwise ``consistent with'', with no attribution \\
A2 & median dR\_O $\leq$ $-$m in $\geq$ 3/4 & Against density-following for this contrast, whose declared direction was $>$ 0. If the paired interval excludes 0 in $\geq$ 3/4: ``C-logP's additional relocation is attributable to the screener's error, in this design''. ``Follows the achieved density'' is kept only for the zones whose density prediction is met \\
A3 & the paired dR\_O interval lies entirely inside ($-$m, m) in $\geq$ 3/4 & Relocation is \textbf{practically unchanged} with known membership; this is the only equivalence reading. With $\rho_L \geq$ 1.10 it is \textbf{evidence against} density-following for this contrast: a density difference of ($\rho_L$ $-$ 1) in L, from stage 1, without a corresponding relocation difference. With both $\rho$ inside the band it is consistent with the thesis. Expected to be unreachable (§9.5) \\
A4 & $|$median dR\_O$|$ $<$ m in $\geq$ 3/4, and A3 does not hold & \textbf{No difference detected.} Neither ``attributable'' nor ``not attributable''. With $\rho_L \geq$ 1.10, the density prediction for dR is \textbf{not borne out}, and this is stated \\
A5 & otherwise (e.g.\ the C-logG pattern: dR +0.12, $-$0.51, +0.14, $-$0.01) & \textbf{Mixed / inconclusive} on dR\_O, with the four medians and intervals shown \\
\bottomrule
\end{tabular}}

\needspace{0.30\textheight}
\noindent\textbf{Group B: the gate fails for C-oracle.}

{\small\noindent
\begin{tabular}{p{0.05\linewidth}p{0.24\linewidth}p{0.60\linewidth}}
\toprule
\# & condition & reading \\
\midrule
B1 & the gate fails for C-logP too & Relocation is not detected on this pool for \textbf{either} arm. This is reported before anything else. The relocation of C-log on the 200k pool is not contradicted, but it is not reproduced here; dR\_O is reported descriptively, with no relocation label \\
B2 & the paired interval of R(C-oracle) lies entirely inside ($-$m, m) in $\geq$ 3/4 & \textbf{Relocation vanishes with known membership}: the only ``vanishes'' reading. Attribution follows §9.4: ``C-logP's relocation is attributable to the screener's error, in this design'' only if the paired dR\_O interval excludes 0 in $\geq$ 3/4 \\
B3 & median R(C-oracle) $\leq$ $-$m in $\geq$ 3/4 & \textbf{Relocation reverses with known membership.} The same attribution rule as B2 \\
B4 & $|$median R(C-oracle)$|$ $<$ m in $\geq$ 3/4, and B2 does not hold & \textbf{Relocation not detected with known membership.} No attribution and no equivalence \\
B5 & otherwise & \textbf{Mixed / inconclusive}: per-zone medians shown, no relocation label \\
\bottomrule
\end{tabular}}

\needspace{0.25\textheight}
\noindent\textbf{Group C: the decision range. Ordered; C1 is checked before C2.}

{\small\noindent
\begin{tabular}{p{0.05\linewidth}p{0.38\linewidth}p{0.46\linewidth}}
\toprule
\# & condition & reading \\
\midrule
C1 & C-oracle at level 3 against B, \textbf{and} the paired interval C-oracle $-$ C-logP on decision-range log-RMSE entirely below 0 in all four cells, \textbf{and} C-logP below level 3 & ``C-oracle has lower decision-range log-RMSE than B in all four cells, in this design'', plus ``the difference in decision-range accuracy between C-oracle and C-logP is attributable to the screener's error, in this design''. ``No demonstrated gain'' is then qualified as holding for the screened arms; the title and headline are revisited \\
C2 & C-oracle at level 3 against B, and C1 does not hold & ``C-oracle has lower decision-range log-RMSE than B in all four cells, in this design'', with no statement about the screener. The result costs the pool; it is not a budget result \\
C3 & C-oracle below level 3 & ``No demonstrated gain'' stands, now also with known bin membership, in this design \\
\bottomrule
\end{tabular}}

\subsubsection*{9.4 Wording about the screener}

\begin{itemize}
\item \textbf{The only permitted attribution.} ``Attributable to the screener's error, in this design'' is used only for a paired C-oracle vs C-logP quantity whose paired interval excludes 0 in $\geq$ 3/4 cells. Where it contains 0, neither ``attributable'' nor ``not attributable'' is written.
\item \textbf{The interval must exclude 0 on the side of the reading.}
  \begin{itemize}
  \item A1 needs it entirely above 0.
  \item A2, B2 and B3 need it entirely below 0. B2 and B3 are reached only after step P passed, so C-logP relocates, and the attributable direction is C-oracle relocating less.
  \item An interval that excludes 0 on the other side never supports the attribution.
  \end{itemize}
  This rule was added on 2026-09-23, after the plan audit and before the declaration commit.
\item \textbf{Equivalence.} ``Practically unchanged'' and similar wording need the paired interval to lie entirely inside ($-$m, m), as in A3. Medians alone never support it.
\item \textbf{The departure is not an empirical attribution.} It is 100\% misbinning by construction (§1), and is never presented as such an attribution.
\item \textbf{Unchanged rules:}
  \begin{itemize}
  \item ``controls'', ``causes'' and ``explains'' stay forbidden for every comparison;
  \item \texttt{achieved $-$ intended} stays net distortion, not prediction error;
  \item the structural analysis is never cited as evidence about the screener;
  \item C-oracle vs B is never read as a statement about the screener except through C1.
  \end{itemize}
\item \textbf{CLAUDE.md} is updated to this effect \textbf{after} stage 2 exists, not before. The update says that the departure's split is definitional and that its \emph{effect} on accuracy and relocation is what the oracle measures.
\item \textbf{Relation to C-logG.} The C-logG reading of 2026-09-23 used a median-only ``similar'' label. That rule was declared and is not revised post hoc. It did not fire: the outcome was mixed. For the oracle the equivalence reading needs the interval, as above.
\end{itemize}

\subsubsection*{9.5 Expected power, declared now}

The analogue already solved is C-logG $-$ C-log: the same pool and the same stream (\texttt{REP\_SEED\_STREAM}), a different screener, 0--4 of 300 rows shared, five realizations. The oracle's two streams are independent; the analogue is structural, not exact. Its per-realization dR gives:
\begin{itemize}
\item standard deviations of 0.11 (rf/natural), 0.32 (rf/stratified), 0.19 (gbm/natural) and 0.34 (gbm/stratified);
\item 95\% paired-interval half-widths of 0.14--0.42 at n = 5 (t = 2.78), against m = 0.095.
\end{itemize}

If dR\_O varies as much, the consequences are these:
\begin{itemize}
\item \textbf{A3 and B2 are practically unreachable.} An interval inside ($-$m, m) needs a per-realization SD below about 0.07.
\item \textbf{Attribution in $\geq$ 3/4 cells is unlikely} unless $|$dR\_O$|$ exceeds about 0.4 in at least one stratified cell: every set of three cells includes one, and their half-widths are 0.39 and 0.42.
\item \textbf{The realistic outcomes are A4, A5 and B4.} ``No difference detected'' is therefore the \textbf{expected} reading of a null or small effect. It is declared here, before any solve, so that it cannot later read as a post hoc failure. The oracle then contributes the stage-1 $\rho$, the point estimates with their intervals, and the group-C levels.
\end{itemize}

Five realizations are kept on purpose, for the same resolution as every other replicated arm in the paper (B, C-log, C-seq, C-logG). Ten or twenty realizations were considered and rejected (2026-09-23), for three reasons:
\begin{itemize}
\item the oracle is meant to strengthen the explanation, not to carry a result the rest of the paper cannot;
\item equivalence stays out of reach even at twenty realizations;
\item the noisiest cells are limited by test points per zone, not by realizations.
\end{itemize}

A larger test set, such as a held-out sub-pool or chunks solved after the stop, was weighed on the same day and rejected on the same resolution argument. It would give the oracle a test no other arm is read on (§12).

\subsubsection*{10. Intervals}

\begin{itemize}
\item \textbf{Paired:} (C-oracle, C-logP), (C-oracle, C0-oracle), (C-logP, C0-logP).
\item \textbf{Welch:} (C-oracle, B), (C-logP, B), (C0-oracle, B), (C0-logP, B), and (C-logP, C-log), the sanity check.
\item \textbf{Learner interaction:} \texttt{Interaction("C-oracle", "C-logP", method="paired")}.
\item \textbf{New, for §9.2--9.3:} per cell, the paired interval over r1..r5 of \texttt{dR\_O} and of \texttt{dZ\_O} for Z $\in$ \{L, H\}. Each is computed with \texttt{evaluate.paired\_interval} on the per-realization values. They sit beside the medians that \texttt{relocation\_reading} already reads.
\item No entries in \texttt{config.COMPARISONS}: there is no r0, so there is no Wilcoxon. Every conclusion comes from the replication intervals, and an interval containing 0 is inconclusive.
\end{itemize}

\subsection{The C-logG reading rule}\label{sec:s-clogg}

The declared form is what the code repository holds at commit 6b6b822 (\cref{tab:s:chronology}, row~21). Its configuration carries the constants of the relocation reading and the predictions, in these comments, reproduced verbatim:

{\footnotesize
\begin{verbatim}
# Relocation readings (spec 2026-09-23 §10.2), declared before any C-logG point
# was generated. A 10% margin on the log scale, read on the median over the five
# realizations, in at least 3 of the 4 learner x test cells.
RELOCATION_MARGIN = 1.10
RELOCATION_MIN_CELLS = 3

# Predictions for the GBM-screener arm, declared 2026-09-23 BEFORE any C-logG
# point was generated (spec 2026-09-23 §10). Every quantity is a paired difference
# between the GBM and the RF design of the same realization, delta = GBM - RF.
# Stage 1 (training data only, before any test evaluation):
#   allocation closer: mean delta of sum |achieved - intended| <= -10 points, and
#     delta < 0 in at least 4 of 5 realizations;
#   secondary: mean delta of the share staying in the bin its prediction implies
#     >= +10 pp, and mean delta of mean |displacement| <= -0.15 bins, each with the
#     anticipated sign in at least 4 of 5 realizations;
#   points realizing below 0.5 and above 6: reported separately and directionally,
#     no threshold.
# Stage 2 (test): the relocation gate and the readings of RELOCATION_MARGIN above.
# Levels for C-logG against B on the decision range: 1 point improvement,
# 2 interval entirely below 0 in one cell, 3 in all four learner x test cells.
# Only level 3 reopens the seed study.
\end{verbatim}}

The ``spec'' these comments name is a working design document that the repository does not track; nothing here rests on it. The gate and the readings are those of the reading function \texttt{relocation\_reading} in \texttt{sgs/evaluate.py}, committed before the declaration (row~20) and unchanged at it. With $L$ and $H$ the log ratios of C-logG's log-RMSE to B's on $[0.5, 1)$ and $[1, 2)$, $R = H - L$, $m = \ln 1.10$, medians over $r_1$ to $r_5$ per learner $\times$ test cell, and ``in $\geq 3/4$'' meaning in at least three of the four cells:
\begin{itemize}
\item \textbf{Gate.} Relocation is present for C-logG only if median $L < 0$ and median $H > 0$ in $\geq 3/4$.
\item \textbf{If the gate passes:} ``larger'' if median $dR \geq m$ in $\geq 3/4$; otherwise ``similar'' if $|\text{median } dR| < m$ in $\geq 3/4$, a label read on the medians alone; otherwise mixed/inconclusive.
\item \textbf{If the gate fails:} ``reverses'' if median $R \leq -m$ in $\geq 3/4$; otherwise ``vanishes'' if $|\text{median } R| < m$ in $\geq 3/4$; otherwise mixed/inconclusive.
\end{itemize}
The quantity $dR$ is the per-realization paired difference $R(\text{C-logG}) - R(\text{C-log})$, each $R$ taken against the same B, which therefore cancels. At the declaration commit the code computed $dR$ as C-logG's own $R$ against B; it was corrected to the paired difference 17 minutes later (row~22), before any C-logG set was committed (rows~23--24). The median-only ``similar'' label is kept as declared; the oracle's equivalence reading, declared later, requires the interval instead (\cref{sec:s-oracle}, §9.4).

\paragraph{Outcome.} Stage 1 met its thresholds. At stage 2 the gate passes, in three of the four cells (GBM on the stratified test set is the exception). The median $dR$ is $+0.118$ and $-0.512$ under GBM, on the natural and the stratified test set, and $+0.136$ and $-0.010$ under RF: $dR \geq m$ in two cells and $|dR| < m$ in one, so no label reaches three of four, and the reading is mixed/inconclusive under the declared rule. Relocation is thus present by the gate; its size relative to C-log is not classified. Against B on the decision range, C-logG reaches level 1 in three cells and no interval lies entirely below 0. C-logG was chosen post hoc (\cref{sec:arms}).

\needspace{10\baselineskip}
\subsection{The structural procedure}\label{sec:s-structural-procedure}

The procedure of the descriptive analysis of achieved against intended density (\cref{sec:res-bins}), from \texttt{docs/predictions-and-outcomes.md}. It is not a preregistration.

\paragraph{Provenance.} The procedure was committed before any code or output of the structural analysis was committed, in a commit that changes only the procedure's document and the repository's instructions (\cref{tab:s:chronology}, row~15); the structural code and its artifact followed within minutes (row~16). Git shows the order of these commits, and nothing about when the code was written or first run. The procedure was fixed on replication data already computed: the bin-level replication results of every arm, and their figures, had been committed the day before (row~14). It is not a preregistration, and it is not presented as one. It takes no outcome and no ``held?'' column, because it is not a prediction. What it records is the procedure of the structural analysis: the weighting, the bin set, the thresholds and the summary statistic. A first wording of the record claimed more than git shows; it was corrected to the present one on the same day (row~19). The rules below are unchanged since the procedure's commit; the statement of the blinding's limit is the corrected one.

\needspace{8\baselineskip}
\paragraph{The rules (verbatim).}\leavevmode
\begin{itemize}
\item \textbf{Bins.} The 15 geometric bins of the selection range, \texttt{config.N\_BINS}. The
  two bins outside it have no defined intended density; they appear in the
  supplementary rows and document out-of-selection-range spillover and tail
  behaviour, never under the name \emph{truncation}.
\item \textbf{Realizations.} r1--r5 confirmatory; r0 reported apart and never averaged
  into a confirmatory mean.
\item \textbf{Thin bins.} Bins with fewer than \texttt{config.STRUCTURAL\_MIN\_TEST\_N = 10} test
  points are marked and never dropped; no statement in the text rests on one.
\item \textbf{No pooled weighted statistic.} The figure quotes no aggregate, so the
  choice between equal and test-count weighting does not arise. The single
  descriptive number is the count of bins in which the change in density and
  the change in log-RMSE have opposite signs, \texttt{$\Delta$density $\times$ $\Delta$log-RMSE $<$ 0}.
  Because no sentence may rest on a marked bin, that count is reported over the
  \textbf{unmarked} bins; the count over all 15 is reported beside it as a
  descriptive sensitivity, together with how many bins were thin. Bins where
  either difference is exactly zero leave the denominator and are reported.
\item \textbf{Status.} No new hypothesis test or inferential claim is introduced. Welch
  intervals are reused to visualize between-design uncertainty at bin level.
  Inference remains with the pre-declared zone contrasts.
\item \textbf{What \texttt{achieved $-$ intended} is.} The net distortion of the requested
  bin-count distribution induced by screening error --- not the screener's
  prediction error. Two points swapping bins leave it untouched. The
  supplementary displacement rows measure the individual predictions instead,
  and then only on the points the screener selected, which are conditioned on
  having been predicted inside the range and inside a deficit bin.
\item \textbf{Binning conventions.} Everything read from \texttt{replications.json} --- the
  figure's densities and errors, and both sides of the outside table --- uses the
  half-open reporting zones \texttt{[lo, hi)}, whose outer bin is FOS $\geq$ 6. The
  allocation and displacement rows follow the generator's \texttt{(lo, hi]}, whose outer
  bin is FOS $>$ 6, because they are compared against the designs' own bins. The
  two agree except on an exact bin edge.
\item \textbf{What none of it does.} It does not separate the allocation rule from the
  screener's error. That needs a reference without screener error; the oracle
  keeps that job.
\end{itemize}

\paragraph{Blinding, and its limit (verbatim, the commit reference replaced by a row of \cref{tab:s:chronology}).} The blinding is partial.
\begin{itemize}
\item \textbf{What existed.} The bin-level errors and mean training counts of every arm
    were computed and committed before this procedure: \texttt{replications.json} and
    \texttt{replications\_\_\{test\}\_\_\{family\}.pdf} (row~14). Those figures
    draw, on the same 17 bins, the mean log-RMSE over r1--r5 of B, C-log, C-seq,
    C$_0$-log and C$_0$-seq.
\item \textbf{What was read.} The zone-level replication results of the guided arms,
    recorded earlier in this document (for instance C-log $-$ B in 0.5--1 across
    the four cells). One bin-level row of \texttt{replications.json} for a non-guided
    arm: arm A, GBM, test-natural, r0. The design columns of
    \texttt{data/replications/designs/}, including the per-bin selected counts of
    \texttt{guided\_log\_r1}, seen while checking that selected equals intended.
\item \textbf{What is stated as not inspected.} The author states that the bin-level
    errors of the guided arms were not inspected before this procedure was
    fixed. That statement cannot be verified from the repository, because
    those results were available. A file existing does not show that it was
    inspected, and the fact that it was available does not show that it was
    not.
\item \textbf{What the procedure therefore protects.} At most, the choice of the
    bins, thresholds, weighting and summary statistic above, to the extent the
    statement holds. It does not protect the direction of the zone-level
    pattern: knowing that 0.5--1 gains and 1--2 loses already implies, roughly,
    what the bins inside those zones will show. This is one reason the
    analysis is descriptive.
\end{itemize}

\paragraph{Amendment 1: the C$_0$ ablations --- an extension decided after the review of Tasks 0--4.}
\textbf{This is not part of the original declaration above.} The independent review
of the implemented Tasks 0--4 found that the procedure, the design and the
module covered only C-log and C-seq against B, leaving out C$_0$-log and C$_0$-seq
contrary to the repository's rule that every C$_0$ variant appears in every
analysis. The omission was in the declaration itself, not only in the code.
\emph{Provenance:} the amendment was committed in a commit that changes only the procedure's document (\cref{tab:s:chronology}, row~17), and the code and artifact rows for C$_0$ about two minutes later (row~18); git shows that order and nothing about when that code was written or run. The bin-level C$_0$ results it reorganizes were not new: they had been computed and drawn, like those of every other arm, since row~14. The author states that they were not inspected before this amendment was fixed. As with
the declaration above, the repository cannot verify that statement. What was
known for certain is the zone- and decision-range-level C$_0$ replication result
recorded earlier in this document. From the main pipeline, it was also known
that the original C$_0$ under-predicts every FOS $\geq$ 6 test point.

\begin{itemize}
\item \textbf{Unchanged.} The main figure and the concordance count stay C-log and
  C-seq against B, the arms with the same budget (600). No C$_0$ row enters
  either.
\item \textbf{Supplementary ablation rows, error.} For C$_0$-log and C$_0$-seq, for each family $\times$
  test cell and each of the 17 reporting bins (the 15 in-range bins and the
  two outside ones, all \texttt{[lo, hi)} as read from \texttt{replications.json}), the
  difference in mean log-RMSE over r1--r5:
  \begin{itemize}
  \item \textbf{against B, by a Welch interval} (\texttt{config.REP\_WELCH}). Every such row is
    labelled as a comparison with \textbf{different budgets}: C$_0$ trains on the
    300 guided points alone, B on 600. The difference mixes the budget with
    the training distribution and is not a like-for-like comparison.
  \item \textbf{against the corresponding C, by a paired interval} (\texttt{config.REP\_PAIRED}:
    C-log $-$ C$_0$-log, C-seq $-$ C$_0$-seq, in that order). Within a realization, C$_0$
    trains on exactly the guided points that C adds to the seed, so the
    realization index is a genuine pairing factor.
  \end{itemize}
  Thin bins are marked by the same \texttt{STRUCTURAL\_MIN\_TEST\_N} rule and never
  dropped. r0 does not enter these rows. The intervals visualize
  between-design variability; they are no test, and an interval containing 0
  is inconclusive, never a tie.
\item \textbf{Supplementary ablation rows, training distribution.} Where the C$_0$ points
  are set against B, the quantity is the fraction \texttt{n\_bin / N\_total} over the
  17 bins, never a difference of raw counts, which would carry the budget
  difference. It is shown per bin as C$_0$ $-$ B with a Welch interval, over
  r1--r5. These fractions do not enter the concordance count or any other
  summary.
\item \textbf{Allocation and displacement.} C$_0$-log and C$_0$-seq use the same guided
  points as C-log and C-seq in each realization. Their allocation, batch and
  displacement rows are those of C, shared rather than duplicated, and the
  artifact says so.
\item \textbf{The original C$_0$} (arm \texttt{C0}, role \texttt{guided}) exists only as r0 in the main
  pipeline. It was never replicated, so it has no replication interval. It
  also predates the log-deficit rule and stores no allocation provenance, so
  it has no allocation row. The artifact names it in a note and computes
  nothing for it.
\end{itemize}

\paragraph{Amendment 2 (2026-09-23 or the date of this commit): the pool arms in the structural analysis.}
The title is the source's; the commit is dated in \cref{tab:s:chronology}, row~36.
C-oracle and C-logP join \texttt{ARMS\_VS\_B}, and C0-oracle and C0-logP join the ablations, under the same
procedure as C-logG: descriptive, Welch against B, paired against the corresponding C. There is no
r0 for any of them. C-oracle's allocation and displacement rows are exact by construction; the
table shows them as a check, not as a finding. The overlap between C-oracle and C-logP is counted
by shared pool row. Recorded after the stage-1 note and the replication metrics of the pool arms
existed, as the git history shows; nothing here is a preregistration.

\subsection{The C-seq predictions, read on the single realization}\label{sec:s-cseq}

Declared with C-seq (\cref{tab:s:chronology}, row~9), before its batches were committed, and read on the single realization $r_0$, which is exploratory; the outcomes are recorded as observed, including the prediction that failed. Outside the replications there is one realization per design, and the paired bootstrap covers test-sampling variability only. The replicated reading of C-seq against C-log is in \cref{sec:res-sensitivity}.

\begin{table}[htbp]
\caption{C-seq predictions on training data, checked before any test evaluation, and their outcomes ($r_0$). ``guided-log'' is the C-log set.}\label{tab:s:cseq-train}
\small
\begin{tabular}{@{}p{0.40\linewidth}p{0.42\linewidth}p{0.10\linewidth}@{}}
\toprule
prediction & outcome & held? \\
\midrule
the three batches put $\leq$ 10 points below FOS 0.5 (guided-log: 45) & 32 (per batch 22 / 6 / 4); the first batch, whose screener saw only the seed, behaves like guided-log & \textbf{no} \\
seed + batches deviate less from the target per-bin counts than seed + guided-log & sum of absolute deviations 66 vs 92 & yes \\
\bottomrule
\end{tabular}
\end{table}

\begin{table}[htbp]
\caption{C-seq predictions on the test sets, log-RMSE against B, and their outcomes ($r_0$, tuned fits; \cref{tab:s:chronology}, row~11). On the decision range [0.5, 2), every C-seq vs B bootstrap interval of $\Delta$RMSE on the log scale contains 0.}\label{tab:s:cseq-test}
\small
\begin{tabular}{@{}p{0.30\linewidth}p{0.52\linewidth}p{0.10\linewidth}@{}}
\toprule
prediction & outcome & held? \\
\midrule
C-log's advantage on test points with FOS $<$ 0.5 shrinks for C-seq & shrinks in 3 of 4 family $\times$ test cells; grows under RF on test-natural & mostly \\
C-seq may gain in [0.5, 1) & gains in 3 of 4 cells; slightly worse in the fourth (GBM, test-stratified: 0.054 vs 0.052, n = 12) & mostly \\
C-seq ties or loses in [1, 2) & loses in 4 of 4 cells & yes \\
no systematic global gain & RF: C-seq below B on both tests; GBM: above B on both tests & yes \\
\bottomrule
\end{tabular}
\end{table}

\clearpage
\section{Selected hyper-parameters}\label{sec:s-params}

The hyper-parameters selected by cross-validation for every arm of the single realization, under the log criterion and under the FOS-scale sensitivity criterion (\cref{sec:tuning}). The pool arms have no single realization and are fitted with the common hyper-parameters only.

\begin{table}[htbp]
\centering
\caption{Hyper-parameters selected by CV under each scoring, one column per hyper-parameter (scikit-learn names: rate: learning\_rate; max feat.: max\_features, a fraction of the features, 1.0 = all; iter.: max\_iter; leaves: max\_leaf\_nodes; min leaf: min\_samples\_leaf). C$_0$, C$_0$-log, C$_0$-seq and C$_0$-logG are ablations: they train on the added points without the seed and carry no budget.}
\label{tab:scorings:params}
\textit{RF}\par\smallskip
\begin{tabular}{lrrrr}
\toprule
 & \multicolumn{2}{c}{log-rmse} & \multicolumn{2}{c}{fos-rmse} \\
\cmidrule(lr){2-3} \cmidrule(lr){4-5}
arm & max feat. & min leaf & max feat. & min leaf \\
\midrule
A & 1.0 & 1 & 1.0 & 2 \\
B & 1.0 & 1 & 1.0 & 1 \\
C & 1.0 & 1 & 1.0 & 1 \\
C$_0$ & 1.0 & 1 & 1.0 & 1 \\
C-log & 1.0 & 1 & 1.0 & 1 \\
C$_0$-log & 1.0 & 1 & 1.0 & 1 \\
C-seq & 1.0 & 1 & 1.0 & 1 \\
C$_0$-seq & 1.0 & 1 & 1.0 & 1 \\
C-logG & 1.0 & 1 & 1.0 & 1 \\
C$_0$-logG & 1.0 & 1 & 1.0 & 1 \\
\bottomrule
\end{tabular}
\par\medskip
\textit{GBM}\par\smallskip
\begin{tabular}{lrrrrrrrr}
\toprule
 & \multicolumn{4}{c}{log-rmse} & \multicolumn{4}{c}{fos-rmse} \\
\cmidrule(lr){2-5} \cmidrule(lr){6-9}
arm & rate & iter. & leaves & min leaf & rate & iter. & leaves & min leaf \\
\midrule
A & 0.06 & 8000 & 3 & 3 & 0.06 & 4000 & 3 & 2 \\
B & 0.06 & 8000 & 3 & 3 & 0.06 & 8000 & 4 & 1 \\
C & 0.06 & 8000 & 3 & 2 & 0.03 & 8000 & 3 & 2 \\
C$_0$ & 0.03 & 8000 & 4 & 5 & 0.1 & 2000 & 3 & 5 \\
C-log & 0.06 & 8000 & 3 & 3 & 0.1 & 8000 & 3 & 1 \\
C$_0$-log & 0.1 & 8000 & 3 & 5 & 0.1 & 2000 & 3 & 5 \\
C-seq & 0.03 & 8000 & 3 & 2 & 0.1 & 4000 & 3 & 2 \\
C$_0$-seq & 0.1 & 8000 & 3 & 5 & 0.03 & 8000 & 3 & 2 \\
C-logG & 0.03 & 8000 & 3 & 3 & 0.03 & 8000 & 3 & 3 \\
C$_0$-logG & 0.06 & 8000 & 3 & 5 & 0.03 & 8000 & 3 & 5 \\
\bottomrule
\end{tabular}
\end{table}

\clearpage
\section{Learning curve}\label{sec:s-learning-curve}

The learning curve covers B and C, the linear-bin design, only, with hyper-parameters fixed at every budget (\cref{sec:tuning}). It does not cover C-log, C-seq, C-logG, the pool arms or any C$_0$ ablation, and it is not used to extrapolate the relocation to other budgets.

\begin{figure}[htbp]\centering
\includegraphics[width=\linewidth]{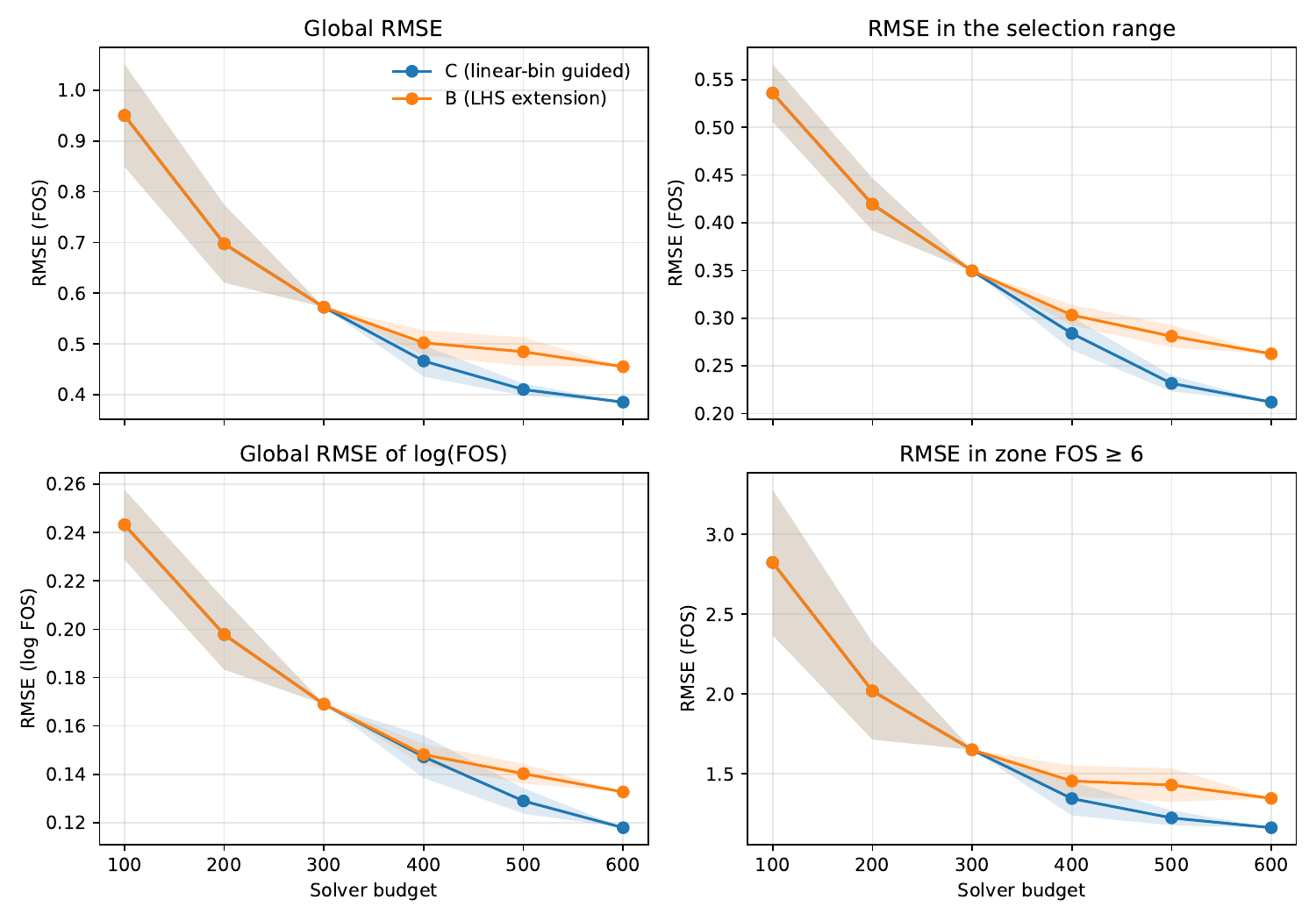}
\caption[Learning curve, natural test set, RF]{Learning curve on test-natural, RF, for B (seed + Latin hypercube extension) and C (seed + the linear-bin guided design); C-log, C-seq, C-logG, the pool arms and every C$_0$ ablation are not covered. Hyper-parameters are fixed at every budget, not tuned per point: scikit-learn defaults. Budgets 100 and 200 are random subsets of the 300-point seed, shared by both curves; 300 is the seed itself; 400 and 500 add a random subset of the arm's 300 added points to the seed, so the subset of C is not itself a guided design at that budget; 600 is the arm's whole training set. Points are means over 10 subsamples, and the band is plus or minus one standard deviation over them; at 300 and 600 there is a single fit and no band. The band covers subsampling only. Panels: global RMSE (FOS), RMSE on the selection range (FOS), global RMSE of log(FOS), and RMSE (FOS) in the highest reporting zone. The curve is not used to extrapolate the relocation to other budgets.}\label{fig:s:lc:natural:rf}
\end{figure}
\begin{figure}[htbp]\centering
\includegraphics[width=\linewidth]{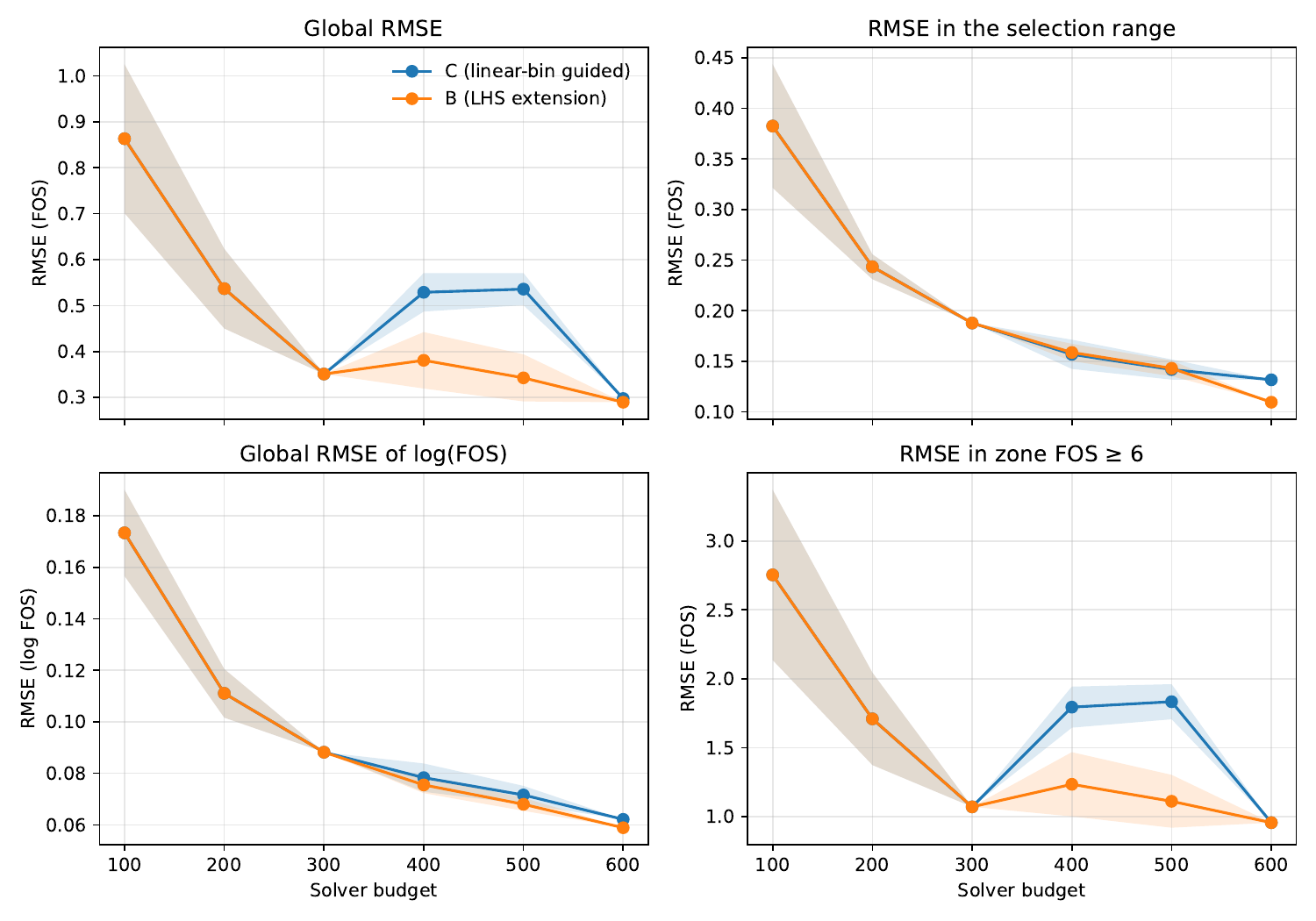}
\caption[Learning curve, natural test set, GBM]{Learning curve on test-natural, GBM, for B (seed + Latin hypercube extension) and C (seed + the linear-bin guided design); C-log, C-seq, C-logG, the pool arms and every C$_0$ ablation are not covered. Hyper-parameters are fixed at every budget, not tuned per point: the grid point selected by cross-validation on the seed alone (learning\_rate = 0.06, max\_iter = 8000, max\_leaf\_nodes = 3, min\_samples\_leaf = 3). Budgets 100 and 200 are random subsets of the 300-point seed, shared by both curves; 300 is the seed itself; 400 and 500 add a random subset of the arm's 300 added points to the seed, so the subset of C is not itself a guided design at that budget; 600 is the arm's whole training set. Points are means over 10 subsamples, and the band is plus or minus one standard deviation over them; at 300 and 600 there is a single fit and no band. The band covers subsampling only. Panels: global RMSE (FOS), RMSE on the selection range (FOS), global RMSE of log(FOS), and RMSE (FOS) in the highest reporting zone. The curve is not used to extrapolate the relocation to other budgets.}\label{fig:s:lc:natural:gbm}
\end{figure}
\begin{figure}[htbp]\centering
\includegraphics[width=\linewidth]{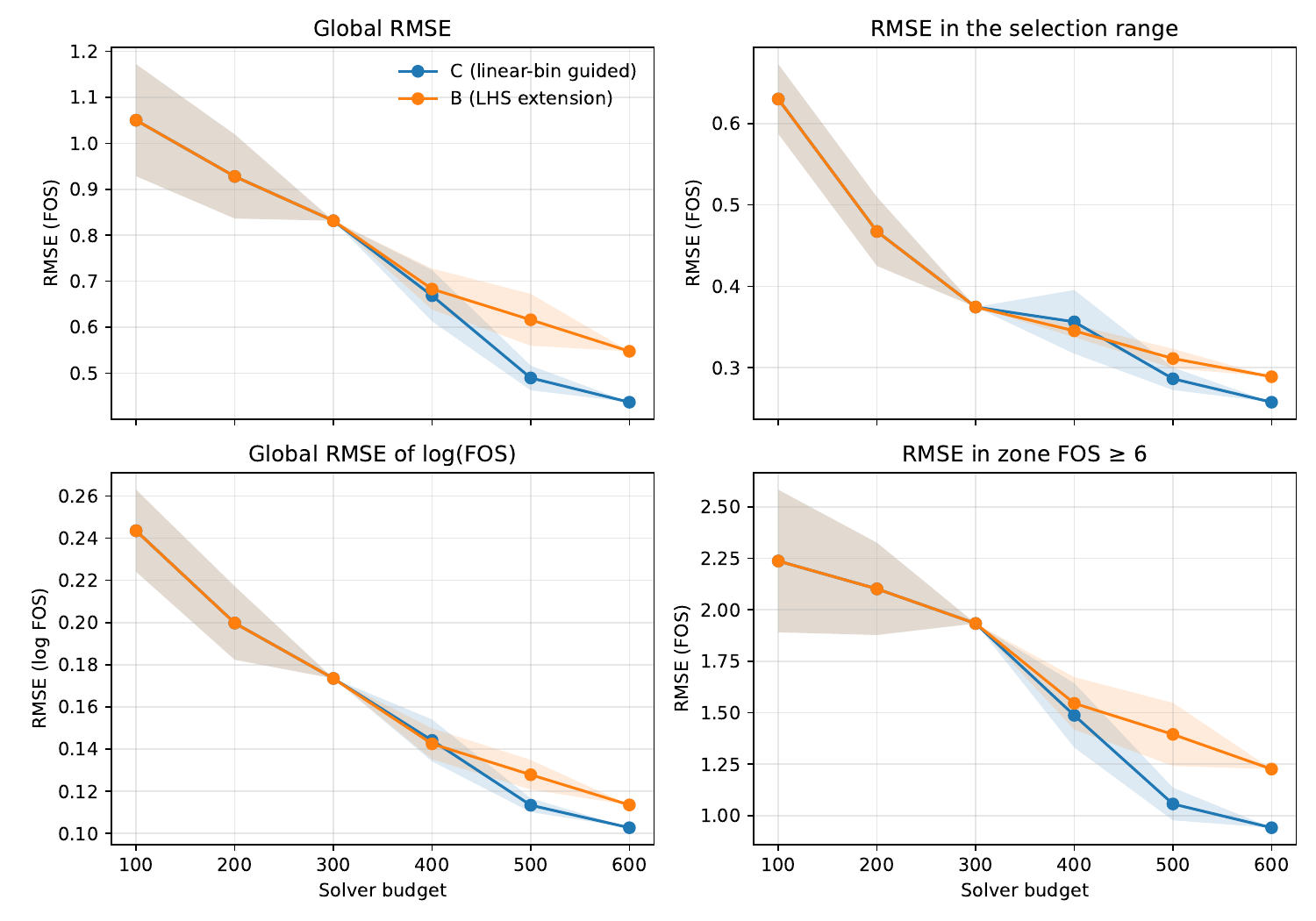}
\caption[Learning curve, stratified test set, RF]{Learning curve on test-stratified, RF, for B (seed + Latin hypercube extension) and C (seed + the linear-bin guided design); C-log, C-seq, C-logG, the pool arms and every C$_0$ ablation are not covered. Hyper-parameters are fixed at every budget, not tuned per point: scikit-learn defaults. Budgets 100 and 200 are random subsets of the 300-point seed, shared by both curves; 300 is the seed itself; 400 and 500 add a random subset of the arm's 300 added points to the seed, so the subset of C is not itself a guided design at that budget; 600 is the arm's whole training set. Points are means over 10 subsamples, and the band is plus or minus one standard deviation over them; at 300 and 600 there is a single fit and no band. The band covers subsampling only. Panels: global RMSE (FOS), RMSE on the selection range (FOS), global RMSE of log(FOS), and RMSE (FOS) in the highest reporting zone. The curve is not used to extrapolate the relocation to other budgets.}\label{fig:s:lc:stratified:rf}
\end{figure}
\begin{figure}[htbp]\centering
\includegraphics[width=\linewidth]{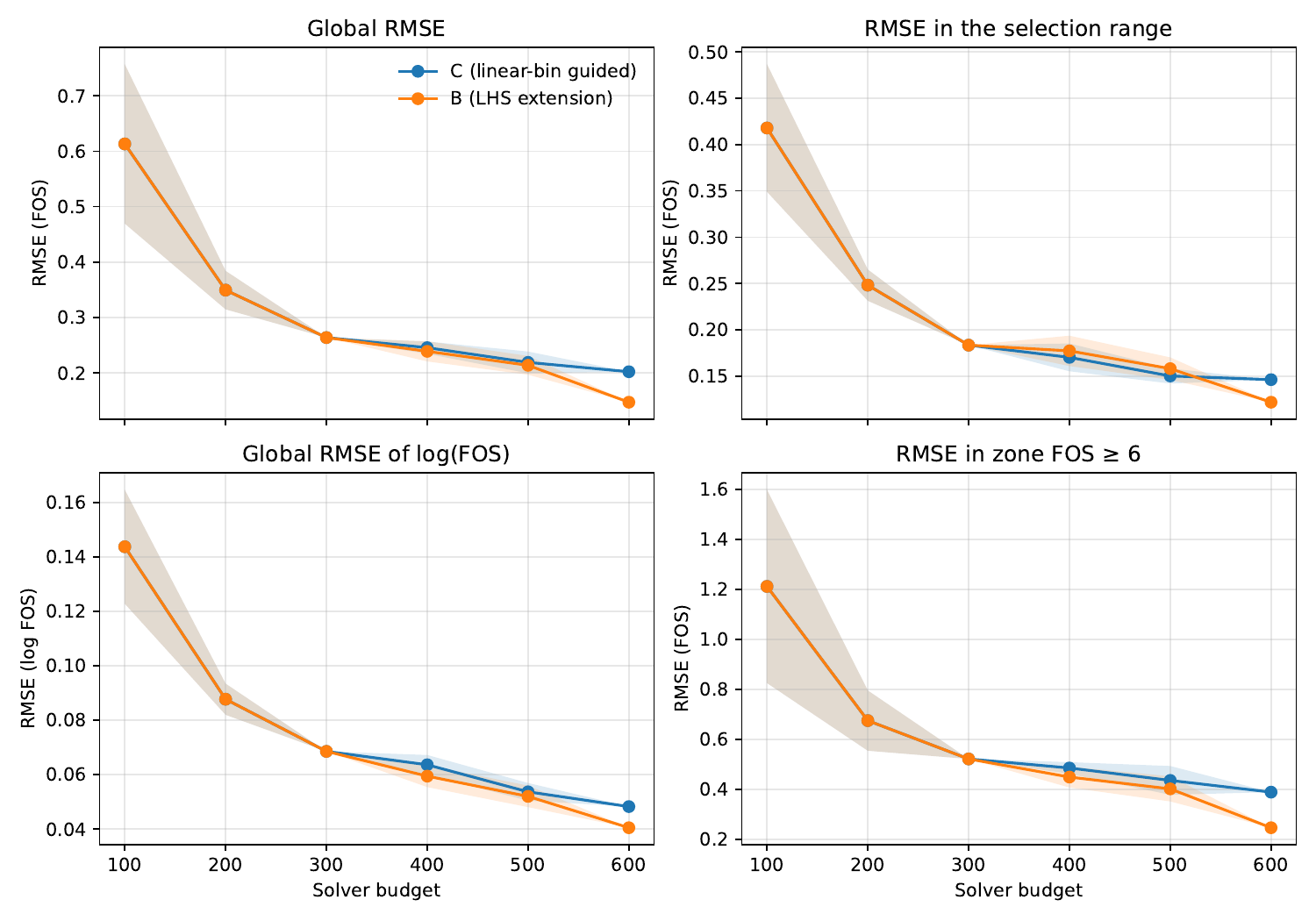}
\caption[Learning curve, stratified test set, GBM]{Learning curve on test-stratified, GBM, for B (seed + Latin hypercube extension) and C (seed + the linear-bin guided design); C-log, C-seq, C-logG, the pool arms and every C$_0$ ablation are not covered. Hyper-parameters are fixed at every budget, not tuned per point: the grid point selected by cross-validation on the seed alone (learning\_rate = 0.06, max\_iter = 8000, max\_leaf\_nodes = 3, min\_samples\_leaf = 3). Budgets 100 and 200 are random subsets of the 300-point seed, shared by both curves; 300 is the seed itself; 400 and 500 add a random subset of the arm's 300 added points to the seed, so the subset of C is not itself a guided design at that budget; 600 is the arm's whole training set. Points are means over 10 subsamples, and the band is plus or minus one standard deviation over them; at 300 and 600 there is a single fit and no band. The band covers subsampling only. Panels: global RMSE (FOS), RMSE on the selection range (FOS), global RMSE of log(FOS), and RMSE (FOS) in the highest reporting zone. The curve is not used to extrapolate the relocation to other budgets.}\label{fig:s:lc:stratified:gbm}
\end{figure}

\clearpage
\section{Replications}\label{sec:s-replications}

Per-realization values on the decision range, every replicated comparison and the learner interactions (\cref{sec:replications,sec:statistics}), then the bin diagnostic of every replicated arm. $r_0$ is shown in the value tables and enters no interval.

\begin{table}[htbp]
\centering
\caption{Log-RMSE on the decision range, test-natural, every realization; r0 = existing sets (exploratory). Common fixed hyper-parameters. C$_0$-log, C$_0$-seq, C$_0$-logG, C$_0$-oracle and C$_0$-logP are ablations without the seed and carry no budget: they train on 300 points against the 600 of B, C-log, C-seq, C-logG, C-oracle and C-logP. C-oracle, C-logP, C$_0$-oracle and C$_0$-logP have no r0 (shown as '--').}
\label{tab:rep:test-natural:values}
\begin{tabular}{lrrrrrrrrr}
\toprule
family & arm & r0 (exploratory) & r1 & r2 & r3 & r4 & r5 & mean & sd \\
\midrule
gbm & A & 0.076 & -- & -- & -- & -- & -- & -- & -- \\
gbm & B & 0.050 & 0.050 & 0.052 & 0.050 & 0.051 & 0.043 & 0.049 & 0.003 \\
gbm & C-log & 0.052 & 0.061 & 0.052 & 0.061 & 0.052 & 0.065 & 0.058 & 0.006 \\
gbm & C-logG & 0.056 & 0.056 & 0.052 & 0.053 & 0.052 & 0.048 & 0.052 & 0.003 \\
gbm & C-logP & -- & 0.054 & 0.055 & 0.044 & 0.055 & 0.056 & 0.053 & 0.005 \\
gbm & C-oracle & -- & 0.047 & 0.055 & 0.051 & 0.053 & 0.047 & 0.051 & 0.003 \\
gbm & C-seq & 0.053 & 0.057 & 0.050 & 0.046 & 0.058 & 0.056 & 0.053 & 0.005 \\
\bottomrule
\end{tabular}
\end{table}

\begin{table}[htbp]
\centering
\caption{Log-RMSE on the decision range, test-stratified, every realization; r0 = existing sets (exploratory). Common fixed hyper-parameters. C$_0$-log, C$_0$-seq, C$_0$-logG, C$_0$-oracle and C$_0$-logP are ablations without the seed and carry no budget: they train on 300 points against the 600 of B, C-log, C-seq, C-logG, C-oracle and C-logP. C-oracle, C-logP, C$_0$-oracle and C$_0$-logP have no r0 (shown as '--').}
\label{tab:rep:test-stratified:values}
\begin{tabular}{lrrrrrrrrr}
\toprule
family & arm & r0 (exploratory) & r1 & r2 & r3 & r4 & r5 & mean & sd \\
\midrule
gbm & A & 0.082 & -- & -- & -- & -- & -- & -- & -- \\
gbm & B & 0.045 & 0.052 & 0.070 & 0.042 & 0.056 & 0.044 & 0.053 & 0.011 \\
gbm & C-log & 0.039 & 0.042 & 0.062 & 0.042 & 0.053 & 0.043 & 0.048 & 0.009 \\
gbm & C-logG & 0.055 & 0.058 & 0.042 & 0.047 & 0.047 & 0.047 & 0.048 & 0.006 \\
gbm & C-logP & -- & 0.046 & 0.042 & 0.054 & 0.039 & 0.043 & 0.045 & 0.005 \\
gbm & C-oracle & -- & 0.039 & 0.048 & 0.046 & 0.041 & 0.047 & 0.044 & 0.004 \\
gbm & C-seq & 0.056 & 0.057 & 0.046 & 0.055 & 0.040 & 0.047 & 0.049 & 0.007 \\
\bottomrule
\end{tabular}
\end{table}

\begin{table}[htbp]
\centering
\caption{Learner interaction on log(log-RMSE\_rf) − log(log-RMSE\_gbm) between arms; Welch, or paired where the two arms share a realization's candidate pool and intended allocation. This table: test-natural.}
\label{tab:rep:interactions:test-natural}
\begin{tabular}{lrrrrrr}
\toprule
measure & contrast & method & diff & CI low & CI high & df \\
\midrule
all & C-log vs B & welch & -0.195 & -0.293 & -0.097 & 6.460 \\
all & C-seq vs B & welch & -0.100 & -0.219 & 0.020 & 7.949 \\
all & C-logG vs B & welch & -0.042 & -0.146 & 0.062 & 7.539 \\
all & C-logG vs C-log & paired & 0.154 & 0.041 & 0.266 & 4.000 \\
all & C-oracle vs C-logP & paired & 0.064 & -0.098 & 0.225 & 4.000 \\
selection-range & C-log vs B & welch & -0.122 & -0.193 & -0.051 & 7.950 \\
selection-range & C-seq vs B & welch & -0.008 & -0.119 & 0.103 & 6.301 \\
selection-range & C-logG vs B & welch & -0.048 & -0.132 & 0.037 & 7.581 \\
selection-range & C-logG vs C-log & paired & 0.074 & -0.015 & 0.163 & 4.000 \\
selection-range & C-oracle vs C-logP & paired & -0.067 & -0.286 & 0.152 & 4.000 \\
decision-range & C-log vs B & welch & -0.249 & -0.350 & -0.148 & 6.188 \\
decision-range & C-seq vs B & welch & -0.066 & -0.177 & 0.045 & 5.836 \\
decision-range & C-logG vs B & welch & -0.131 & -0.234 & -0.027 & 6.096 \\
decision-range & C-logG vs C-log & paired & 0.118 & -0.032 & 0.268 & 4.000 \\
decision-range & C-oracle vs C-logP & paired & -0.093 & -0.337 & 0.152 & 4.000 \\
0.5–1 & C-log vs B & welch & -0.234 & -0.434 & -0.033 & 7.433 \\
0.5–1 & C-seq vs B & welch & 0.012 & -0.237 & 0.260 & 6.405 \\
0.5–1 & C-logG vs B & welch & -0.148 & -0.334 & 0.038 & 7.766 \\
0.5–1 & C-logG vs C-log & paired & 0.086 & -0.232 & 0.403 & 4.000 \\
0.5–1 & C-oracle vs C-logP & paired & -0.369 & -0.809 & 0.071 & 4.000 \\
1–2 & C-log vs B & welch & -0.172 & -0.285 & -0.059 & 7.812 \\
1–2 & C-seq vs B & welch & -0.021 & -0.147 & 0.105 & 7.319 \\
1–2 & C-logG vs B & welch & -0.037 & -0.154 & 0.080 & 7.673 \\
1–2 & C-logG vs C-log & paired & 0.135 & 0.004 & 0.266 & 4.000 \\
1–2 & C-oracle vs C-logP & paired & 0.014 & -0.249 & 0.276 & 4.000 \\
\bottomrule
\end{tabular}
\end{table}
\begin{table}[htbp]
\centering
\caption{Learner interaction on log(log-RMSE\_rf) − log(log-RMSE\_gbm) between arms; Welch, or paired where the two arms share a realization's candidate pool and intended allocation. This table: test-stratified.}
\label{tab:rep:interactions:test-stratified}
\begin{tabular}{lrrrrrr}
\toprule
measure & contrast & method & diff & CI low & CI high & df \\
\midrule
all & C-log vs B & welch & -0.172 & -0.401 & 0.056 & 5.160 \\
all & C-seq vs B & welch & -0.167 & -0.412 & 0.079 & 7.464 \\
all & C-logG vs B & welch & 0.030 & -0.199 & 0.259 & 4.752 \\
all & C-logG vs C-log & paired & 0.203 & 0.112 & 0.293 & 4.000 \\
all & C-oracle vs C-logP & paired & 0.278 & 0.129 & 0.426 & 4.000 \\
selection-range & C-log vs B & welch & -0.050 & -0.261 & 0.161 & 6.420 \\
selection-range & C-seq vs B & welch & -0.054 & -0.262 & 0.153 & 5.675 \\
selection-range & C-logG vs B & welch & -0.032 & -0.239 & 0.175 & 4.939 \\
selection-range & C-logG vs C-log & paired & 0.018 & -0.072 & 0.109 & 4.000 \\
selection-range & C-oracle vs C-logP & paired & 0.005 & -0.167 & 0.178 & 4.000 \\
decision-range & C-log vs B & welch & 0.027 & -0.263 & 0.317 & 7.088 \\
decision-range & C-seq vs B & welch & 0.042 & -0.191 & 0.274 & 7.984 \\
decision-range & C-logG vs B & welch & -0.009 & -0.218 & 0.201 & 7.724 \\
decision-range & C-logG vs C-log & paired & -0.035 & -0.295 & 0.224 & 4.000 \\
decision-range & C-oracle vs C-logP & paired & -0.162 & -0.365 & 0.040 & 4.000 \\
0.5–1 & C-log vs B & welch & 0.081 & -0.229 & 0.391 & 6.881 \\
0.5–1 & C-seq vs B & welch & -0.094 & -0.421 & 0.233 & 7.657 \\
0.5–1 & C-logG vs B & welch & -0.201 & -0.514 & 0.113 & 7.106 \\
0.5–1 & C-logG vs C-log & paired & -0.282 & -0.629 & 0.065 & 4.000 \\
0.5–1 & C-oracle vs C-logP & paired & -0.243 & -0.496 & 0.009 & 4.000 \\
1–2 & C-log vs B & welch & 0.109 & -0.257 & 0.474 & 7.971 \\
1–2 & C-seq vs B & welch & 0.186 & -0.144 & 0.515 & 7.039 \\
1–2 & C-logG vs B & welch & 0.174 & -0.216 & 0.564 & 7.961 \\
1–2 & C-logG vs C-log & paired & 0.065 & -0.308 & 0.439 & 4.000 \\
1–2 & C-oracle vs C-logP & paired & -0.106 & -0.458 & 0.246 & 4.000 \\
\bottomrule
\end{tabular}
\end{table}

\begingroup
\small
\setlength{\tabcolsep}{4pt}
\setlength{\LTcapwidth}{\textwidth}
\begin{longtable}{llllrrrrr}
\caption{Replicated comparisons on test-natural: Welch (independent arms) or paired intervals, log-RMSE and relative scale; P(first<second) is descriptive. C$_0$-log, C$_0$-seq, C$_0$-logG, C$_0$-oracle and C$_0$-logP are ablations without the seed and carry no budget: they train on 300 points against the 600 of B, C-log, C-seq, C-logG, C-oracle and C-logP. C-oracle, C-logP, C$_0$-oracle and C$_0$-logP have no r0 (shown as '--').}\label{tab:rep:test-natural:comparisons} \\
\toprule
measure & comparison & method & scale & diff & CI low & CI high & df & P(first<second) \\
\midrule
\endfirsthead
\multicolumn{9}{l}{\textit{Table \thetable, continued}} \\
\toprule
measure & comparison & method & scale & diff & CI low & CI high & df & P(first<second) \\
\midrule
\endhead
\midrule
\multicolumn{9}{r}{\textit{continued on the next page}} \\
\endfoot
\bottomrule
\endlastfoot
\multicolumn{9}{l}{\textit{GBM}} \\
all & C-log vs B & welch & log-rmse & 0.007 & 0.002 & 0.013 & 7.135 & 0.040 \\
all & C-log vs B & welch & relative & 0.124 & 0.025 & 0.223 & 6.558 & 0.040 \\
all & C-seq vs B & welch & log-rmse & 0.007 & 0.001 & 0.012 & 7.073 & 0.080 \\
all & C-seq vs B & welch & relative & 0.116 & 0.017 & 0.216 & 6.668 & 0.080 \\
all & C$_0$-log vs B & welch & log-rmse & 0.116 & 0.109 & 0.123 & 7.661 & 0.000 \\
all & C$_0$-log vs B & welch & relative & 1.133 & 1.037 & 1.230 & 5.233 & 0.000 \\
all & C$_0$-seq vs B & welch & log-rmse & 0.122 & 0.110 & 0.134 & 5.452 & 0.000 \\
\end{longtable}
\endgroup

\begingroup
\small
\setlength{\tabcolsep}{4pt}
\setlength{\LTcapwidth}{\textwidth}
\begin{longtable}{llllrrrrr}
\caption{Replicated comparisons on test-stratified: Welch (independent arms) or paired intervals, log-RMSE and relative scale; P(first<second) is descriptive. C$_0$-log, C$_0$-seq, C$_0$-logG, C$_0$-oracle and C$_0$-logP are ablations without the seed and carry no budget: they train on 300 points against the 600 of B, C-log, C-seq, C-logG, C-oracle and C-logP. C-oracle, C-logP, C$_0$-oracle and C$_0$-logP have no r0 (shown as '--').}\label{tab:rep:test-stratified:comparisons} \\
\toprule
measure & comparison & method & scale & diff & CI low & CI high & df & P(first<second) \\
\midrule
\endfirsthead
\multicolumn{9}{l}{\textit{Table \thetable, continued}} \\
\toprule
measure & comparison & method & scale & diff & CI low & CI high & df & P(first<second) \\
\midrule
\endhead
\midrule
\multicolumn{9}{r}{\textit{continued on the next page}} \\
\endfoot
\bottomrule
\endlastfoot
\multicolumn{9}{l}{\textit{GBM}} \\
all & C-log vs B & welch & log-rmse & 0.000 & -0.007 & 0.007 & 6.056 & 0.520 \\
all & C-log vs B & welch & relative & 0.005 & -0.138 & 0.147 & 6.089 & 0.520 \\
all & C-seq vs B & welch & log-rmse & 0.004 & -0.003 & 0.011 & 4.538 & 0.280 \\
all & C-seq vs B & welch & relative & 0.085 & -0.056 & 0.227 & 4.443 & 0.280 \\
all & C$_0$-log vs B & welch & log-rmse & 0.032 & 0.021 & 0.044 & 6.691 & 0.000 \\
all & C$_0$-log vs B & welch & relative & 0.502 & 0.335 & 0.669 & 7.998 & 0.000 \\
all & C$_0$-seq vs B & welch & log-rmse & 0.047 & 0.031 & 0.063 & 5.460 & 0.000 \\
\end{longtable}
\endgroup

\begin{figure}[htbp]\centering
\includegraphics[width=\linewidth]{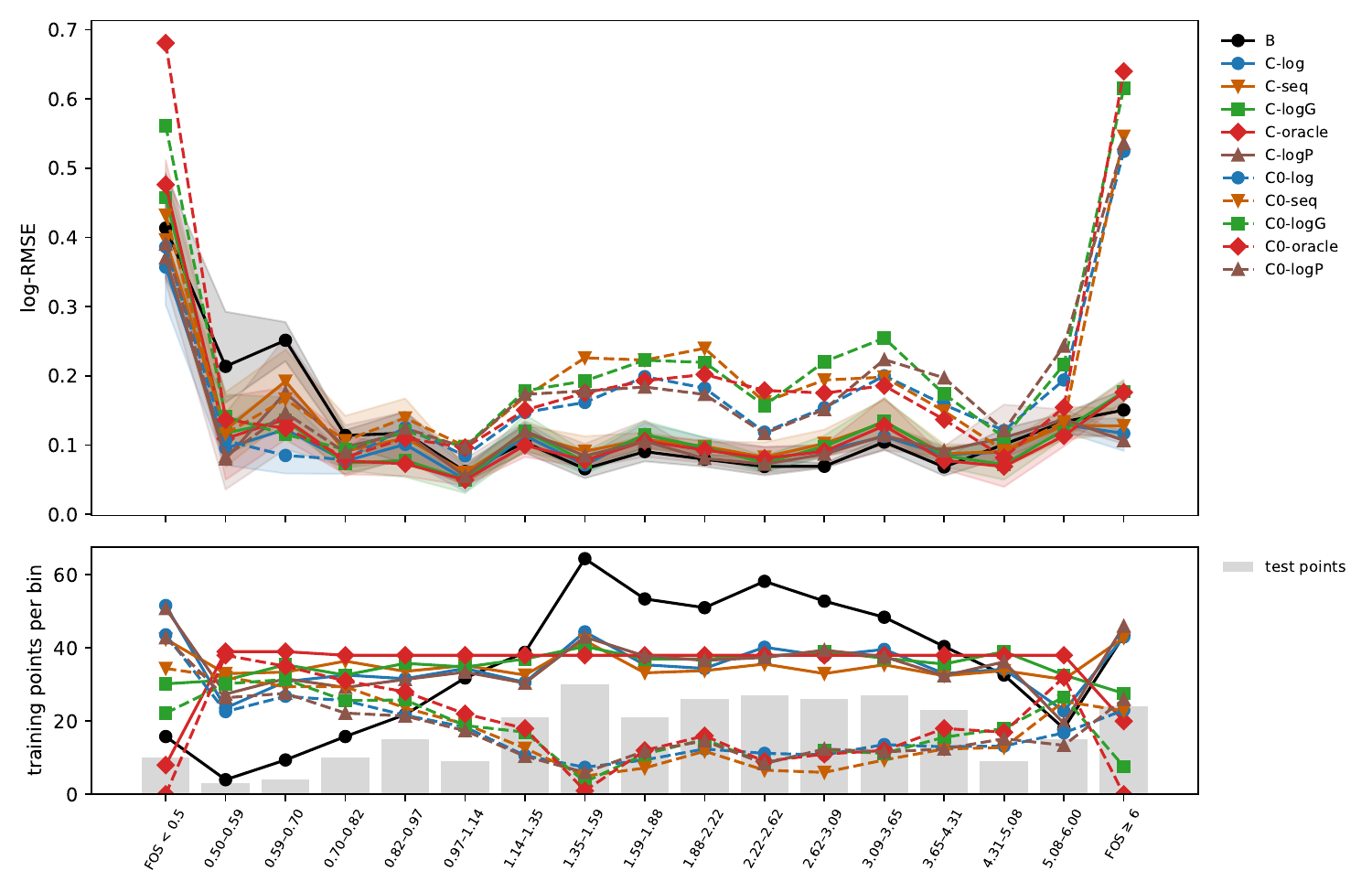}
\caption[Bin diagnostic, natural test set, RF]{Bin diagnostic on test-natural, RF, over the confirmatory realizations r1-r5: the 15 bins of the selection range and the two zones outside it, half-open, [lo, hi), as everywhere the replication metrics are read; edges are rounded to two decimals. Top: log-RMSE per bin, mean over the realizations, with the min-max range over them shaded for B, C-log, C-seq, C-logG, C-oracle and C-logP. Bottom: mean training points per bin, each arm in its style above, and the test points per bin (grey bars). The shaded range is no interval and no test. C$_0$-log, C$_0$-seq, C$_0$-logG, C$_0$-oracle and C$_0$-logP are ablations, drawn dashed in the colour of their arm: they train on the 300 added points without the seed, against the 600 of the other arms, so they join different budgets and are not like-for-like comparisons. A, C and the original C$_0$ have no replications and are not drawn; r0 is exploratory and is not drawn. Common hyper-parameters fixed from the seed. Descriptive only; bins with few test points are not marked here (see the structural analysis).}\label{fig:s:rep:natural:rf}
\end{figure}
\begin{figure}[htbp]\centering
\includegraphics[width=\linewidth]{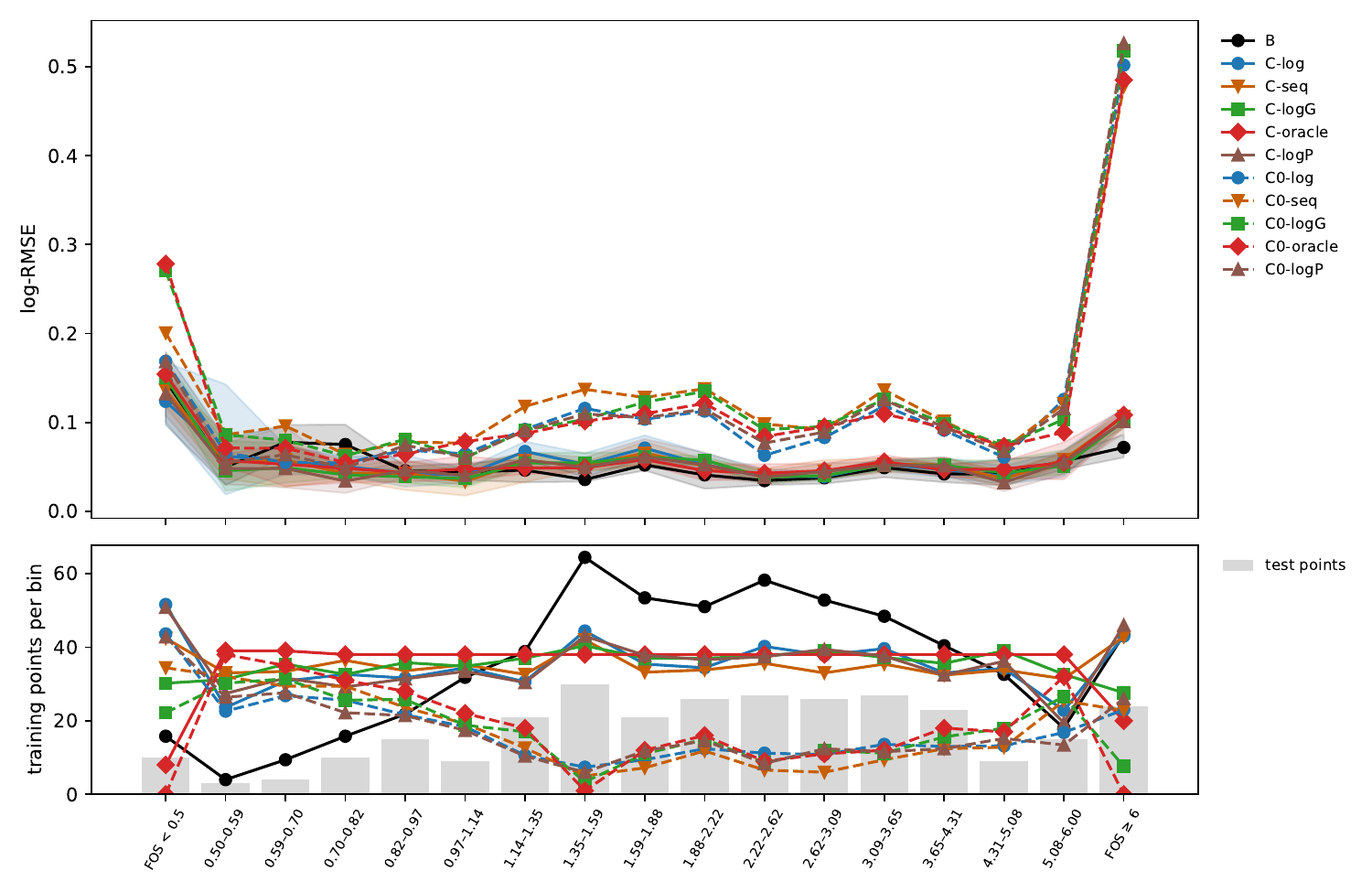}
\caption[Bin diagnostic, natural test set, GBM]{Bin diagnostic on test-natural, GBM, over the confirmatory realizations r1-r5: the 15 bins of the selection range and the two zones outside it, half-open, [lo, hi), as everywhere the replication metrics are read; edges are rounded to two decimals. Top: log-RMSE per bin, mean over the realizations, with the min-max range over them shaded for B, C-log, C-seq, C-logG, C-oracle and C-logP. Bottom: mean training points per bin, each arm in its style above, and the test points per bin (grey bars). The shaded range is no interval and no test. C$_0$-log, C$_0$-seq, C$_0$-logG, C$_0$-oracle and C$_0$-logP are ablations, drawn dashed in the colour of their arm: they train on the 300 added points without the seed, against the 600 of the other arms, so they join different budgets and are not like-for-like comparisons. A, C and the original C$_0$ have no replications and are not drawn; r0 is exploratory and is not drawn. Common hyper-parameters fixed from the seed. Descriptive only; bins with few test points are not marked here (see the structural analysis).}\label{fig:s:rep:natural:gbm}
\end{figure}
\begin{figure}[htbp]\centering
\includegraphics[width=\linewidth]{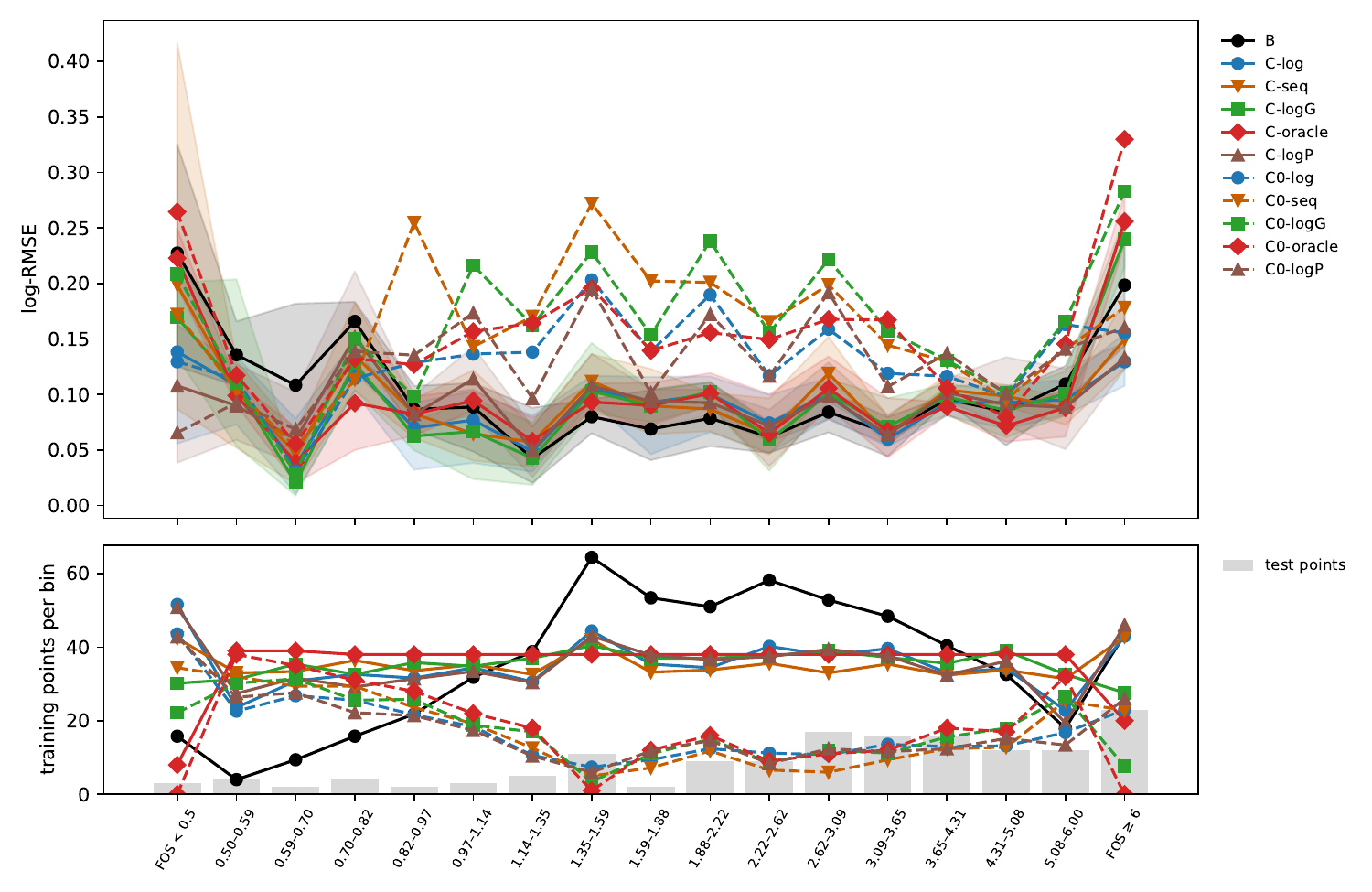}
\caption[Bin diagnostic, stratified test set, RF]{Bin diagnostic on test-stratified, RF, over the confirmatory realizations r1-r5: the 15 bins of the selection range and the two zones outside it, half-open, [lo, hi), as everywhere the replication metrics are read; edges are rounded to two decimals. Top: log-RMSE per bin, mean over the realizations, with the min-max range over them shaded for B, C-log, C-seq, C-logG, C-oracle and C-logP. Bottom: mean training points per bin, each arm in its style above, and the test points per bin (grey bars). The shaded range is no interval and no test. C$_0$-log, C$_0$-seq, C$_0$-logG, C$_0$-oracle and C$_0$-logP are ablations, drawn dashed in the colour of their arm: they train on the 300 added points without the seed, against the 600 of the other arms, so they join different budgets and are not like-for-like comparisons. A, C and the original C$_0$ have no replications and are not drawn; r0 is exploratory and is not drawn. Common hyper-parameters fixed from the seed. Descriptive only; bins with few test points are not marked here (see the structural analysis).}\label{fig:s:rep:stratified:rf}
\end{figure}
\begin{figure}[htbp]\centering
\includegraphics[width=\linewidth]{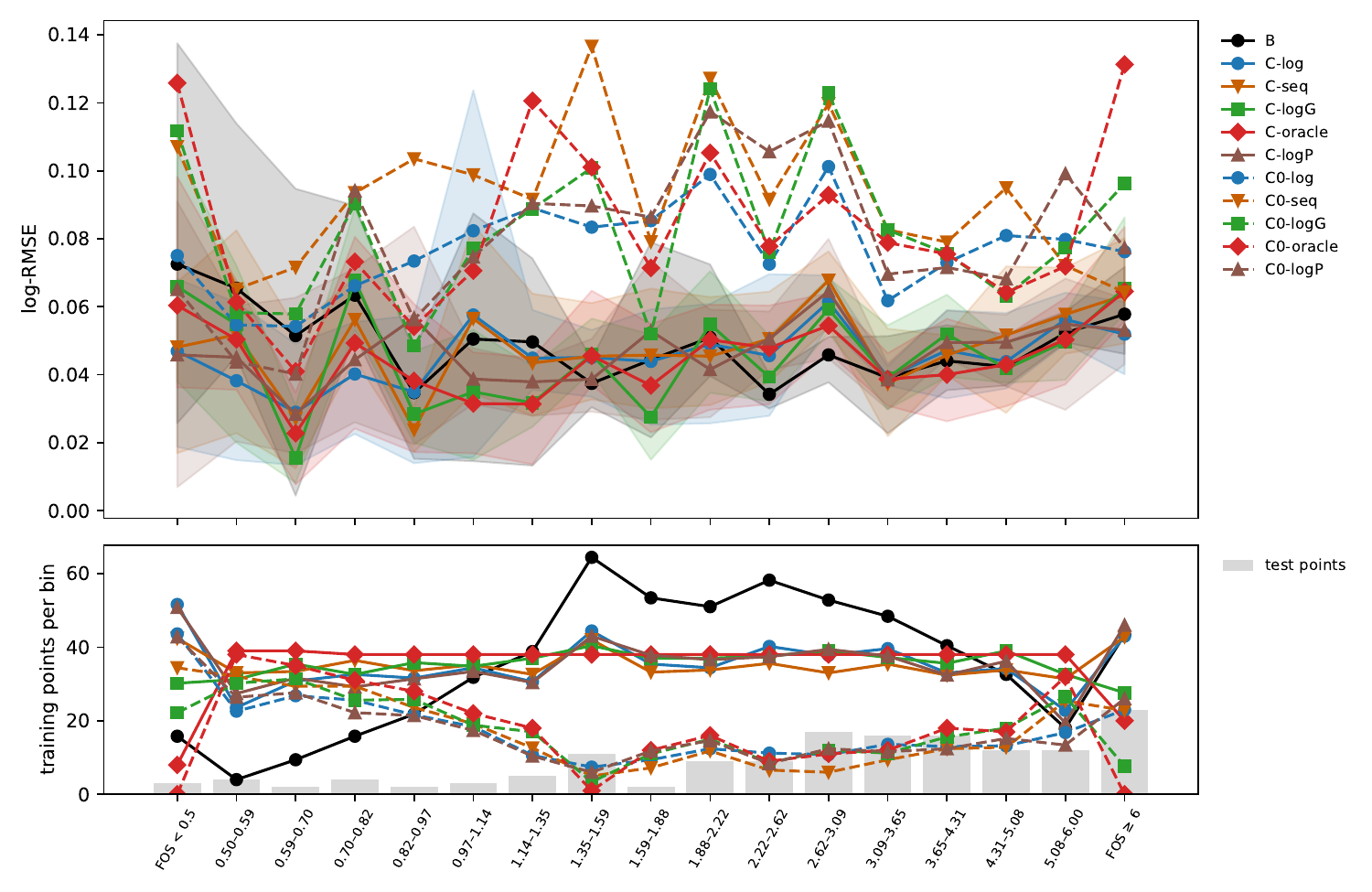}
\caption[Bin diagnostic, stratified test set, GBM]{Bin diagnostic on test-stratified, GBM, over the confirmatory realizations r1-r5: the 15 bins of the selection range and the two zones outside it, half-open, [lo, hi), as everywhere the replication metrics are read; edges are rounded to two decimals. Top: log-RMSE per bin, mean over the realizations, with the min-max range over them shaded for B, C-log, C-seq, C-logG, C-oracle and C-logP. Bottom: mean training points per bin, each arm in its style above, and the test points per bin (grey bars). The shaded range is no interval and no test. C$_0$-log, C$_0$-seq, C$_0$-logG, C$_0$-oracle and C$_0$-logP are ablations, drawn dashed in the colour of their arm: they train on the 300 added points without the seed, against the 600 of the other arms, so they join different budgets and are not like-for-like comparisons. A, C and the original C$_0$ have no replications and are not drawn; r0 is exploratory and is not drawn. Common hyper-parameters fixed from the seed. Descriptive only; bins with few test points are not marked here (see the structural analysis).}\label{fig:s:rep:stratified:gbm}
\end{figure}

\clearpage
\section{The structural analysis}\label{sec:s-structural}

The tables and figures of the descriptive analysis of \cref{sec:res-bins}, under the procedure of \cref{sec:s-structural-procedure}. Its pattern is read only as consistent with density-driven relocation, never causally, and it is never cited as evidence about the screener; $\text{achieved} - \text{intended}$ is the net distortion of the requested distribution, not a prediction error. The effect of the departure on accuracy is measured by the oracle's paired contrast (\cref{sec:res-oracle}), not here.

\subsection{Concordance and the bins outside the selection range}\label{sec:s-structural-concordance}

\begin{table}[htbp]
\centering
\small
\setlength{\tabcolsep}{4pt}
\caption{Concordance: the number of in-range bins in which the difference in training density and the difference in log-RMSE against B have opposite signs, k of n. Primary over the bins with at least 10 test points; over all 15 as a descriptive sensitivity. Bins where either difference is exactly zero leave the denominator. A count, with no test attached.}
\label{tab:structural:concordance}
\begin{tabular}{lrrrrrrr}
\toprule
family & test & arm & k / n (unmarked) & zero (unmarked) & thin & k / n (all 15) & zero (all 15) \\
\midrule
rf & test-natural & C-log & 11 / 11 & 0 & 4 & 15 / 15 & 0 \\
rf & test-natural & C-seq & 11 / 11 & 0 & 4 & 15 / 15 & 0 \\
rf & test-natural & C-logG & 11 / 11 & 0 & 4 & 15 / 15 & 0 \\
rf & test-natural & C-oracle & 10 / 11 & 0 & 4 & 14 / 15 & 0 \\
rf & test-natural & C-logP & 9 / 11 & 0 & 4 & 12 / 15 & 0 \\
rf & test-stratified & C-log & 3 / 6 & 0 & 9 & 12 / 15 & 0 \\
rf & test-stratified & C-seq & 4 / 6 & 0 & 9 & 13 / 15 & 0 \\
rf & test-stratified & C-logG & 5 / 6 & 0 & 9 & 13 / 15 & 0 \\
rf & test-stratified & C-oracle & 5 / 6 & 0 & 9 & 13 / 15 & 0 \\
rf & test-stratified & C-logP & 4 / 6 & 0 & 9 & 12 / 15 & 0 \\
gbm & test-natural & C-log & 10 / 11 & 0 & 4 & 13 / 15 & 0 \\
gbm & test-natural & C-seq & 10 / 11 & 0 & 4 & 13 / 15 & 0 \\
gbm & test-natural & C-logG & 11 / 11 & 0 & 4 & 14 / 15 & 0 \\
gbm & test-natural & C-oracle & 11 / 11 & 0 & 4 & 12 / 15 & 0 \\
gbm & test-natural & C-logP & 10 / 11 & 0 & 4 & 14 / 15 & 0 \\
gbm & test-stratified & C-log & 4 / 6 & 0 & 9 & 8 / 15 & 0 \\
gbm & test-stratified & C-seq & 3 / 6 & 0 & 9 & 9 / 15 & 0 \\
gbm & test-stratified & C-logG & 5 / 6 & 0 & 9 & 11 / 15 & 0 \\
gbm & test-stratified & C-oracle & 3 / 6 & 0 & 9 & 8 / 15 & 0 \\
gbm & test-stratified & C-logP & 3 / 6 & 0 & 9 & 9 / 15 & 0 \\
\bottomrule
\end{tabular}
\end{table}

\begin{table}[htbp]
\centering
\small
\setlength{\tabcolsep}{4pt}
\caption{Outside the selection range, where nothing was allocated. 'density' rows are the spillover itself: training points that landed there because the gate is on a predicted FOS. 'error' rows are the test error in those bins, with the number of test points; rows with fewer than 10 are thin and carry no statement in the text. Differences against B with descriptive 95\% Welch intervals over r1-r5. Both sides are read from the replication metrics, so both use the half-open reporting zones and the outer bin is FOS >= 6. This table: the 'density' rows.}
\label{tab:structural:outside:density}
\begin{tabular}{lrrrr}
\toprule
arm & side & diff & ci lo & ci hi \\
\midrule
C-log & at or above 6 & -0.400 & -3.364 & 2.564 \\
C-log & below 0.5 & 35.800 & 29.077 & 42.523 \\
C-logG & at or above 6 & -15.800 & -20.143 & -11.457 \\
C-logG & below 0.5 & 14.400 & 7.526 & 21.274 \\
C-logP & at or above 6 & 2.600 & -0.364 & 5.564 \\
C-logP & below 0.5 & 35.000 & 26.931 & 43.069 \\
C-oracle & at or above 6 & -23.400 & -26.390 & -20.410 \\
C-oracle & below 0.5 & -7.800 & -10.631 & -4.969 \\
C-seq & at or above 6 & -0.600 & -5.427 & 4.227 \\
C-seq & below 0.5 & 26.600 & 22.385 & 30.815 \\
\bottomrule
\end{tabular}
\end{table}
\begin{table}[htbp]
\centering
\small
\setlength{\tabcolsep}{4pt}
\caption{Outside the selection range, where nothing was allocated. 'density' rows are the spillover itself: training points that landed there because the gate is on a predicted FOS. 'error' rows are the test error in those bins, with the number of test points; rows with fewer than 10 are thin and carry no statement in the text. Differences against B with descriptive 95\% Welch intervals over r1-r5. Both sides are read from the replication metrics, so both use the half-open reporting zones and the outer bin is FOS >= 6. This table: the 'error' rows.}
\label{tab:structural:outside:error}
\begin{tabular}{lrrrrrrrr}
\toprule
arm & side & family & test & test n & thin & diff & ci lo & ci hi \\
\midrule
C-log & at or above 6 & rf & test-natural & 24 & no & -0.034 & -0.066 & -0.001 \\
C-log & at or above 6 & rf & test-stratified & 23 & no & -0.069 & -0.112 & -0.026 \\
C-log & at or above 6 & gbm & test-natural & 24 & no & 0.035 & 0.023 & 0.048 \\
C-log & at or above 6 & gbm & test-stratified & 23 & no & -0.006 & -0.019 & 0.007 \\
C-log & below 0.5 & rf & test-natural & 10 & no & -0.056 & -0.145 & 0.032 \\
C-log & below 0.5 & rf & test-stratified & 3 & yes & -0.089 & -0.196 & 0.018 \\
C-log & below 0.5 & gbm & test-natural & 10 & no & -0.023 & -0.065 & 0.018 \\
C-log & below 0.5 & gbm & test-stratified & 3 & yes & -0.026 & -0.080 & 0.029 \\
C-logG & at or above 6 & rf & test-natural & 24 & no & 0.027 & -0.004 & 0.057 \\
C-logG & at or above 6 & rf & test-stratified & 23 & no & 0.041 & -0.001 & 0.084 \\
C-logG & at or above 6 & gbm & test-natural & 24 & no & 0.030 & 0.018 & 0.041 \\
C-logG & at or above 6 & gbm & test-stratified & 23 & no & 0.008 & -0.008 & 0.024 \\
C-logG & below 0.5 & rf & test-natural & 10 & no & 0.044 & -0.045 & 0.133 \\
C-logG & below 0.5 & rf & test-stratified & 3 & yes & -0.058 & -0.145 & 0.029 \\
C-logG & below 0.5 & gbm & test-natural & 10 & no & 0.004 & -0.033 & 0.041 \\
C-logG & below 0.5 & gbm & test-stratified & 3 & yes & -0.007 & -0.063 & 0.050 \\
C-logP & at or above 6 & rf & test-natural & 24 & no & -0.044 & -0.074 & -0.014 \\
C-logP & at or above 6 & rf & test-stratified & 23 & no & -0.066 & -0.108 & -0.023 \\
C-logP & at or above 6 & gbm & test-natural & 24 & no & 0.029 & 0.016 & 0.043 \\
C-logP & at or above 6 & gbm & test-stratified & 23 & no & -0.005 & -0.018 & 0.008 \\
C-logP & below 0.5 & rf & test-natural & 10 & no & -0.043 & -0.132 & 0.045 \\
C-logP & below 0.5 & rf & test-stratified & 3 & yes & -0.120 & -0.214 & -0.025 \\
C-logP & below 0.5 & gbm & test-natural & 10 & no & -0.014 & -0.052 & 0.023 \\
C-logP & below 0.5 & gbm & test-stratified & 3 & yes & -0.027 & -0.087 & 0.034 \\
C-oracle & at or above 6 & rf & test-natural & 24 & no & 0.026 & -0.005 & 0.056 \\
C-oracle & at or above 6 & rf & test-stratified & 23 & no & 0.057 & 0.015 & 0.100 \\
C-oracle & at or above 6 & gbm & test-natural & 24 & no & 0.037 & 0.024 & 0.049 \\
C-oracle & at or above 6 & gbm & test-stratified & 23 & no & 0.007 & -0.009 & 0.022 \\
C-oracle & below 0.5 & rf & test-natural & 10 & no & 0.063 & -0.026 & 0.151 \\
C-oracle & below 0.5 & rf & test-stratified & 3 & yes & -0.005 & -0.092 & 0.083 \\
C-oracle & below 0.5 & gbm & test-natural & 10 & no & 0.007 & -0.030 & 0.044 \\
C-oracle & below 0.5 & gbm & test-stratified & 3 & yes & -0.012 & -0.068 & 0.043 \\
C-seq & at or above 6 & rf & test-natural & 24 & no & -0.023 & -0.054 & 0.008 \\
C-seq & at or above 6 & rf & test-stratified & 23 & no & -0.049 & -0.094 & -0.003 \\
C-seq & at or above 6 & gbm & test-natural & 24 & no & 0.035 & 0.022 & 0.048 \\
C-seq & at or above 6 & gbm & test-stratified & 23 & no & 0.006 & -0.010 & 0.021 \\
C-seq & below 0.5 & rf & test-natural & 10 & no & -0.017 & -0.106 & 0.072 \\
C-seq & below 0.5 & rf & test-stratified & 3 & yes & -0.029 & -0.193 & 0.136 \\
C-seq & below 0.5 & gbm & test-natural & 10 & no & -0.011 & -0.048 & 0.026 \\
C-seq & below 0.5 & gbm & test-stratified & 3 & yes & -0.024 & -0.079 & 0.031 \\
\bottomrule
\end{tabular}
\end{table}

\clearpage
\subsection{Allocation, batches and displacement}\label{sec:s-structural-allocation}

\begin{table}[htbp]
\centering
\small
\setlength{\tabcolsep}{4pt}
\caption{Allocation fidelity for C-log: mean ± sd over the confirmatory realizations r1-r5, with r0 (exploratory) in its own columns. 'deviation' is achieved minus intended, the net distortion of the requested bin-count distribution and not the screener's prediction error. The two outside rows count the points whose true FOS left the selection range, where nothing was intended; the last row sums |deviation| over the 15 in-range bins. Bins follow the generator, (lo, hi], so the outer bin is FOS > 6. Selected equals intended by construction and is not shown. C-oracle bins on the solved FOS and meets the intended allocation by construction; its rows are a check, not a finding. What known bin membership changes in test accuracy is read only from the paired contrast C-oracle against C-logP, in this design. C$_0$-log, C$_0$-seq, C$_0$-logG, C$_0$-oracle and C$_0$-logP train on the same guided points as C-log, C-seq, C-logG, C-oracle and C-logP in every realization, so their allocation, batch and displacement rows are those of C, shared rather than duplicated.}
\label{tab:structural:allocation:c-log}
\begin{tabular}{lrrrrrrr}
\toprule
bin & seed & intended & achieved & deviation & intended r0 & achieved r0 & deviation r0 \\
\midrule
0.50-0.59 & 1 & 38 & 22.6 ± 3.5 & -15.4 ± 3.5 & 38 & 18 & -20 \\
0.59-0.70 & 4 & 35 & 26.8 ± 3.6 & -8.2 ± 3.6 & 35 & 28 & -7 \\
0.70-0.82 & 7 & 31 & 25.6 ± 3.8 & -5.4 ± 3.8 & 31 & 25 & -6 \\
0.82-0.97 & 10 & 28 & 21.6 ± 3.8 & -6.4 ± 3.8 & 28 & 24 & -4 \\
0.97-1.14 & 16 & 22 & 18.4 ± 3.5 & -3.6 ± 3.5 & 22 & 16 & -6 \\
1.14-1.35 & 20 & 18 & 10.6 ± 2.4 & -7.4 ± 2.4 & 18 & 9 & -9 \\
1.35-1.59 & 37 & 1 & 7.4 ± 1.3 & 6.4 ± 1.3 & 1 & 11 & 10 \\
1.59-1.88 & 26 & 12 & 9.4 ± 0.9 & -2.6 ± 0.9 & 12 & 10 & -2 \\
1.88-2.22 & 22 & 16 & 12.4 ± 3.0 & -3.6 ± 3.0 & 16 & 12 & -4 \\
2.22-2.62 & 29 & 9 & 11.2 ± 2.7 & 2.2 ± 2.7 & 9 & 10 & 1 \\
2.62-3.09 & 27 & 11 & 10.8 ± 2.4 & -0.2 ± 2.4 & 11 & 9 & -2 \\
3.09-3.65 & 26 & 12 & 13.6 ± 1.9 & 1.6 ± 1.9 & 12 & 15 & 3 \\
3.65-4.31 & 20 & 18 & 13.0 ± 3.2 & -5.0 ± 3.2 & 18 & 15 & -3 \\
4.31-5.08 & 21 & 17 & 13.2 ± 3.9 & -3.8 ± 3.9 & 17 & 17 & 0 \\
5.08-6.00 & 6 & 32 & 16.8 ± 2.3 & -15.2 ± 2.3 & 32 & 17 & -15 \\
< 0.5 (outside) & -- & 0 & 43.6 ± 5.5 & 43.6 ± 5.5 & 0 & 45 & 45 \\
> 6 (outside) & -- & 0 & 23.0 ± 1.2 & 23.0 ± 1.2 & 0 & 19 & 19 \\
Σ|deviation|, 15 bins &  &  &  & 90.2 ± 9.0 &  &  & 92 \\
\bottomrule
\end{tabular}
\end{table}

\begin{table}[htbp]
\centering
\small
\setlength{\tabcolsep}{4pt}
\caption{Allocation fidelity for C-seq: mean ± sd over the confirmatory realizations r1-r5, with r0 (exploratory) in its own columns. 'deviation' is achieved minus intended, the net distortion of the requested bin-count distribution and not the screener's prediction error. The two outside rows count the points whose true FOS left the selection range, where nothing was intended; the last row sums |deviation| over the 15 in-range bins. Bins follow the generator, (lo, hi], so the outer bin is FOS > 6. Selected equals intended by construction and is not shown. C-oracle bins on the solved FOS and meets the intended allocation by construction; its rows are a check, not a finding. What known bin membership changes in test accuracy is read only from the paired contrast C-oracle against C-logP, in this design. C$_0$-log, C$_0$-seq, C$_0$-logG, C$_0$-oracle and C$_0$-logP train on the same guided points as C-log, C-seq, C-logG, C-oracle and C-logP in every realization, so their allocation, batch and displacement rows are those of C, shared rather than duplicated.}
\label{tab:structural:allocation:c-seq}
\begin{tabular}{lrrrrrrr}
\toprule
bin & seed & intended & achieved & deviation & intended r0 & achieved r0 & deviation r0 \\
\midrule
0.50-0.59 & 1 & 49.8 ± 5.4 & 32.0 ± 1.0 & -17.8 ± 5.9 & 47 & 32 & -15 \\
0.59-0.70 & 4 & 39.2 ± 3.2 & 29.4 ± 1.7 & -9.8 ± 4.8 & 30 & 30 & 0 \\
0.70-0.82 & 7 & 25.2 ± 2.9 & 29.4 ± 1.7 & 4.2 ± 2.4 & 34 & 24 & -10 \\
0.82-0.97 & 10 & 31.8 ± 4.5 & 23.6 ± 1.1 & -8.2 ± 5.4 & 38 & 24 & -14 \\
0.97-1.14 & 16 & 23.6 ± 1.8 & 19.2 ± 2.2 & -4.4 ± 3.0 & 19 & 20 & 1 \\
1.14-1.35 & 20 & 14.2 ± 3.6 & 12.6 ± 2.6 & -1.6 ± 5.4 & 15 & 16 & 1 \\
1.35-1.59 & 37 & 0 & 5.0 ± 1.4 & 5.0 ± 1.4 & 0 & 6 & 6 \\
1.59-1.88 & 26 & 8.0 ± 1.2 & 7.2 ± 2.5 & -0.8 ± 3.6 & 8 & 7 & -1 \\
1.88-2.22 & 22 & 14.0 ± 1.2 & 11.8 ± 1.9 & -2.2 ± 2.2 & 15 & 11 & -4 \\
2.22-2.62 & 29 & 4.2 ± 0.4 & 6.6 ± 2.6 & 2.4 ± 2.6 & 4 & 7 & 3 \\
2.62-3.09 & 27 & 7.2 ± 1.1 & 6.0 ± 2.1 & -1.2 ± 3.1 & 6 & 7 & 1 \\
3.09-3.65 & 26 & 7.4 ± 1.5 & 9.4 ± 2.6 & 2.0 ± 4.1 & 6 & 11 & 5 \\
3.65-4.31 & 20 & 17.2 ± 3.0 & 12.4 ± 2.2 & -4.8 ± 5.0 & 19 & 9 & -10 \\
4.31-5.08 & 21 & 12.6 ± 1.9 & 12.8 ± 2.8 & 0.2 ± 4.4 & 15 & 11 & -4 \\
5.08-6.00 & 6 & 45.6 ± 3.0 & 25.4 ± 2.9 & -20.2 ± 5.1 & 44 & 29 & -15 \\
< 0.5 (outside) & -- & 0 & 34.4 ± 3.3 & 34.4 ± 3.3 & 0 & 32 & 32 \\
> 6 (outside) & -- & 0 & 22.8 ± 3.8 & 22.8 ± 3.8 & 0 & 24 & 24 \\
Σ|deviation|, 15 bins &  &  &  & 96.0 ± 8.0 &  &  & 90 \\
\bottomrule
\end{tabular}
\end{table}

\begin{table}[htbp]
\centering
\small
\setlength{\tabcolsep}{4pt}
\caption{Allocation fidelity for C-logG: mean ± sd over the confirmatory realizations r1-r5, with r0 (exploratory) in its own columns. 'deviation' is achieved minus intended, the net distortion of the requested bin-count distribution and not the screener's prediction error. The two outside rows count the points whose true FOS left the selection range, where nothing was intended; the last row sums |deviation| over the 15 in-range bins. Bins follow the generator, (lo, hi], so the outer bin is FOS > 6. Selected equals intended by construction and is not shown. C-oracle bins on the solved FOS and meets the intended allocation by construction; its rows are a check, not a finding. What known bin membership changes in test accuracy is read only from the paired contrast C-oracle against C-logP, in this design. C$_0$-log, C$_0$-seq, C$_0$-logG, C$_0$-oracle and C$_0$-logP train on the same guided points as C-log, C-seq, C-logG, C-oracle and C-logP in every realization, so their allocation, batch and displacement rows are those of C, shared rather than duplicated.}
\label{tab:structural:allocation:c-logg}
\begin{tabular}{lrrrrrrr}
\toprule
bin & seed & intended & achieved & deviation & intended r0 & achieved r0 & deviation r0 \\
\midrule
0.50-0.59 & 1 & 38 & 30.2 ± 5.0 & -7.8 ± 5.0 & 38 & 35 & -3 \\
0.59-0.70 & 4 & 35 & 31.4 ± 3.8 & -3.6 ± 3.8 & 35 & 27 & -8 \\
0.70-0.82 & 7 & 31 & 25.6 ± 3.0 & -5.4 ± 3.0 & 31 & 29 & -2 \\
0.82-0.97 & 10 & 28 & 25.8 ± 2.3 & -2.2 ± 2.3 & 28 & 24 & -4 \\
0.97-1.14 & 16 & 22 & 18.8 ± 2.8 & -3.2 ± 2.8 & 22 & 21 & -1 \\
1.14-1.35 & 20 & 18 & 17.0 ± 2.5 & -1.0 ± 2.5 & 18 & 14 & -4 \\
1.35-1.59 & 37 & 1 & 3.4 ± 1.1 & 2.4 ± 1.1 & 1 & 7 & 6 \\
1.59-1.88 & 26 & 12 & 11.0 ± 2.0 & -1.0 ± 2.0 & 12 & 7 & -5 \\
1.88-2.22 & 22 & 16 & 15.0 ± 4.9 & -1.0 ± 4.9 & 16 & 15 & -1 \\
2.22-2.62 & 29 & 9 & 8.6 ± 1.5 & -0.4 ± 1.5 & 9 & 10 & 1 \\
2.62-3.09 & 27 & 11 & 12.0 ± 3.2 & 1.0 ± 3.2 & 11 & 11 & 0 \\
3.09-3.65 & 26 & 12 & 11.2 ± 2.2 & -0.8 ± 2.2 & 12 & 11 & -1 \\
3.65-4.31 & 20 & 18 & 15.6 ± 3.2 & -2.4 ± 3.2 & 18 & 19 & 1 \\
4.31-5.08 & 21 & 17 & 18.0 ± 3.7 & 1.0 ± 3.7 & 17 & 14 & -3 \\
5.08-6.00 & 6 & 32 & 26.6 ± 4.4 & -5.4 ± 4.4 & 32 & 28 & -4 \\
< 0.5 (outside) & -- & 0 & 22.2 ± 5.6 & 22.2 ± 5.6 & 0 & 20 & 20 \\
> 6 (outside) & -- & 0 & 7.6 ± 3.4 & 7.6 ± 3.4 & 0 & 8 & 8 \\
Σ|deviation|, 15 bins &  &  &  & 49.4 ± 4.3 &  &  & 44 \\
\bottomrule
\end{tabular}
\end{table}

\begin{table}[htbp]
\centering
\small
\setlength{\tabcolsep}{4pt}
\caption{Allocation fidelity for C-oracle: mean ± sd over the confirmatory realizations r1-r5, with r0 (exploratory) in its own columns. 'deviation' is achieved minus intended, the net distortion of the requested bin-count distribution and not the screener's prediction error. The two outside rows count the points whose true FOS left the selection range, where nothing was intended; the last row sums |deviation| over the 15 in-range bins. Bins follow the generator, (lo, hi], so the outer bin is FOS > 6. Selected equals intended by construction and is not shown. C-oracle bins on the solved FOS and meets the intended allocation by construction; its rows are a check, not a finding. What known bin membership changes in test accuracy is read only from the paired contrast C-oracle against C-logP, in this design. C$_0$-log, C$_0$-seq, C$_0$-logG, C$_0$-oracle and C$_0$-logP train on the same guided points as C-log, C-seq, C-logG, C-oracle and C-logP in every realization, so their allocation, batch and displacement rows are those of C, shared rather than duplicated. C-oracle bins on the solved FOS: every deviation is 0 by construction, a check, not a finding.}
\label{tab:structural:allocation:c-oracle}
\begin{tabular}{lrrrrrrr}
\toprule
bin & seed & intended & achieved & deviation & intended r0 & achieved r0 & deviation r0 \\
\midrule
0.50-0.59 & 1 & 38 & 38 & 0 & -- & -- & -- \\
0.59-0.70 & 4 & 35 & 35 & 0 & -- & -- & -- \\
0.70-0.82 & 7 & 31 & 31 & 0 & -- & -- & -- \\
0.82-0.97 & 10 & 28 & 28 & 0 & -- & -- & -- \\
0.97-1.14 & 16 & 22 & 22 & 0 & -- & -- & -- \\
1.14-1.35 & 20 & 18 & 18 & 0 & -- & -- & -- \\
1.35-1.59 & 37 & 1 & 1 & 0 & -- & -- & -- \\
1.59-1.88 & 26 & 12 & 12 & 0 & -- & -- & -- \\
1.88-2.22 & 22 & 16 & 16 & 0 & -- & -- & -- \\
2.22-2.62 & 29 & 9 & 9 & 0 & -- & -- & -- \\
2.62-3.09 & 27 & 11 & 11 & 0 & -- & -- & -- \\
3.09-3.65 & 26 & 12 & 12 & 0 & -- & -- & -- \\
3.65-4.31 & 20 & 18 & 18 & 0 & -- & -- & -- \\
4.31-5.08 & 21 & 17 & 17 & 0 & -- & -- & -- \\
5.08-6.00 & 6 & 32 & 32 & 0 & -- & -- & -- \\
< 0.5 (outside) & -- & 0 & 0 & 0 & -- & -- & -- \\
> 6 (outside) & -- & 0 & 0 & 0 & -- & -- & -- \\
Σ|deviation|, 15 bins &  &  &  & 0 & -- & -- & -- \\
\bottomrule
\end{tabular}
\end{table}

\begin{table}[htbp]
\centering
\small
\setlength{\tabcolsep}{4pt}
\caption{Allocation fidelity for C-logP: mean ± sd over the confirmatory realizations r1-r5, with r0 (exploratory) in its own columns. 'deviation' is achieved minus intended, the net distortion of the requested bin-count distribution and not the screener's prediction error. The two outside rows count the points whose true FOS left the selection range, where nothing was intended; the last row sums |deviation| over the 15 in-range bins. Bins follow the generator, (lo, hi], so the outer bin is FOS > 6. Selected equals intended by construction and is not shown. C-oracle bins on the solved FOS and meets the intended allocation by construction; its rows are a check, not a finding. What known bin membership changes in test accuracy is read only from the paired contrast C-oracle against C-logP, in this design. C$_0$-log, C$_0$-seq, C$_0$-logG, C$_0$-oracle and C$_0$-logP train on the same guided points as C-log, C-seq, C-logG, C-oracle and C-logP in every realization, so their allocation, batch and displacement rows are those of C, shared rather than duplicated.}
\label{tab:structural:allocation:c-logp}
\begin{tabular}{lrrrrrrr}
\toprule
bin & seed & intended & achieved & deviation & intended r0 & achieved r0 & deviation r0 \\
\midrule
0.50-0.59 & 1 & 38 & 26.4 ± 2.9 & -11.6 ± 2.9 & -- & -- & -- \\
0.59-0.70 & 4 & 35 & 27.6 ± 4.9 & -7.4 ± 4.9 & -- & -- & -- \\
0.70-0.82 & 7 & 31 & 22.2 ± 4.3 & -8.8 ± 4.3 & -- & -- & -- \\
0.82-0.97 & 10 & 28 & 21.4 ± 4.2 & -6.6 ± 4.2 & -- & -- & -- \\
0.97-1.14 & 16 & 22 & 17.4 ± 4.9 & -4.6 ± 4.9 & -- & -- & -- \\
1.14-1.35 & 20 & 18 & 10.4 ± 1.7 & -7.6 ± 1.7 & -- & -- & -- \\
1.35-1.59 & 37 & 1 & 6.0 ± 4.2 & 5.0 ± 4.2 & -- & -- & -- \\
1.59-1.88 & 26 & 12 & 11.8 ± 1.3 & -0.2 ± 1.3 & -- & -- & -- \\
1.88-2.22 & 22 & 16 & 14.6 ± 1.7 & -1.4 ± 1.7 & -- & -- & -- \\
2.22-2.62 & 29 & 9 & 8.4 ± 1.1 & -0.6 ± 1.1 & -- & -- & -- \\
2.62-3.09 & 27 & 11 & 12.4 ± 0.5 & 1.4 ± 0.5 & -- & -- & -- \\
3.09-3.65 & 26 & 12 & 11.6 ± 2.8 & -0.4 ± 2.8 & -- & -- & -- \\
3.65-4.31 & 20 & 18 & 12.4 ± 1.3 & -5.6 ± 1.3 & -- & -- & -- \\
4.31-5.08 & 21 & 17 & 15.2 ± 3.3 & -1.8 ± 3.3 & -- & -- & -- \\
5.08-6.00 & 6 & 32 & 13.4 ± 1.5 & -18.6 ± 1.5 & -- & -- & -- \\
< 0.5 (outside) & -- & 0 & 42.8 ± 6.6 & 42.8 ± 6.6 & -- & -- & -- \\
> 6 (outside) & -- & 0 & 26.0 ± 1.2 & 26.0 ± 1.2 & -- & -- & -- \\
Σ|deviation|, 15 bins &  &  &  & 87.2 ± 9.0 & -- & -- & -- \\
\bottomrule
\end{tabular}
\end{table}

\begin{table}[htbp]
\centering
\small
\setlength{\tabcolsep}{4pt}
\caption{C-seq batch by batch: the intended count per bin and the deviation, achieved minus intended, then the cumulative deviation after batches 2 and 3 (cum.); mean ± sd over r1-r5, r0 (exploratory) apart. Each batch reallocates against what the earlier batches actually returned; the cumulative columns show whether retraining corrects a deviation or moves it. The outside rows count points whose true FOS left the selection range, where nothing was intended (generator bins, outer bin FOS > 6). C-oracle bins on the solved FOS and meets the intended allocation by construction; its rows are a check, not a finding. What known bin membership changes in test accuracy is read only from the paired contrast C-oracle against C-logP, in this design. C$_0$-log, C$_0$-seq, C$_0$-logG, C$_0$-oracle and C$_0$-logP train on the same guided points as C-log, C-seq, C-logG, C-oracle and C-logP in every realization, so their allocation, batch and displacement rows are those of C, shared rather than duplicated.}
\label{tab:structural:batches}
\begin{tabular}{lrrrrrrrr}
\toprule
bin & int. 1 & dev. 1 & int. 2 & dev. 2 & int. 3 & dev. 3 & cum. 2 & cum. 3 \\
\midrule
0.50-0.59 & 22 & -10.2 ± 2.8 & 16.2 ± 2.8 & -5.6 ± 3.4 & 11.6 ± 3.4 & -2.0 ± 1.0 & -15.8 ± 5.4 & -17.8 ± 5.9 \\
0.59-0.70 & 19 & -6.6 ± 2.1 & 12.6 ± 2.1 & -1.6 ± 1.5 & 7.6 ± 1.5 & -1.6 ± 1.7 & -8.2 ± 3.2 & -9.8 ± 4.8 \\
0.70-0.82 & 16 & -0.6 ± 2.3 & 6.4 ± 2.1 & 3.4 ± 0.9 & 2.8 ± 1.3 & 1.4 ± 1.7 & 2.8 ± 2.9 & 4.2 ± 2.4 \\
0.82-0.97 & 13 & -4.4 ± 1.1 & 10.0 ± 1.4 & -2.4 ± 4.0 & 8.8 ± 4.1 & -1.4 ± 1.1 & -6.8 ± 4.5 & -8.2 ± 5.4 \\
0.97-1.14 & 7 & -2.6 ± 0.9 & 8.2 ± 0.8 & -2.2 ± 1.5 & 8.4 ± 1.1 & 0.4 ± 2.1 & -4.8 ± 1.9 & -4.4 ± 3.0 \\
1.14-1.35 & 3 & 0.2 ± 1.3 & 5.4 ± 1.5 & 0.2 ± 3.0 & 5.8 ± 2.9 & -2.0 ± 2.8 & 0.4 ± 3.9 & -1.6 ± 5.4 \\
1.35-1.59 & 0 & 0.8 ± 0.8 & 0 & 1.0 ± 0.7 & 0 & 3.2 ± 1.1 & 1.8 ± 1.1 & 5.0 ± 1.4 \\
1.59-1.88 & 0 & 0.4 ± 0.5 & 2.2 ± 0.8 & 0.2 ± 0.8 & 5.8 ± 1.1 & -1.4 ± 2.7 & 0.6 ± 0.9 & -0.8 ± 3.6 \\
1.88-2.22 & 1 & -0.2 ± 0.8 & 5.4 ± 0.9 & -1.4 ± 0.9 & 7.6 ± 0.9 & -0.6 ± 1.5 & -1.6 ± 1.1 & -2.2 ± 2.2 \\
2.22-2.62 & 0 & 0 & 0 & 1.2 ± 0.4 & 4.2 ± 0.4 & 1.2 ± 2.3 & 1.2 ± 0.4 & 2.4 ± 2.6 \\
2.62-3.09 & 0 & 0.2 ± 0.4 & 1.0 ± 0.7 & 0.0 ± 0.7 & 6.2 ± 0.8 & -1.4 ± 2.4 & 0.2 ± 0.8 & -1.2 ± 3.1 \\
3.09-3.65 & 0 & 0.4 ± 0.9 & 1.8 ± 1.1 & 0.4 ± 0.5 & 5.6 ± 0.5 & 1.2 ± 2.8 & 0.8 ± 1.3 & 2.0 ± 4.1 \\
3.65-4.31 & 2 & -1.4 ± 0.9 & 7.4 ± 0.9 & -1.6 ± 2.3 & 7.8 ± 2.3 & -1.8 ± 2.2 & -3.0 ± 3.0 & -4.8 ± 5.0 \\
4.31-5.08 & 1 & 0.0 ± 0.7 & 6.0 ± 0.7 & 0.6 ± 1.1 & 5.6 ± 1.3 & -0.4 ± 2.9 & 0.6 ± 1.8 & 0.2 ± 4.4 \\
5.08-6.00 & 16 & -11.4 ± 1.5 & 17.4 ± 1.5 & -6.2 ± 3.8 & 12.2 ± 3.8 & -2.6 ± 2.9 & -17.6 ± 3.0 & -20.2 ± 5.1 \\
< 0.5 (outside) & 0 & 23.2 ± 2.8 & 0 & 7.6 ± 1.1 & 0 & 3.6 ± 1.1 & 30.8 ± 2.5 & 34.4 ± 3.3 \\
> 6 (outside) & 0 & 12.2 ± 0.8 & 0 & 6.4 ± 2.7 & 0 & 4.2 ± 1.9 & 18.6 ± 2.3 & 22.8 ± 3.8 \\
\bottomrule
\end{tabular}
\end{table}

\begin{table}[htbp]
\centering
\small
\setlength{\tabcolsep}{4pt}
\caption{C-seq batch by batch: the intended count per bin and the deviation, achieved minus intended, then the cumulative deviation after batches 2 and 3 (cum.); mean ± sd over r1-r5, r0 (exploratory) apart. Each batch reallocates against what the earlier batches actually returned; the cumulative columns show whether retraining corrects a deviation or moves it. The outside rows count points whose true FOS left the selection range, where nothing was intended (generator bins, outer bin FOS > 6). C-oracle bins on the solved FOS and meets the intended allocation by construction; its rows are a check, not a finding. What known bin membership changes in test accuracy is read only from the paired contrast C-oracle against C-logP, in this design. This table is r0 alone: exploratory, never averaged in. C$_0$-log, C$_0$-seq, C$_0$-logG, C$_0$-oracle and C$_0$-logP train on the same guided points as C-log, C-seq, C-logG, C-oracle and C-logP in every realization, so their allocation, batch and displacement rows are those of C, shared rather than duplicated.}
\label{tab:structural:batches:r0}
\begin{tabular}{lrrrrrrrr}
\toprule
bin & int. 1 & dev. 1 & int. 2 & dev. 2 & int. 3 & dev. 3 & cum. 2 & cum. 3 \\
\midrule
0.50-0.59 & 22 & -7 & 13 & -6 & 12 & -2 & -13 & -15 \\
0.59-0.70 & 19 & -1 & 7 & 2 & 4 & -1 & 1 & 0 \\
0.70-0.82 & 16 & -6 & 12 & 0 & 6 & -4 & -6 & -10 \\
0.82-0.97 & 13 & -8 & 14 & -5 & 11 & -1 & -13 & -14 \\
0.97-1.14 & 7 & -2 & 7 & 2 & 5 & 1 & 0 & 1 \\
1.14-1.35 & 3 & 0 & 5 & -1 & 7 & 2 & -1 & 1 \\
1.35-1.59 & 0 & 2 & 0 & 3 & 0 & 1 & 5 & 6 \\
1.59-1.88 & 0 & 0 & 2 & 0 & 6 & -1 & 0 & -1 \\
1.88-2.22 & 1 & -1 & 6 & -2 & 8 & -1 & -3 & -4 \\
2.22-2.62 & 0 & 0 & 0 & 1 & 4 & 2 & 1 & 3 \\
2.62-3.09 & 0 & 1 & 0 & 0 & 6 & 0 & 1 & 1 \\
3.09-3.65 & 0 & 1 & 1 & 1 & 5 & 3 & 2 & 5 \\
3.65-4.31 & 2 & -2 & 8 & -3 & 9 & -5 & -5 & -10 \\
4.31-5.08 & 1 & 0 & 6 & -2 & 8 & -2 & -2 & -4 \\
5.08-6.00 & 16 & -13 & 19 & -3 & 9 & 1 & -16 & -15 \\
< 0.5 (outside) & 0 & 22 & 0 & 6 & 0 & 4 & 28 & 32 \\
> 6 (outside) & 0 & 14 & 0 & 7 & 0 & 3 & 21 & 24 \\
\bottomrule
\end{tabular}
\end{table}

\begin{table}[htbp]
\centering
\small
\setlength{\tabcolsep}{4pt}
\caption{The screener on the points it selected: share staying in the predicted bin, absolute displacement in bins, share landing outside [0.5, 6]; mean ± sd over r1-r5, r0 (exploratory) apart. Conditioned on selection, so this is its accuracy where it was used, not over the pool. The displacement columns cover only points whose true FOS stayed inside the selection range; the rest are counted by 'outside'. C-oracle's row is a construction check (stayed = 1, shift 0), not a description of a screener. C-oracle bins on the solved FOS and meets the intended allocation by construction; its rows are a check, not a finding. What known bin membership changes in test accuracy is read only from the paired contrast C-oracle against C-logP, in this design. C$_0$-log, C$_0$-seq, C$_0$-logG, C$_0$-oracle and C$_0$-logP train on the same guided points as C-log, C-seq, C-logG, C-oracle and C-logP in every realization, so their allocation, batch and displacement rows are those of C, shared rather than duplicated.}
\label{tab:structural:displacement}
\begin{tabular}{lrrrrrr}
\toprule
arm & set & n & stayed & mean |Δbin| & median |Δbin| & outside \\
\midrule
C-log & r1-r5 & 300 & 0.35 ± 0.03 & 0.68 ± 0.04 & 1 & 0.22 ± 0.02 \\
C-log & r0 (exploratory) & 300 & 0.36 & 0.66 & 1.00 & 0.21 \\
C-logG & r1-r5 & 300 & 0.62 ± 0.03 & 0.33 ± 0.02 & 0 & 0.10 ± 0.03 \\
C-logG & r0 (exploratory) & 300 & 0.63 & 0.32 & 0.00 & 0.09 \\
C-logP & r1-r5 & 300 & 0.31 ± 0.03 & 0.73 ± 0.04 & 1 & 0.23 ± 0.02 \\
C-logP & r0 (exploratory) & -- & -- & -- & -- & -- \\
C-oracle & r1-r5 & 300 & 1 & 0 & 0 & 0 \\
C-oracle & r0 (exploratory) & -- & -- & -- & -- & -- \\
C-seq & r1-r5 & 300 & 0.43 ± 0.03 & 0.54 ± 0.06 & 0.40 ± 0.55 & 0.19 ± 0.01 \\
C-seq & r0 (exploratory) & 300 & 0.43 & 0.52 & 0.00 & 0.19 \\
\bottomrule
\end{tabular}
\end{table}

\begin{table}[htbp]
\centering
\caption{Shared candidates (same pool, different screener): how many of guided-log-g's points guided-log also drew, per realization, by normalized distance. Both screeners choose, in each realization, from the same candidate pool against the same intended allocation; only the screener differs.}
\label{tab:structural:shared_candidates}
\begin{tabular}{lrr}
\toprule
realization & n & shared candidates (same pool, different screener) \\
\midrule
r0 (exploratory) & 300 & 4 \\
r1 & 300 & 2 \\
r2 & 300 & 2 \\
r3 & 300 & 2 \\
r4 & 300 & 0 \\
r5 & 300 & 3 \\
\bottomrule
\end{tabular}
\end{table}

\begin{figure}[htbp]\centering
\includegraphics[width=0.40\linewidth]{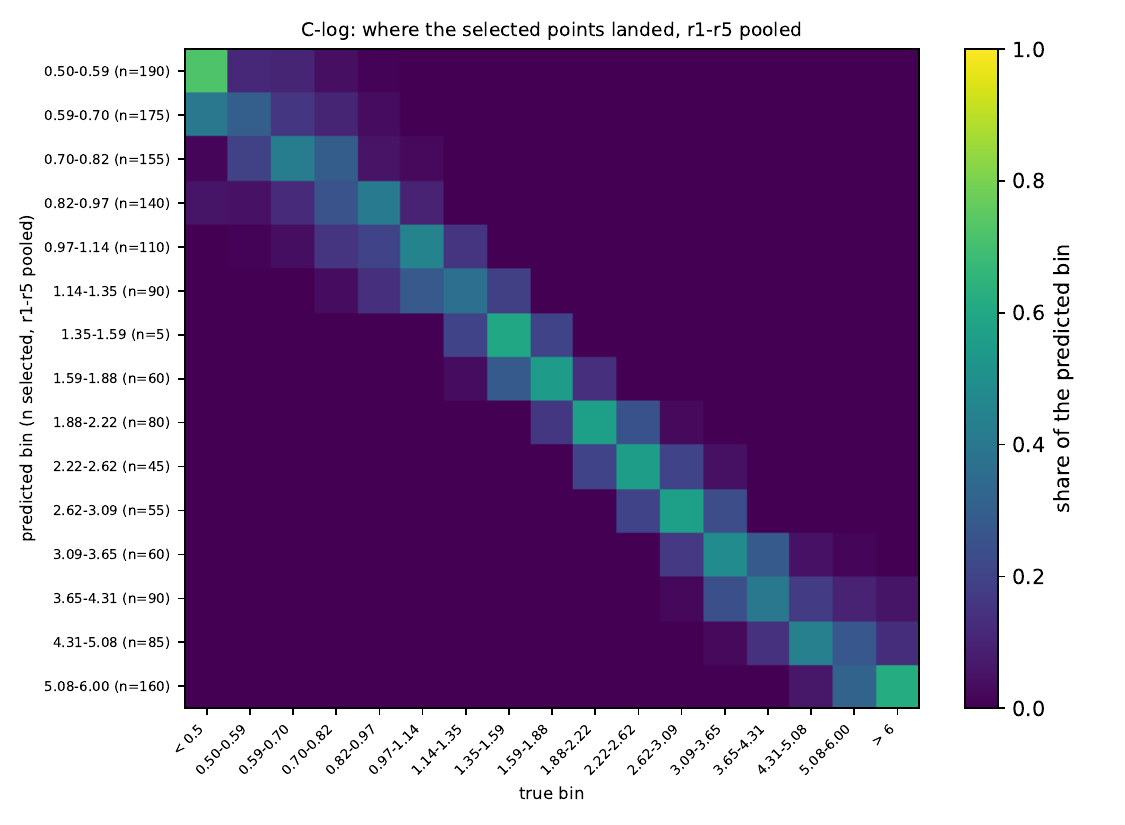}\hspace{0.04\linewidth}%
\includegraphics[width=0.40\linewidth]{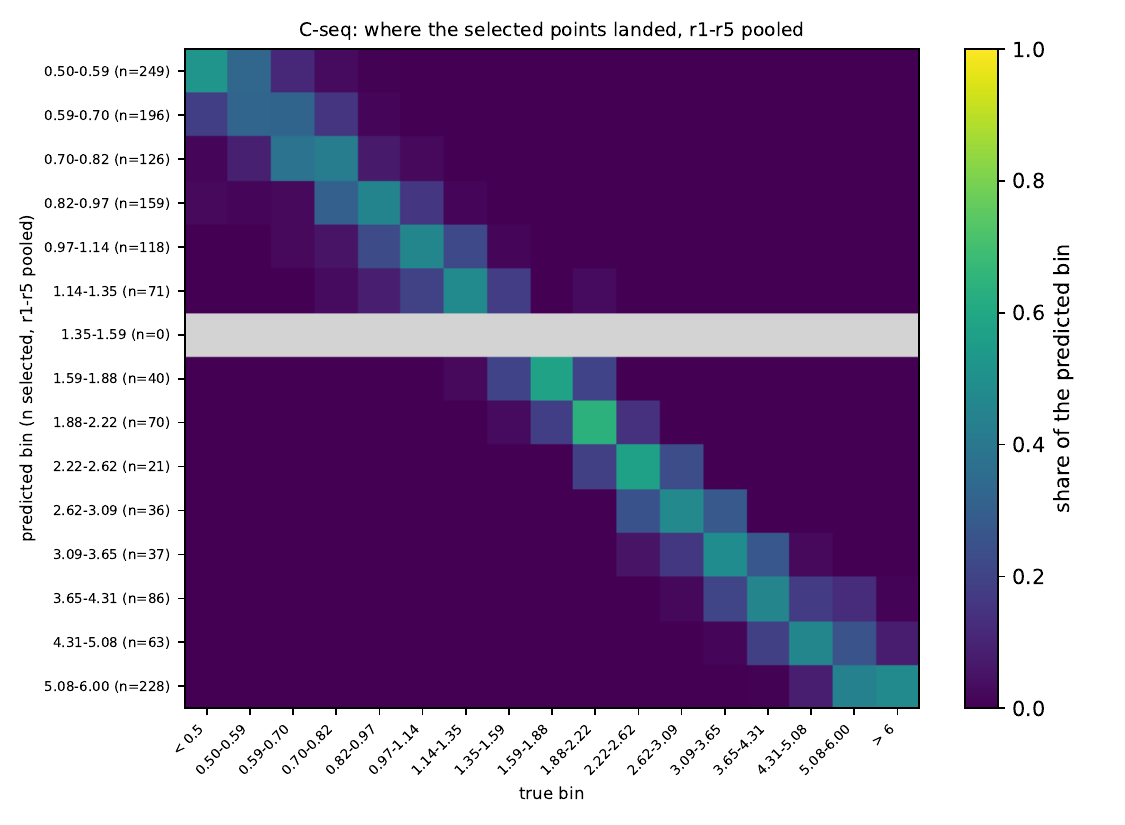}\\[4pt]
\includegraphics[width=0.40\linewidth]{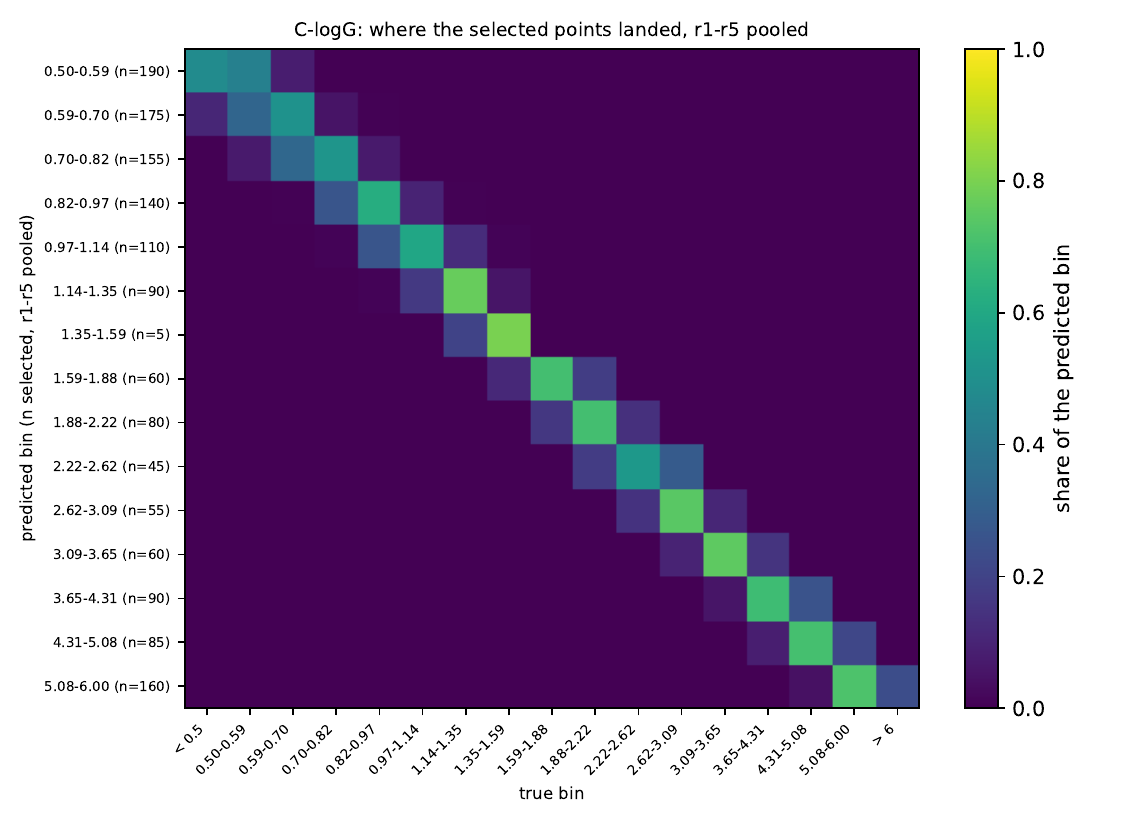}\hspace{0.04\linewidth}%
\includegraphics[width=0.40\linewidth]{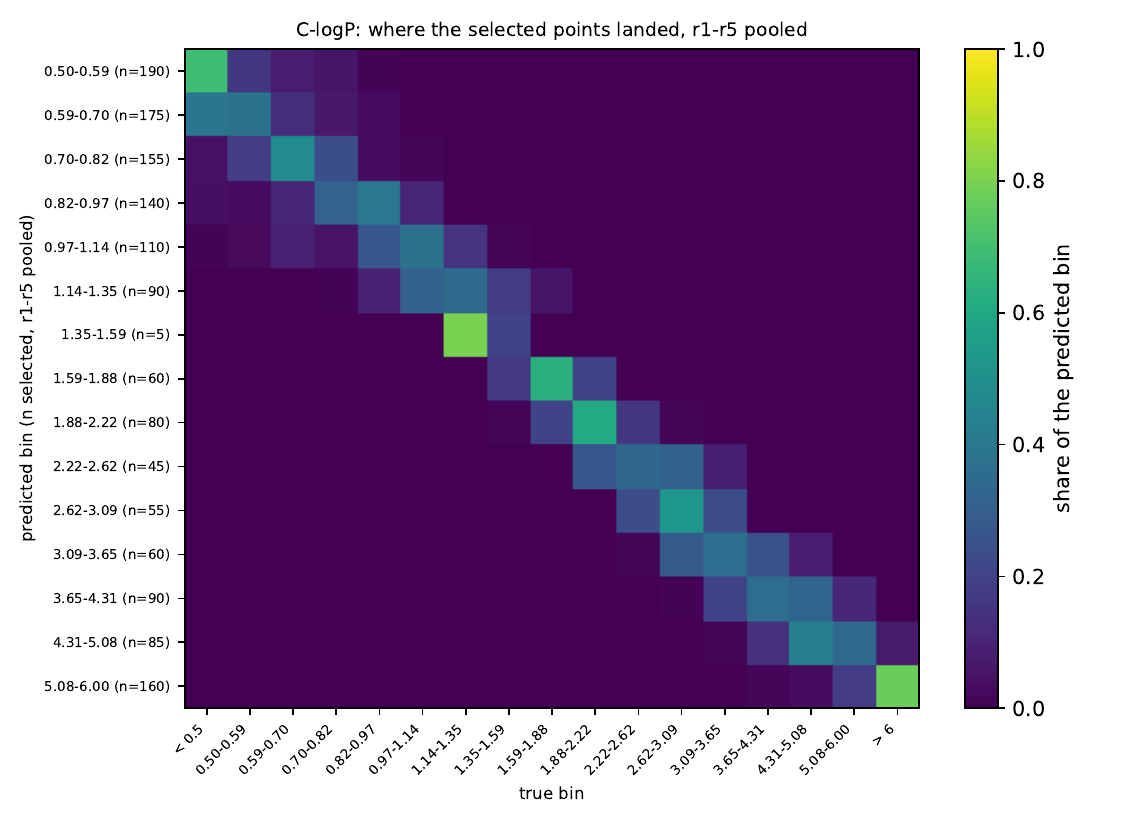}
\caption[Predicted against true bin]{Predicted bin (rows) against true bin (columns) for the points the screener selected, C-log, C-seq, C-logG and C-logP, pooled over the confirmatory realizations r1-r5; r0 is excluded. Each row is normalized by the number of points selected into that predicted bin, given beside its label; a predicted bin into which nothing was selected is shown grey. Bins follow the generator, (lo, hi], so the outer columns are FOS < 0.5 and FOS > 6. The matrices describe the screener only on the points it selected, which were predicted inside the selection range and inside a deficit bin; they are not its accuracy over the candidate pool. Off-diagonal mass includes compensating errors, which leave achieved minus intended untouched. C-oracle has no matrix: it bins on the solved FOS, so every selected point stays in its bin by construction. C-oracle bins on the solved FOS and meets the intended allocation by construction; its rows are a check, not a finding. What known bin membership changes in test accuracy is read only from the paired contrast C-oracle against C-logP, in this design.}\label{fig:s:matrix}
\end{figure}

\clearpage
\subsection{The ablations}\label{sec:s-structural-ablation}

\paragraph{The original C$_0$.} The original C$_0$ (arm C$_0$, role guided) exists only as r0 in the main pipeline. It was never replicated, so it has no replication interval, and it predates the log-deficit rule and stores no allocation provenance, so it has no allocation row. Nothing is computed for it here.

\begin{figure}[!htbp]\centering
\includegraphics[width=\textwidth,height=0.60\textheight,keepaspectratio]{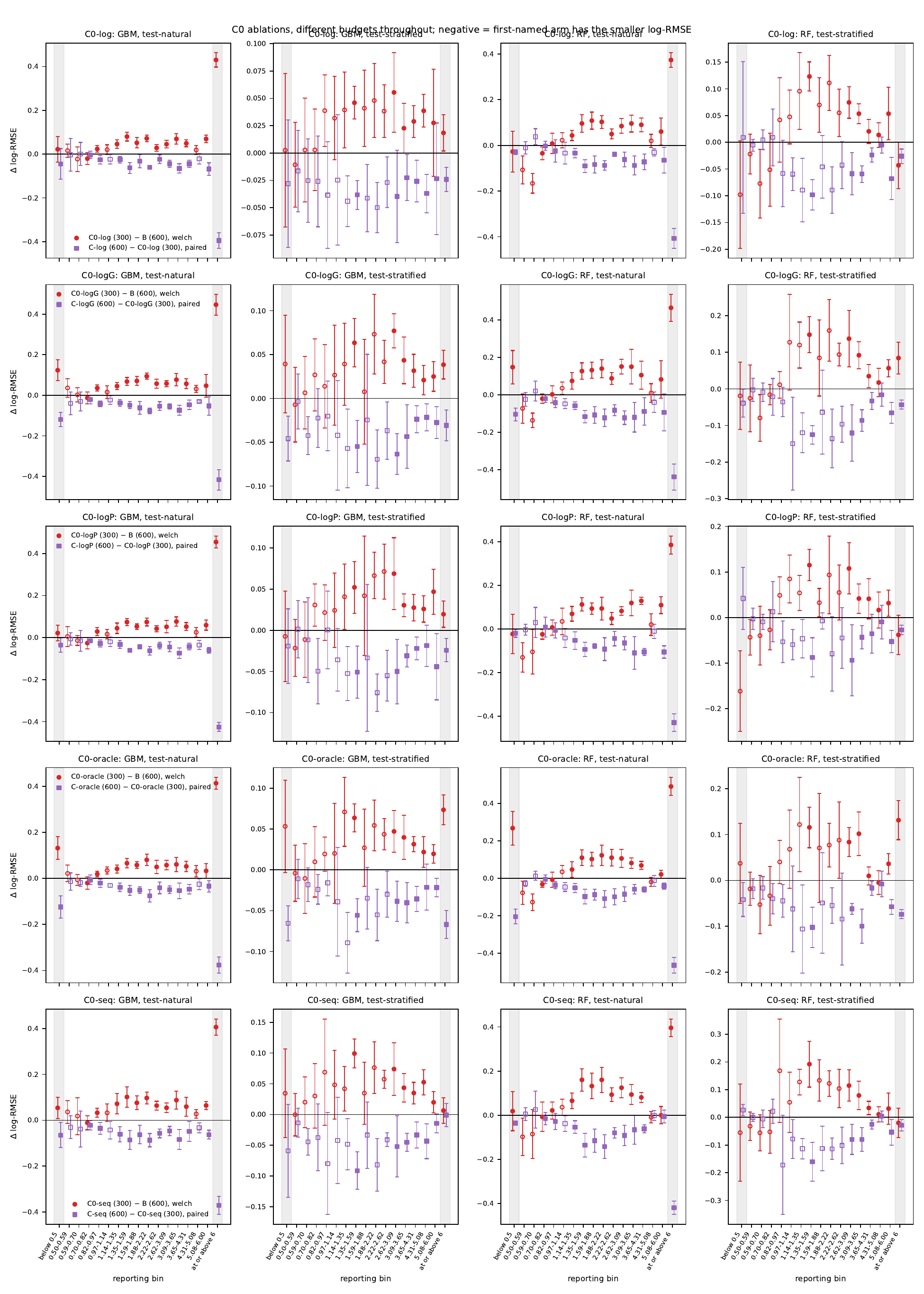}
\caption[The C$_0$ ablations per bin]{Supplementary ablation. C$_0$-log, C$_0$-seq, C$_0$-logG, C$_0$-oracle and C$_0$-logP train on the guided points alone, without the seed. Training-set sizes: B 600, C-log 600, C-seq 600, C-logG 600, C-oracle 600, C-logP 600, C$_0$-log 300, C$_0$-seq 300, C$_0$-logG 300, C$_0$-oracle 300, C$_0$-logP 300. Red: C$_0$ minus B, 95\% Welch interval. Purple: C minus the corresponding C$_0$, 95\% paired interval over the realizations, because within a realization C$_0$ trains on exactly the guided points that C adds to the seed. The two differences are oriented oppositely; in both, a negative value means that the first-named arm has the smaller log-RMSE. Every comparison joins arms trained on different budgets and is not a like-for-like comparison. Means over the five confirmatory realizations, on the half-open reporting zones; grey shading marks the two bins outside the selection range, and hollow markers the bins with fewer than 10 test points. The intervals visualize variability between designs and constitute no test; an interval containing 0 is inconclusive, not a tie. C-oracle, C-logP and their C$_0$ ablations exist only in r1-r5; they have no r0.}\label{fig:s:ablation}
\end{figure}

{\small\begingroup
\small
\setlength{\tabcolsep}{4pt}
\setlength{\LTcapwidth}{\textwidth}
\begin{longtable}{llrrrrr}
\caption{C$_0$ training distribution against B, as the fraction of the training set in each reporting bin, with descriptive 95\% Welch intervals over r1-r5. Fractions only, never raw counts, because C$_0$ trains on 300 points and B on 600.}\label{tab:structural:ablation:fractions} \\
\toprule
arm & bin & C$_0$ fraction & B fraction & diff & ci lo & ci hi \\
\midrule
\endfirsthead
\multicolumn{7}{l}{\textit{Table \thetable, continued}} \\
\toprule
arm & bin & C$_0$ fraction & B fraction & diff & ci lo & ci hi \\
\midrule
\endhead
\midrule
\multicolumn{7}{r}{\textit{continued on the next page}} \\
\endfoot
\bottomrule
\endlastfoot
C$_0$-log & below 0.5 & 0.145 & 0.026 & 0.119 & 0.096 & 0.142 \\
C$_0$-log & 0.50-0.59 & 0.075 & 0.007 & 0.069 & 0.054 & 0.083 \\
C$_0$-log & 0.59-0.70 & 0.089 & 0.016 & 0.074 & 0.059 & 0.088 \\
C$_0$-log & 0.70-0.82 & 0.085 & 0.026 & 0.059 & 0.044 & 0.074 \\
C$_0$-log & 0.82-0.97 & 0.072 & 0.036 & 0.036 & 0.020 & 0.051 \\
C$_0$-log & 0.97-1.14 & 0.061 & 0.053 & 0.008 & -0.006 & 0.023 \\
C$_0$-log & 1.14-1.35 & 0.035 & 0.065 & -0.029 & -0.039 & -0.019 \\
C$_0$-log & 1.35-1.59 & 0.025 & 0.107 & -0.083 & -0.088 & -0.077 \\
C$_0$-log & 1.59-1.88 & 0.031 & 0.089 & -0.058 & -0.064 & -0.051 \\
C$_0$-log & 1.88-2.22 & 0.041 & 0.085 & -0.044 & -0.056 & -0.031 \\
C$_0$-log & 2.22-2.62 & 0.037 & 0.097 & -0.060 & -0.071 & -0.048 \\
C$_0$-log & 2.62-3.09 & 0.036 & 0.088 & -0.052 & -0.064 & -0.040 \\
C$_0$-log & 3.09-3.65 & 0.045 & 0.081 & -0.035 & -0.047 & -0.024 \\
C$_0$-log & 3.65-4.31 & 0.043 & 0.067 & -0.024 & -0.038 & -0.010 \\
C$_0$-log & 4.31-5.08 & 0.044 & 0.054 & -0.010 & -0.026 & 0.006 \\
C$_0$-log & 5.08-6.00 & 0.056 & 0.030 & 0.026 & 0.016 & 0.036 \\
C$_0$-log & at or above 6 & 0.077 & 0.072 & 0.004 & -0.002 & 0.010 \\
C$_0$-seq & below 0.5 & 0.115 & 0.026 & 0.088 & 0.075 & 0.102 \\
C$_0$-seq & 0.50-0.59 & 0.107 & 0.007 & 0.100 & 0.096 & 0.104 \\
C$_0$-seq & 0.59-0.70 & 0.098 & 0.016 & 0.082 & 0.075 & 0.089 \\
C$_0$-seq & 0.70-0.82 & 0.098 & 0.026 & 0.072 & 0.064 & 0.079 \\
C$_0$-seq & 0.82-0.97 & 0.079 & 0.036 & 0.042 & 0.037 & 0.048 \\
C$_0$-seq & 0.97-1.14 & 0.064 & 0.053 & 0.011 & 0.000 & 0.022 \\
C$_0$-seq & 1.14-1.35 & 0.042 & 0.065 & -0.023 & -0.034 & -0.012 \\
C$_0$-seq & 1.35-1.59 & 0.017 & 0.107 & -0.091 & -0.096 & -0.085 \\
C$_0$-seq & 1.59-1.88 & 0.024 & 0.089 & -0.065 & -0.075 & -0.055 \\
C$_0$-seq & 1.88-2.22 & 0.039 & 0.085 & -0.046 & -0.056 & -0.036 \\
C$_0$-seq & 2.22-2.62 & 0.022 & 0.097 & -0.075 & -0.087 & -0.063 \\
C$_0$-seq & 2.62-3.09 & 0.020 & 0.088 & -0.068 & -0.079 & -0.057 \\
C$_0$-seq & 3.09-3.65 & 0.031 & 0.081 & -0.049 & -0.062 & -0.037 \\
C$_0$-seq & 3.65-4.31 & 0.041 & 0.067 & -0.026 & -0.038 & -0.014 \\
C$_0$-seq & 4.31-5.08 & 0.043 & 0.054 & -0.012 & -0.023 & -0.000 \\
C$_0$-seq & 5.08-6.00 & 0.085 & 0.030 & 0.055 & 0.042 & 0.067 \\
C$_0$-seq & at or above 6 & 0.076 & 0.072 & 0.004 & -0.012 & 0.019 \\
C$_0$-logG & below 0.5 & 0.074 & 0.026 & 0.048 & 0.025 & 0.071 \\
C$_0$-logG & 0.50-0.59 & 0.101 & 0.007 & 0.094 & 0.073 & 0.115 \\
C$_0$-logG & 0.59-0.70 & 0.105 & 0.016 & 0.089 & 0.074 & 0.104 \\
C$_0$-logG & 0.70-0.82 & 0.085 & 0.026 & 0.059 & 0.047 & 0.071 \\
C$_0$-logG & 0.82-0.97 & 0.086 & 0.036 & 0.050 & 0.040 & 0.059 \\
C$_0$-logG & 0.97-1.14 & 0.063 & 0.053 & 0.010 & -0.003 & 0.022 \\
C$_0$-logG & 1.14-1.35 & 0.057 & 0.065 & -0.008 & -0.019 & 0.003 \\
C$_0$-logG & 1.35-1.59 & 0.011 & 0.107 & -0.096 & -0.101 & -0.091 \\
C$_0$-logG & 1.59-1.88 & 0.037 & 0.089 & -0.052 & -0.061 & -0.044 \\
C$_0$-logG & 1.88-2.22 & 0.050 & 0.085 & -0.035 & -0.055 & -0.015 \\
C$_0$-logG & 2.22-2.62 & 0.029 & 0.097 & -0.068 & -0.077 & -0.059 \\
C$_0$-logG & 2.62-3.09 & 0.040 & 0.088 & -0.048 & -0.062 & -0.034 \\
C$_0$-logG & 3.09-3.65 & 0.037 & 0.081 & -0.043 & -0.055 & -0.031 \\
C$_0$-logG & 3.65-4.31 & 0.052 & 0.067 & -0.015 & -0.030 & -0.001 \\
C$_0$-logG & 4.31-5.08 & 0.060 & 0.054 & 0.006 & -0.010 & 0.021 \\
C$_0$-logG & 5.08-6.00 & 0.089 & 0.030 & 0.059 & 0.041 & 0.077 \\
C$_0$-logG & at or above 6 & 0.025 & 0.072 & -0.047 & -0.061 & -0.033 \\
C$_0$-oracle & below 0.5 & 0.000 & 0.026 & -0.026 & -0.031 & -0.022 \\
C$_0$-oracle & 0.50-0.59 & 0.127 & 0.007 & 0.120 & 0.117 & 0.123 \\
C$_0$-oracle & 0.59-0.70 & 0.117 & 0.016 & 0.101 & 0.096 & 0.106 \\
C$_0$-oracle & 0.70-0.82 & 0.103 & 0.026 & 0.077 & 0.071 & 0.083 \\
C$_0$-oracle & 0.82-0.97 & 0.093 & 0.036 & 0.057 & 0.052 & 0.062 \\
C$_0$-oracle & 0.97-1.14 & 0.073 & 0.053 & 0.020 & 0.011 & 0.029 \\
C$_0$-oracle & 1.14-1.35 & 0.060 & 0.065 & -0.005 & -0.011 & 0.002 \\
C$_0$-oracle & 1.35-1.59 & 0.003 & 0.107 & -0.104 & -0.107 & -0.101 \\
C$_0$-oracle & 1.59-1.88 & 0.040 & 0.089 & -0.049 & -0.055 & -0.043 \\
C$_0$-oracle & 1.88-2.22 & 0.053 & 0.085 & -0.032 & -0.041 & -0.023 \\
C$_0$-oracle & 2.22-2.62 & 0.030 & 0.097 & -0.067 & -0.075 & -0.059 \\
C$_0$-oracle & 2.62-3.09 & 0.037 & 0.088 & -0.051 & -0.062 & -0.041 \\
C$_0$-oracle & 3.09-3.65 & 0.040 & 0.081 & -0.041 & -0.052 & -0.030 \\
C$_0$-oracle & 3.65-4.31 & 0.060 & 0.067 & -0.007 & -0.018 & 0.003 \\
C$_0$-oracle & 4.31-5.08 & 0.057 & 0.054 & 0.002 & 0.001 & 0.003 \\
C$_0$-oracle & 5.08-6.00 & 0.107 & 0.030 & 0.077 & 0.068 & 0.085 \\
C$_0$-oracle & at or above 6 & 0.000 & 0.072 & -0.072 & -0.077 & -0.067 \\
C$_0$-logP & below 0.5 & 0.143 & 0.026 & 0.116 & 0.089 & 0.143 \\
C$_0$-logP & 0.50-0.59 & 0.088 & 0.007 & 0.081 & 0.070 & 0.093 \\
C$_0$-logP & 0.59-0.70 & 0.092 & 0.016 & 0.076 & 0.056 & 0.096 \\
C$_0$-logP & 0.70-0.82 & 0.074 & 0.026 & 0.048 & 0.030 & 0.065 \\
C$_0$-logP & 0.82-0.97 & 0.071 & 0.036 & 0.035 & 0.018 & 0.052 \\
C$_0$-logP & 0.97-1.14 & 0.058 & 0.053 & 0.005 & -0.015 & 0.025 \\
C$_0$-logP & 1.14-1.35 & 0.035 & 0.065 & -0.030 & -0.038 & -0.022 \\
C$_0$-logP & 1.35-1.59 & 0.020 & 0.107 & -0.087 & -0.105 & -0.070 \\
C$_0$-logP & 1.59-1.88 & 0.039 & 0.089 & -0.050 & -0.057 & -0.043 \\
C$_0$-logP & 1.88-2.22 & 0.049 & 0.085 & -0.036 & -0.046 & -0.027 \\
C$_0$-logP & 2.22-2.62 & 0.028 & 0.097 & -0.069 & -0.077 & -0.061 \\
C$_0$-logP & 2.62-3.09 & 0.041 & 0.088 & -0.047 & -0.057 & -0.036 \\
C$_0$-logP & 3.09-3.65 & 0.039 & 0.081 & -0.042 & -0.055 & -0.029 \\
C$_0$-logP & 3.65-4.31 & 0.041 & 0.067 & -0.026 & -0.037 & -0.015 \\
C$_0$-logP & 4.31-5.08 & 0.051 & 0.054 & -0.004 & -0.017 & 0.010 \\
C$_0$-logP & 5.08-6.00 & 0.045 & 0.030 & 0.015 & 0.006 & 0.023 \\
C$_0$-logP & at or above 6 & 0.087 & 0.072 & 0.014 & 0.008 & 0.020 \\
\end{longtable}
\endgroup
}

\clearpage
\section{The single realization}\label{sec:s-r0}

The tuned fits of the single realization $r_0$ (\cref{sec:res-sensitivity}). They are exploratory: $r_0$ shaped the later variants and their predictions, and it enters no interval. The pool arms have no $r_0$. Under the log criterion, which is the primary one: the global metrics, the RMSE by subset, the log-RMSE by reporting zone, the declared paired comparisons and the cross-conformal-style intervals. Under the FOS-scale criterion, the sensitivity check: the comparison of the two criteria and the paired comparisons. The one-sided direction was declared for the FOS scale on the whole test set and the selection range; on the log scale and on the decision range the one-sided $p$-values are post hoc (\cref{sec:statistics}). The paired bootstrap covers test-sampling variability only.

\subsection{Natural test set, log criterion}

\begin{table}[htbp]
\centering
\small
\setlength{\tabcolsep}{4pt}
\caption{Global metrics on test-natural. C$_0$, C$_0$-log, C$_0$-seq and C$_0$-logG are ablations: they train on the added points without the seed and carry no budget.}
\label{tab:test-natural__log-rmse:global}
\begin{tabular}{lrrrrrrrrr}
\toprule
arm & $R^2$ & RMSE & log-RMSE & MAE & bias & max |e| & minimax & worst zone & minimax log \\
\midrule
\multicolumn{10}{l}{\textit{RF}} \\
A & 0.951 & 0.572 & 0.169 & 0.335 & -0.064 & 3.756 & 1.650 & FOS ≥ 6 & 0.528 \\
B & 0.969 & 0.455 & 0.133 & 0.234 & -0.046 & 3.394 & 1.345 & FOS ≥ 6 & 0.483 \\
C & 0.978 & 0.385 & 0.118 & 0.218 & -0.030 & 2.284 & 1.162 & FOS ≥ 6 & 0.386 \\
C$_0$ & 0.682 & 1.453 & 0.222 & 0.455 & -0.289 & 14.399 & 5.072 & FOS ≥ 6 & 0.668 \\
C-log & 0.971 & 0.438 & 0.135 & 0.260 & -0.029 & 2.248 & 1.159 & FOS ≥ 6 & 0.396 \\
C$_0$-log & 0.642 & 1.542 & 0.247 & 0.605 & -0.317 & 14.236 & 5.217 & FOS ≥ 6 & 0.553 \\
C-seq & 0.975 & 0.409 & 0.121 & 0.239 & -0.027 & 2.448 & 1.117 & FOS ≥ 6 & 0.367 \\
C$_0$-seq & 0.637 & 1.553 & 0.276 & 0.617 & -0.295 & 14.686 & 5.165 & FOS ≥ 6 & 0.535 \\
C-logG & 0.967 & 0.471 & 0.130 & 0.259 & -0.057 & 3.614 & 1.452 & FOS ≥ 6 & 0.437 \\
C$_0$-logG & 0.615 & 1.598 & 0.261 & 0.599 & -0.388 & 14.530 & 5.455 & FOS ≥ 6 & 0.597 \\
\addlinespace
\multicolumn{10}{l}{\textit{GBM}} \\
A & 0.981 & 0.351 & 0.088 & 0.178 & -0.082 & 3.166 & 1.071 & FOS ≥ 6 & 0.270 \\
B & 0.987 & 0.290 & 0.059 & 0.116 & -0.046 & 4.051 & 0.956 & FOS ≥ 6 & 0.197 \\
C & 0.989 & 0.271 & 0.062 & 0.140 & -0.055 & 1.935 & 0.853 & FOS ≥ 6 & 0.130 \\
C$_0$ & 0.712 & 1.383 & 0.177 & 0.391 & -0.266 & 13.804 & 4.842 & FOS ≥ 6 & 0.485 \\
C-log & 0.986 & 0.305 & 0.056 & 0.138 & -0.056 & 2.389 & 0.989 & FOS ≥ 6 & 0.096 \\
C$_0$-log & 0.697 & 1.419 & 0.169 & 0.402 & -0.269 & 14.508 & 4.951 & FOS ≥ 6 & 0.508 \\
C-seq & 0.978 & 0.380 & 0.063 & 0.158 & -0.068 & 3.440 & 1.272 & FOS ≥ 6 & 0.131 \\
C$_0$-seq & 0.701 & 1.409 & 0.189 & 0.432 & -0.259 & 14.078 & 4.861 & FOS ≥ 6 & 0.490 \\
C-logG & 0.977 & 0.391 & 0.070 & 0.161 & -0.069 & 3.530 & 1.301 & FOS ≥ 6 & 0.183 \\
C$_0$-logG & 0.697 & 1.419 & 0.179 & 0.427 & -0.284 & 14.009 & 4.933 & FOS ≥ 6 & 0.503 \\
\bottomrule
\end{tabular}
\end{table}

\begin{table}[htbp]
\centering
\caption{RMSE on test-natural by subset, log and FOS scale. The decision range [0.5, 2) carries the primary criterion (log scale). C$_0$, C$_0$-log, C$_0$-seq and C$_0$-logG are ablations: they train on the added points without the seed and carry no budget.}
\label{tab:test-natural__log-rmse:subsets}
\begin{tabular}{lrrrrrr}
\toprule
 & \multicolumn{3}{c}{log-RMSE} & \multicolumn{3}{c}{RMSE} \\
\cmidrule(lr){2-4} \cmidrule(lr){5-7}
arm & all & selection range & decision range & all & selection range & decision range \\
\midrule
\multicolumn{7}{l}{\textit{RF}} \\
A & 0.169 & 0.133 & 0.140 & 0.572 & 0.350 & 0.155 \\
B & 0.133 & 0.096 & 0.098 & 0.455 & 0.263 & 0.120 \\
C & 0.118 & 0.093 & 0.111 & 0.385 & 0.212 & 0.139 \\
C$_0$ & 0.222 & 0.121 & 0.152 & 1.453 & 0.240 & 0.179 \\
C-log & 0.135 & 0.116 & 0.119 & 0.438 & 0.308 & 0.156 \\
C$_0$-log & 0.247 & 0.183 & 0.195 & 1.542 & 0.472 & 0.290 \\
C-seq & 0.121 & 0.102 & 0.095 & 0.409 & 0.275 & 0.137 \\
C$_0$-seq & 0.276 & 0.229 & 0.208 & 1.553 & 0.559 & 0.339 \\
C-logG & 0.130 & 0.095 & 0.096 & 0.471 & 0.243 & 0.132 \\
C$_0$-logG & 0.261 & 0.180 & 0.160 & 1.598 & 0.440 & 0.229 \\
\addlinespace
\multicolumn{7}{l}{\textit{GBM}} \\
A & 0.088 & 0.071 & 0.076 & 0.351 & 0.188 & 0.096 \\
B & 0.059 & 0.044 & 0.050 & 0.290 & 0.109 & 0.064 \\
C & 0.062 & 0.053 & 0.057 & 0.271 & 0.131 & 0.075 \\
C$_0$ & 0.177 & 0.088 & 0.098 & 1.383 & 0.203 & 0.141 \\
C-log & 0.056 & 0.049 & 0.052 & 0.305 & 0.129 & 0.071 \\
C$_0$-log & 0.169 & 0.089 & 0.084 & 1.419 & 0.243 & 0.121 \\
C-seq & 0.063 & 0.051 & 0.054 & 0.380 & 0.131 & 0.079 \\
C$_0$-seq & 0.189 & 0.129 & 0.115 & 1.409 & 0.328 & 0.173 \\
C-logG & 0.070 & 0.056 & 0.061 & 0.391 & 0.139 & 0.087 \\
C$_0$-logG & 0.179 & 0.106 & 0.108 & 1.419 & 0.275 & 0.162 \\
\bottomrule
\end{tabular}
\end{table}

\begin{table}[htbp]
\centering
\small
\setlength{\tabcolsep}{4pt}
\caption{Zone RMSE (log FOS) on test-natural, rf. C$_0$, C$_0$-log, C$_0$-seq and C$_0$-logG are ablations: they train on the added points without the seed and carry no budget.}
\label{tab:test-natural__log-rmse:zones_log:rf}
\begin{tabular}{lrrrrrrrrrrr}
\toprule
zone & n & A & B & C & C$_0$ & C-log & C$_0$-log & C-seq & C$_0$-seq & C-logG & C$_0$-logG \\
\midrule
FOS < 0.5 & 10 & 0.528 & 0.483 & 0.386 & 0.668 & 0.396 & 0.456 & 0.367 & 0.454 & 0.437 & 0.570 \\
0.5–1 & 33 & 0.222 & 0.136 & 0.147 & 0.211 & 0.162 & 0.213 & 0.081 & 0.113 & 0.102 & 0.111 \\
1–2 & 94 & 0.096 & 0.080 & 0.095 & 0.125 & 0.100 & 0.188 & 0.099 & 0.233 & 0.093 & 0.174 \\
2–6 & 139 & 0.127 & 0.094 & 0.073 & 0.082 & 0.112 & 0.171 & 0.109 & 0.246 & 0.095 & 0.196 \\
FOS ≥ 6 & 24 & 0.209 & 0.145 & 0.124 & 0.520 & 0.120 & 0.553 & 0.111 & 0.535 & 0.173 & 0.597 \\
\bottomrule
\end{tabular}
\end{table}

\begin{table}[htbp]
\centering
\small
\setlength{\tabcolsep}{4pt}
\caption{Zone RMSE (log FOS) on test-natural, gbm. C$_0$, C$_0$-log, C$_0$-seq and C$_0$-logG are ablations: they train on the added points without the seed and carry no budget.}
\label{tab:test-natural__log-rmse:zones_log:gbm}
\begin{tabular}{lrrrrrrrrrrr}
\toprule
zone & n & A & B & C & C$_0$ & C-log & C$_0$-log & C-seq & C$_0$-seq & C-logG & C$_0$-logG \\
\midrule
FOS < 0.5 & 10 & 0.270 & 0.197 & 0.130 & 0.414 & 0.096 & 0.158 & 0.131 & 0.241 & 0.183 & 0.242 \\
0.5–1 & 33 & 0.092 & 0.064 & 0.065 & 0.101 & 0.052 & 0.083 & 0.038 & 0.064 & 0.056 & 0.072 \\
1–2 & 94 & 0.069 & 0.044 & 0.054 & 0.097 & 0.052 & 0.085 & 0.059 & 0.128 & 0.062 & 0.118 \\
2–6 & 139 & 0.066 & 0.037 & 0.049 & 0.079 & 0.046 & 0.093 & 0.048 & 0.140 & 0.051 & 0.104 \\
FOS ≥ 6 & 24 & 0.108 & 0.078 & 0.097 & 0.485 & 0.095 & 0.508 & 0.118 & 0.490 & 0.114 & 0.503 \\
\bottomrule
\end{tabular}
\end{table}

\begingroup
\small
\setlength{\tabcolsep}{3pt}
\setlength{\LTcapwidth}{\textwidth}
\begin{longtable}{llllrrrrrrr}
\caption{Paired comparisons on test-natural: Wilcoxon on d = |e1| - |e2| (smaller absolute errors) and bootstrap ΔRMSE. C$_0$, C$_0$-log, C$_0$-seq and C$_0$-logG are ablations: they train on the added points without the seed and carry no budget.}\label{tab:test-natural__log-rmse:comparisons} \\
\toprule
 &  &  &  & \multicolumn{4}{c}{Wilcoxon on d} & \multicolumn{3}{c}{bootstrap} \\
\cmidrule(lr){5-8} \cmidrule(lr){9-11}
subset & scale & comparison & alternative & n & median & P(d<0) & p & ΔRMSE & CI low & CI high \\
\midrule
\endfirsthead
\multicolumn{11}{l}{\textit{Table \thetable, continued}} \\
\toprule
 &  &  &  & \multicolumn{4}{c}{Wilcoxon on d} & \multicolumn{3}{c}{bootstrap} \\
\cmidrule(lr){5-8} \cmidrule(lr){9-11}
subset & scale & comparison & alternative & n & median & P(d<0) & p & ΔRMSE & CI low & CI high \\
\midrule
\endhead
\midrule
\multicolumn{11}{r}{\textit{continued on the next page}} \\
\endfoot
\bottomrule
\endlastfoot
\multicolumn{11}{l}{\textit{GBM}} \\
all & log & C vs B & less & 300 & 0.005 & 0.440 & 1.000 & 0.003 & -0.009 & 0.015 \\
all & fos & C vs B & less & 300 & 0.010 & 0.437 & 0.999 & -0.018 & -0.195 & 0.145 \\
selection-range & log & C vs B & less & 266 & 0.004 & 0.455 & 0.998 & 0.010 & 0.004 & 0.015 \\
selection-range & fos & C vs B & less & 266 & 0.007 & 0.451 & 0.983 & 0.022 & 0.005 & 0.039 \\
decision-range & log & C vs B & less & 127 & 0.004 & 0.425 & 0.996 & 0.008 & -0.000 & 0.015 \\
decision-range & fos & C vs B & less & 127 & 0.006 & 0.425 & 0.994 & 0.011 & 0.001 & 0.020 \\
all & log & C vs A & less & 300 & -0.008 & 0.600 & <0.001 & -0.026 & -0.041 & -0.013 \\
\end{longtable}
\endgroup

\begin{table}[htbp]
\centering
\caption{Cross-conformal-style intervals on test-natural, empirical coverage. Coverage over the whole test set and by reporting zone. C$_0$, C$_0$-log, C$_0$-seq and C$_0$-logG are ablations: they train on the added points without the seed and carry no budget.}
\label{tab:test-natural__log-rmse:conformal}
\begin{tabular}{lrrrrrr}
\toprule
 &  & \multicolumn{5}{c}{by zone} \\
\cmidrule(lr){3-7}
arm & all & FOS < 0.5 & 0.5–1 & 1–2 & 2–6 & FOS ≥ 6 \\
\midrule
\multicolumn{7}{l}{\textit{RF}} \\
A & 0.923 & 0.300 & 0.727 & 1.000 & 0.964 & 0.917 \\
B & 0.897 & 0.200 & 0.788 & 0.957 & 0.942 & 0.833 \\
C & 0.933 & 0.100 & 0.848 & 0.979 & 0.978 & 0.958 \\
C$_0$ & 0.880 & 0.000 & 0.788 & 0.947 & 1.000 & 0.417 \\
C-log & 0.893 & 0.400 & 0.848 & 0.926 & 0.914 & 0.917 \\
C$_0$-log & 0.813 & 0.500 & 0.848 & 0.862 & 0.871 & 0.375 \\
C-seq & 0.927 & 0.500 & 0.970 & 0.947 & 0.928 & 0.958 \\
C$_0$-seq & 0.780 & 0.500 & 0.970 & 0.766 & 0.827 & 0.417 \\
C-logG & 0.913 & 0.200 & 0.909 & 0.979 & 0.942 & 0.792 \\
C$_0$-logG & 0.810 & 0.100 & 0.970 & 0.840 & 0.885 & 0.333 \\
\addlinespace
\multicolumn{7}{l}{\textit{GBM}} \\
A & 0.920 & 0.700 & 0.818 & 0.957 & 0.964 & 0.750 \\
B & 0.930 & 0.500 & 0.788 & 0.957 & 0.978 & 0.917 \\
C & 0.917 & 0.700 & 0.848 & 0.957 & 0.957 & 0.708 \\
C$_0$ & 0.897 & 0.300 & 0.879 & 0.968 & 0.971 & 0.458 \\
C-log & 0.927 & 0.600 & 0.879 & 0.957 & 0.971 & 0.750 \\
C$_0$-log & 0.833 & 0.700 & 0.939 & 0.894 & 0.842 & 0.458 \\
C-seq & 0.917 & 0.700 & 0.970 & 0.947 & 0.957 & 0.583 \\
C$_0$-seq & 0.787 & 0.700 & 0.939 & 0.755 & 0.842 & 0.417 \\
C-logG & 0.867 & 0.400 & 0.909 & 0.883 & 0.935 & 0.542 \\
C$_0$-logG & 0.810 & 0.500 & 0.970 & 0.787 & 0.878 & 0.417 \\
\bottomrule
\end{tabular}
\end{table}
\begin{table}[htbp]
\centering
\caption{Cross-conformal-style intervals on test-natural, empirical coverage. Mean interval width over the whole test set and by reporting zone; the interval is [prediction / factor, prediction x factor]. C$_0$, C$_0$-log, C$_0$-seq and C$_0$-logG are ablations: they train on the added points without the seed and carry no budget.}
\label{tab:test-natural__log-rmse:conformal:width}
\begin{tabular}{lrrrrrrr}
\toprule
 &  &  & \multicolumn{5}{c}{by zone} \\
\cmidrule(lr){4-8}
arm & factor & all & FOS < 0.5 & 0.5–1 & 1–2 & 2–6 & FOS ≥ 6 \\
\midrule
\multicolumn{8}{l}{\textit{RF}} \\
A & 1.330 & 1.656 & 0.300 & 0.545 & 0.884 & 1.886 & 5.441 \\
B & 1.200 & 1.058 & 0.182 & 0.320 & 0.558 & 1.225 & 3.431 \\
C & 1.213 & 1.127 & 0.184 & 0.340 & 0.589 & 1.299 & 3.712 \\
C$_0$ & 1.277 & 1.303 & 0.307 & 0.431 & 0.755 & 1.656 & 3.024 \\
C-log & 1.205 & 1.087 & 0.172 & 0.327 & 0.575 & 1.241 & 3.633 \\
C$_0$-log & 1.293 & 1.358 & 0.256 & 0.451 & 0.843 & 1.706 & 3.068 \\
C-seq & 1.203 & 1.078 & 0.166 & 0.304 & 0.558 & 1.242 & 3.611 \\
C$_0$-seq & 1.296 & 1.379 & 0.244 & 0.417 & 0.774 & 1.790 & 3.169 \\
C-logG & 1.213 & 1.118 & 0.184 & 0.315 & 0.581 & 1.293 & 3.698 \\
C$_0$-logG & 1.305 & 1.367 & 0.278 & 0.424 & 0.772 & 1.792 & 2.979 \\
\addlinespace
\multicolumn{8}{l}{\textit{GBM}} \\
A & 1.149 & 0.794 & 0.109 & 0.233 & 0.423 & 0.924 & 2.553 \\
B & 1.090 & 0.497 & 0.067 & 0.140 & 0.262 & 0.580 & 1.605 \\
C & 1.106 & 0.579 & 0.074 & 0.164 & 0.305 & 0.681 & 1.842 \\
C$_0$ & 1.199 & 0.970 & 0.167 & 0.301 & 0.546 & 1.232 & 2.366 \\
C-log & 1.109 & 0.594 & 0.076 & 0.166 & 0.314 & 0.695 & 1.907 \\
C$_0$-log & 1.138 & 0.689 & 0.098 & 0.208 & 0.398 & 0.869 & 1.695 \\
C-seq & 1.110 & 0.599 & 0.079 & 0.168 & 0.317 & 0.706 & 1.888 \\
C$_0$-seq & 1.150 & 0.749 & 0.109 & 0.221 & 0.427 & 0.946 & 1.868 \\
C-logG & 1.099 & 0.542 & 0.072 & 0.152 & 0.284 & 0.640 & 1.718 \\
C$_0$-logG & 1.167 & 0.818 & 0.123 & 0.245 & 0.454 & 1.043 & 2.019 \\
\bottomrule
\end{tabular}
\end{table}

\clearpage
\subsection{Stratified test set, log criterion}

\begin{table}[htbp]
\centering
\small
\setlength{\tabcolsep}{4pt}
\caption{Global metrics on test-stratified. C$_0$, C$_0$-log, C$_0$-seq and C$_0$-logG are ablations: they train on the added points without the seed and carry no budget.}
\label{tab:test-stratified__log-rmse:global}
\begin{tabular}{lrrrrrrrrr}
\toprule
arm & $R^2$ & RMSE & log-RMSE & MAE & bias & max |e| & minimax & worst zone & minimax log \\
\midrule
\multicolumn{10}{l}{\textit{RF}} \\
A & 0.844 & 0.832 & 0.174 & 0.505 & -0.306 & 3.756 & 1.934 & FOS ≥ 6 & 0.515 \\
B & 0.932 & 0.548 & 0.114 & 0.330 & -0.178 & 2.708 & 1.226 & FOS ≥ 6 & 0.273 \\
C & 0.957 & 0.437 & 0.103 & 0.279 & -0.150 & 1.782 & 0.941 & FOS ≥ 6 & 0.222 \\
C$_0$ & 0.951 & 0.464 & 0.131 & 0.312 & -0.130 & 1.840 & 0.982 & FOS ≥ 6 & 0.342 \\
C-log & 0.937 & 0.530 & 0.114 & 0.344 & -0.191 & 2.055 & 1.157 & FOS ≥ 6 & 0.166 \\
C$_0$-log & 0.888 & 0.704 & 0.172 & 0.470 & -0.223 & 2.624 & 1.357 & FOS ≥ 6 & 0.215 \\
C-seq & 0.960 & 0.421 & 0.099 & 0.284 & -0.118 & 2.013 & 0.859 & FOS ≥ 6 & 0.229 \\
C$_0$-seq & 0.910 & 0.631 & 0.188 & 0.448 & -0.102 & 2.141 & 1.048 & FOS ≥ 6 & 0.278 \\
C-logG & 0.899 & 0.667 & 0.119 & 0.371 & -0.256 & 3.246 & 1.599 & FOS ≥ 6 & 0.230 \\
C$_0$-logG & 0.840 & 0.841 & 0.192 & 0.530 & -0.302 & 3.583 & 1.847 & FOS ≥ 6 & 0.278 \\
\addlinespace
\multicolumn{10}{l}{\textit{GBM}} \\
A & 0.984 & 0.264 & 0.069 & 0.178 & -0.089 & 0.870 & 0.521 & FOS ≥ 6 & 0.133 \\
B & 0.995 & 0.147 & 0.040 & 0.104 & -0.027 & 0.498 & 0.246 & FOS ≥ 6 & 0.106 \\
C & 0.991 & 0.198 & 0.047 & 0.137 & -0.046 & 0.707 & 0.377 & FOS ≥ 6 & 0.074 \\
C$_0$ & 0.975 & 0.333 & 0.097 & 0.244 & -0.050 & 1.063 & 0.615 & FOS ≥ 6 & 0.201 \\
C-log & 0.987 & 0.241 & 0.049 & 0.148 & -0.052 & 1.002 & 0.524 & FOS ≥ 6 & 0.083 \\
C$_0$-log & 0.979 & 0.308 & 0.076 & 0.206 & -0.096 & 1.303 & 0.558 & FOS ≥ 6 & 0.087 \\
C-seq & 0.990 & 0.216 & 0.057 & 0.149 & -0.019 & 0.792 & 0.355 & FOS ≥ 6 & 0.093 \\
C$_0$-seq & 0.977 & 0.320 & 0.096 & 0.225 & -0.054 & 1.029 & 0.453 & FOS ≥ 6 & 0.123 \\
C-logG & 0.988 & 0.231 & 0.052 & 0.153 & -0.044 & 0.939 & 0.470 & FOS ≥ 6 & 0.066 \\
C$_0$-logG & 0.966 & 0.391 & 0.094 & 0.257 & -0.134 & 1.837 & 0.791 & FOS ≥ 6 & 0.108 \\
\bottomrule
\end{tabular}
\end{table}

\begin{table}[htbp]
\centering
\caption{RMSE on test-stratified by subset, log and FOS scale. The decision range [0.5, 2) carries the primary criterion (log scale). C$_0$, C$_0$-log, C$_0$-seq and C$_0$-logG are ablations: they train on the added points without the seed and carry no budget.}
\label{tab:test-stratified__log-rmse:subsets}
\begin{tabular}{lrrrrrr}
\toprule
 & \multicolumn{3}{c}{log-RMSE} & \multicolumn{3}{c}{RMSE} \\
\cmidrule(lr){2-4} \cmidrule(lr){5-7}
arm & all & selection range & decision range & all & selection range & decision range \\
\midrule
\multicolumn{7}{l}{\textit{RF}} \\
A & 0.174 & 0.119 & 0.124 & 0.832 & 0.375 & 0.138 \\
B & 0.114 & 0.089 & 0.098 & 0.548 & 0.289 & 0.102 \\
C & 0.103 & 0.090 & 0.117 & 0.437 & 0.258 & 0.124 \\
C$_0$ & 0.131 & 0.119 & 0.177 & 0.464 & 0.284 & 0.204 \\
C-log & 0.114 & 0.102 & 0.123 & 0.530 & 0.301 & 0.164 \\
C$_0$-log & 0.172 & 0.168 & 0.207 & 0.704 & 0.508 & 0.313 \\
C-seq & 0.099 & 0.091 & 0.101 & 0.421 & 0.277 & 0.137 \\
C$_0$-seq & 0.188 & 0.196 & 0.240 & 0.631 & 0.527 & 0.432 \\
C-logG & 0.119 & 0.083 & 0.083 & 0.667 & 0.253 & 0.103 \\
C$_0$-logG & 0.192 & 0.170 & 0.174 & 0.841 & 0.472 & 0.247 \\
\addlinespace
\multicolumn{7}{l}{\textit{GBM}} \\
A & 0.069 & 0.064 & 0.082 & 0.264 & 0.183 & 0.096 \\
B & 0.040 & 0.038 & 0.045 & 0.147 & 0.122 & 0.054 \\
C & 0.047 & 0.045 & 0.046 & 0.198 & 0.144 & 0.054 \\
C$_0$ & 0.097 & 0.095 & 0.127 & 0.333 & 0.253 & 0.174 \\
C-log & 0.049 & 0.041 & 0.039 & 0.241 & 0.139 & 0.056 \\
C$_0$-log & 0.076 & 0.076 & 0.085 & 0.308 & 0.238 & 0.117 \\
C-seq & 0.057 & 0.057 & 0.056 & 0.216 & 0.181 & 0.074 \\
C$_0$-seq & 0.096 & 0.100 & 0.105 & 0.320 & 0.293 & 0.155 \\
C-logG & 0.052 & 0.048 & 0.050 & 0.231 & 0.153 & 0.070 \\
C$_0$-logG & 0.094 & 0.090 & 0.095 & 0.391 & 0.262 & 0.133 \\
\bottomrule
\end{tabular}
\end{table}

\begin{table}[htbp]
\centering
\small
\setlength{\tabcolsep}{4pt}
\caption{Zone RMSE (log FOS) on test-stratified, rf. C$_0$, C$_0$-log, C$_0$-seq and C$_0$-logG are ablations: they train on the added points without the seed and carry no budget.}
\label{tab:test-stratified__log-rmse:zones_log:rf}
\begin{tabular}{lrrrrrrrrrrr}
\toprule
zone & n & A & B & C & C$_0$ & C-log & C$_0$-log & C-seq & C$_0$-seq & C-logG & C$_0$-logG \\
\midrule
FOS < 0.5 & 3 & 0.515 & 0.273 & 0.222 & 0.342 & 0.114 & 0.046 & 0.229 & 0.120 & 0.154 & 0.211 \\
0.5–1 & 12 & 0.169 & 0.143 & 0.169 & 0.225 & 0.143 & 0.190 & 0.109 & 0.138 & 0.102 & 0.121 \\
1–2 & 24 & 0.094 & 0.064 & 0.080 & 0.148 & 0.113 & 0.215 & 0.097 & 0.278 & 0.072 & 0.196 \\
2–6 & 88 & 0.117 & 0.085 & 0.077 & 0.084 & 0.092 & 0.149 & 0.086 & 0.174 & 0.082 & 0.169 \\
FOS ≥ 6 & 23 & 0.292 & 0.178 & 0.136 & 0.141 & 0.166 & 0.206 & 0.115 & 0.151 & 0.230 & 0.278 \\
\bottomrule
\end{tabular}
\end{table}

\begin{table}[htbp]
\centering
\small
\setlength{\tabcolsep}{4pt}
\caption{Zone RMSE (log FOS) on test-stratified, gbm. C$_0$, C$_0$-log, C$_0$-seq and C$_0$-logG are ablations: they train on the added points without the seed and carry no budget.}
\label{tab:test-stratified__log-rmse:zones_log:gbm}
\begin{tabular}{lrrrrrrrrrrr}
\toprule
zone & n & A & B & C & C$_0$ & C-log & C$_0$-log & C-seq & C$_0$-seq & C-logG & C$_0$-logG \\
\midrule
FOS < 0.5 & 3 & 0.133 & 0.106 & 0.074 & 0.201 & 0.083 & 0.045 & 0.093 & 0.107 & 0.062 & 0.097 \\
0.5–1 & 12 & 0.103 & 0.052 & 0.057 & 0.110 & 0.026 & 0.087 & 0.054 & 0.056 & 0.031 & 0.072 \\
1–2 & 24 & 0.069 & 0.042 & 0.039 & 0.135 & 0.044 & 0.085 & 0.058 & 0.123 & 0.057 & 0.104 \\
2–6 & 88 & 0.055 & 0.035 & 0.045 & 0.078 & 0.043 & 0.072 & 0.057 & 0.098 & 0.048 & 0.089 \\
FOS ≥ 6 & 23 & 0.079 & 0.035 & 0.053 & 0.087 & 0.076 & 0.080 & 0.050 & 0.066 & 0.066 & 0.108 \\
\bottomrule
\end{tabular}
\end{table}

\begingroup
\small
\setlength{\tabcolsep}{3pt}
\setlength{\LTcapwidth}{\textwidth}
\begin{longtable}{llllrrrrrrr}
\caption{Paired comparisons on test-stratified: Wilcoxon on d = |e1| - |e2| (smaller absolute errors) and bootstrap ΔRMSE. C$_0$, C$_0$-log, C$_0$-seq and C$_0$-logG are ablations: they train on the added points without the seed and carry no budget.}\label{tab:test-stratified__log-rmse:comparisons} \\
\toprule
 &  &  &  & \multicolumn{4}{c}{Wilcoxon on d} & \multicolumn{3}{c}{bootstrap} \\
\cmidrule(lr){5-8} \cmidrule(lr){9-11}
subset & scale & comparison & alternative & n & median & P(d<0) & p & ΔRMSE & CI low & CI high \\
\midrule
\endfirsthead
\multicolumn{11}{l}{\textit{Table \thetable, continued}} \\
\toprule
 &  &  &  & \multicolumn{4}{c}{Wilcoxon on d} & \multicolumn{3}{c}{bootstrap} \\
\cmidrule(lr){5-8} \cmidrule(lr){9-11}
subset & scale & comparison & alternative & n & median & P(d<0) & p & ΔRMSE & CI low & CI high \\
\midrule
\endhead
\midrule
\multicolumn{11}{r}{\textit{continued on the next page}} \\
\endfoot
\bottomrule
\endlastfoot
\multicolumn{11}{l}{\textit{GBM}} \\
all & log & C vs B & less & 150 & 0.008 & 0.367 & 0.999 & 0.007 & -0.000 & 0.013 \\
all & fos & C vs B & less & 150 & 0.020 & 0.367 & 1.000 & 0.051 & 0.017 & 0.085 \\
selection-range & log & C vs B & less & 124 & 0.008 & 0.363 & 0.999 & 0.007 & 0.002 & 0.013 \\
selection-range & fos & C vs B & less & 124 & 0.020 & 0.363 & 1.000 & 0.023 & 0.000 & 0.045 \\
decision-range & log & C vs B & less & 36 & 0.003 & 0.472 & 0.671 & 0.001 & -0.009 & 0.012 \\
decision-range & fos & C vs B & less & 36 & 0.004 & 0.472 & 0.625 & -0.000 & -0.014 & 0.015 \\
all & log & C vs A & less & 150 & -0.011 & 0.600 & <0.001 & -0.021 & -0.029 & -0.013 \\
\end{longtable}
\endgroup

\begin{table}[htbp]
\centering
\caption{Cross-conformal-style intervals on test-stratified, empirical coverage. Coverage over the whole test set and by reporting zone. C$_0$, C$_0$-log, C$_0$-seq and C$_0$-logG are ablations: they train on the added points without the seed and carry no budget.}
\label{tab:test-stratified__log-rmse:conformal}
\begin{tabular}{lrrrrrr}
\toprule
 &  & \multicolumn{5}{c}{by zone} \\
\cmidrule(lr){3-7}
arm & all & FOS < 0.5 & 0.5–1 & 1–2 & 2–6 & FOS ≥ 6 \\
\midrule
\multicolumn{7}{l}{\textit{RF}} \\
A & 0.920 & 0.000 & 0.917 & 1.000 & 1.000 & 0.652 \\
B & 0.900 & 0.667 & 0.667 & 1.000 & 0.977 & 0.652 \\
C & 0.920 & 0.667 & 0.667 & 1.000 & 0.966 & 0.826 \\
C$_0$ & 0.927 & 0.000 & 0.583 & 0.917 & 0.989 & 1.000 \\
C-log & 0.913 & 1.000 & 0.833 & 0.958 & 0.966 & 0.696 \\
C$_0$-log & 0.880 & 1.000 & 0.917 & 0.750 & 0.943 & 0.739 \\
C-seq & 0.953 & 0.667 & 0.917 & 0.917 & 0.989 & 0.913 \\
C$_0$-seq & 0.913 & 1.000 & 0.917 & 0.875 & 0.909 & 0.957 \\
C-logG & 0.907 & 1.000 & 1.000 & 1.000 & 0.966 & 0.522 \\
C$_0$-logG & 0.840 & 1.000 & 0.917 & 0.792 & 0.909 & 0.565 \\
\addlinespace
\multicolumn{7}{l}{\textit{GBM}} \\
A & 0.947 & 0.667 & 0.917 & 0.917 & 0.966 & 0.957 \\
B & 0.960 & 0.667 & 0.917 & 0.958 & 0.966 & 1.000 \\
C & 0.980 & 0.667 & 1.000 & 1.000 & 0.989 & 0.957 \\
C$_0$ & 0.953 & 0.667 & 1.000 & 0.833 & 0.977 & 1.000 \\
C-log & 0.947 & 0.667 & 1.000 & 0.958 & 0.989 & 0.783 \\
C$_0$-log & 0.907 & 1.000 & 0.833 & 0.875 & 0.932 & 0.870 \\
C-seq & 0.920 & 0.667 & 0.917 & 0.917 & 0.920 & 0.957 \\
C$_0$-seq & 0.907 & 1.000 & 1.000 & 0.833 & 0.909 & 0.913 \\
C-logG & 0.927 & 1.000 & 1.000 & 0.875 & 0.966 & 0.783 \\
C$_0$-logG & 0.927 & 1.000 & 1.000 & 0.875 & 0.966 & 0.783 \\
\bottomrule
\end{tabular}
\end{table}
\begin{table}[htbp]
\centering
\caption{Cross-conformal-style intervals on test-stratified, empirical coverage. Mean interval width over the whole test set and by reporting zone; the interval is [prediction / factor, prediction x factor]. C$_0$, C$_0$-log, C$_0$-seq and C$_0$-logG are ablations: they train on the added points without the seed and carry no budget.}
\label{tab:test-stratified__log-rmse:conformal:width}
\begin{tabular}{lrrrrrrr}
\toprule
 &  &  & \multicolumn{5}{c}{by zone} \\
\cmidrule(lr){4-8}
arm & factor & all & FOS < 0.5 & 0.5–1 & 1–2 & 2–6 & FOS ≥ 6 \\
\midrule
\multicolumn{8}{l}{\textit{RF}} \\
A & 1.330 & 1.880 & 0.437 & 0.464 & 0.862 & 2.064 & 3.165 \\
B & 1.200 & 1.241 & 0.213 & 0.279 & 0.541 & 1.318 & 2.313 \\
C & 1.213 & 1.325 & 0.210 & 0.315 & 0.576 & 1.394 & 2.516 \\
C$_0$ & 1.277 & 1.693 & 0.314 & 0.416 & 0.718 & 1.793 & 3.177 \\
C-log & 1.205 & 1.263 & 0.169 & 0.283 & 0.565 & 1.347 & 2.325 \\
C$_0$-log & 1.293 & 1.735 & 0.230 & 0.395 & 0.818 & 1.840 & 3.183 \\
C-seq & 1.203 & 1.279 & 0.171 & 0.282 & 0.546 & 1.354 & 2.421 \\
C$_0$-seq & 1.296 & 1.810 & 0.261 & 0.396 & 0.777 & 1.948 & 3.302 \\
C-logG & 1.213 & 1.286 & 0.192 & 0.284 & 0.569 & 1.399 & 2.262 \\
C$_0$-logG & 1.305 & 1.751 & 0.301 & 0.385 & 0.744 & 1.940 & 2.983 \\
\addlinespace
\multicolumn{8}{l}{\textit{GBM}} \\
A & 1.149 & 0.968 & 0.138 & 0.209 & 0.410 & 1.006 & 1.910 \\
B & 1.090 & 0.609 & 0.082 & 0.124 & 0.253 & 0.625 & 1.238 \\
C & 1.106 & 0.708 & 0.097 & 0.146 & 0.299 & 0.733 & 1.413 \\
C$_0$ & 1.199 & 1.277 & 0.200 & 0.277 & 0.511 & 1.349 & 2.463 \\
C-log & 1.109 & 0.725 & 0.090 & 0.146 & 0.303 & 0.752 & 1.446 \\
C$_0$-log & 1.138 & 0.897 & 0.121 & 0.188 & 0.384 & 0.930 & 1.776 \\
C-seq & 1.110 & 0.741 & 0.101 & 0.148 & 0.303 & 0.765 & 1.498 \\
C$_0$-seq & 1.150 & 0.984 & 0.142 & 0.202 & 0.401 & 1.026 & 1.948 \\
C-logG & 1.099 & 0.666 & 0.091 & 0.134 & 0.272 & 0.694 & 1.321 \\
C$_0$-logG & 1.167 & 1.060 & 0.153 & 0.226 & 0.435 & 1.115 & 2.052 \\
\bottomrule
\end{tabular}
\end{table}

\clearpage
\subsection{The FOS-scale selection criterion}

\begin{table}[htbp]
\centering
\caption{Selection-criterion sensitivity on test-natural: global RMSE, minimax and within-family rank under each scoring. C$_0$, C$_0$-log, C$_0$-seq and C$_0$-logG are ablations: they train on the added points without the seed and carry no budget.}
\label{tab:test-natural:scorings}
\begin{tabular}{lrrrrrr}
\toprule
 & \multicolumn{3}{c}{log-rmse} & \multicolumn{3}{c}{fos-rmse} \\
\cmidrule(lr){2-4} \cmidrule(lr){5-7}
arm & RMSE & minimax & rank & RMSE & minimax & rank \\
\midrule
\multicolumn{7}{l}{\textit{RF}} \\
A & 0.572 & 1.650 & 6 & 0.587 & 1.703 & 6 \\
B & 0.455 & 1.345 & 4 & 0.455 & 1.345 & 4 \\
C & 0.385 & 1.162 & 1 & 0.385 & 1.162 & 1 \\
C$_0$ & 1.453 & 5.072 & 7 & 1.453 & 5.072 & 7 \\
C-log & 0.438 & 1.159 & 3 & 0.438 & 1.159 & 3 \\
C$_0$-log & 1.542 & 5.217 & 8 & 1.542 & 5.217 & 8 \\
C-seq & 0.409 & 1.117 & 2 & 0.409 & 1.117 & 2 \\
C$_0$-seq & 1.553 & 5.165 & 9 & 1.553 & 5.165 & 9 \\
C-logG & 0.471 & 1.452 & 5 & 0.471 & 1.452 & 5 \\
C$_0$-logG & 1.598 & 5.455 & 10 & 1.598 & 5.455 & 10 \\
\addlinespace
\multicolumn{7}{l}{\textit{GBM}} \\
A & 0.351 & 1.071 & 4 & 0.328 & 0.986 & 5 \\
B & 0.290 & 0.956 & 2 & 0.327 & 1.067 & 4 \\
C & 0.271 & 0.853 & 1 & 0.278 & 0.893 & 1 \\
C$_0$ & 1.383 & 4.842 & 7 & 1.377 & 4.829 & 8 \\
C-log & 0.305 & 0.989 & 3 & 0.309 & 0.975 & 2 \\
C$_0$-log & 1.419 & 4.951 & 10 & 1.417 & 4.945 & 10 \\
C-seq & 0.380 & 1.272 & 5 & 0.310 & 1.007 & 3 \\
C$_0$-seq & 1.409 & 4.861 & 8 & 1.336 & 4.576 & 7 \\
C-logG & 0.391 & 1.301 & 6 & 0.391 & 1.301 & 6 \\
C$_0$-logG & 1.419 & 4.933 & 9 & 1.406 & 4.882 & 9 \\
\bottomrule
\end{tabular}
\end{table}

\begin{table}[htbp]
\centering
\caption{Selection-criterion sensitivity on test-stratified: global RMSE, minimax and within-family rank under each scoring. C$_0$, C$_0$-log, C$_0$-seq and C$_0$-logG are ablations: they train on the added points without the seed and carry no budget.}
\label{tab:test-stratified:scorings}
\begin{tabular}{lrrrrrr}
\toprule
 & \multicolumn{3}{c}{log-rmse} & \multicolumn{3}{c}{fos-rmse} \\
\cmidrule(lr){2-4} \cmidrule(lr){5-7}
arm & RMSE & minimax & rank & RMSE & minimax & rank \\
\midrule
\multicolumn{7}{l}{\textit{RF}} \\
A & 0.832 & 1.934 & 9 & 0.843 & 1.956 & 10 \\
B & 0.548 & 1.226 & 5 & 0.548 & 1.226 & 5 \\
C & 0.437 & 0.941 & 2 & 0.437 & 0.941 & 2 \\
C$_0$ & 0.464 & 0.982 & 3 & 0.464 & 0.982 & 3 \\
C-log & 0.530 & 1.157 & 4 & 0.530 & 1.157 & 4 \\
C$_0$-log & 0.704 & 1.357 & 8 & 0.704 & 1.357 & 8 \\
C-seq & 0.421 & 0.859 & 1 & 0.421 & 0.859 & 1 \\
C$_0$-seq & 0.631 & 1.048 & 6 & 0.631 & 1.048 & 6 \\
C-logG & 0.667 & 1.599 & 7 & 0.667 & 1.599 & 7 \\
C$_0$-logG & 0.841 & 1.847 & 10 & 0.841 & 1.847 & 9 \\
\addlinespace
\multicolumn{7}{l}{\textit{GBM}} \\
A & 0.264 & 0.521 & 6 & 0.277 & 0.545 & 6 \\
B & 0.147 & 0.246 & 1 & 0.214 & 0.434 & 2 \\
C & 0.198 & 0.377 & 2 & 0.192 & 0.365 & 1 \\
C$_0$ & 0.333 & 0.615 & 9 & 0.304 & 0.600 & 7 \\
C-log & 0.241 & 0.524 & 5 & 0.259 & 0.542 & 5 \\
C$_0$-log & 0.308 & 0.558 & 7 & 0.317 & 0.580 & 8 \\
C-seq & 0.216 & 0.355 & 3 & 0.215 & 0.370 & 3 \\
C$_0$-seq & 0.320 & 0.453 & 8 & 0.348 & 0.478 & 9 \\
C-logG & 0.231 & 0.470 & 4 & 0.231 & 0.470 & 4 \\
C$_0$-logG & 0.391 & 0.791 & 10 & 0.417 & 0.865 & 10 \\
\bottomrule
\end{tabular}
\end{table}

\begingroup
\small
\setlength{\tabcolsep}{3pt}
\setlength{\LTcapwidth}{\textwidth}
\begin{longtable}{llllrrrrrrr}
\caption{Paired comparisons on test-natural: Wilcoxon on d = |e1| - |e2| (smaller absolute errors) and bootstrap ΔRMSE. C$_0$, C$_0$-log, C$_0$-seq and C$_0$-logG are ablations: they train on the added points without the seed and carry no budget.}\label{tab:test-natural__fos-rmse:comparisons} \\
\toprule
 &  &  &  & \multicolumn{4}{c}{Wilcoxon on d} & \multicolumn{3}{c}{bootstrap} \\
\cmidrule(lr){5-8} \cmidrule(lr){9-11}
subset & scale & comparison & alternative & n & median & P(d<0) & p & ΔRMSE & CI low & CI high \\
\midrule
\endfirsthead
\multicolumn{11}{l}{\textit{Table \thetable, continued}} \\
\toprule
 &  &  &  & \multicolumn{4}{c}{Wilcoxon on d} & \multicolumn{3}{c}{bootstrap} \\
\cmidrule(lr){5-8} \cmidrule(lr){9-11}
subset & scale & comparison & alternative & n & median & P(d<0) & p & ΔRMSE & CI low & CI high \\
\midrule
\endhead
\midrule
\multicolumn{11}{r}{\textit{continued on the next page}} \\
\endfoot
\bottomrule
\endlastfoot
\multicolumn{11}{l}{\textit{GBM}} \\
all & log & C vs B & less & 300 & 0.001 & 0.483 & 0.770 & -0.004 & -0.013 & 0.005 \\
all & fos & C vs B & less & 300 & 0.004 & 0.480 & 0.734 & -0.049 & -0.218 & 0.103 \\
selection-range & log & C vs B & less & 266 & 0.000 & 0.492 & 0.662 & -0.000 & -0.006 & 0.006 \\
selection-range & fos & C vs B & less & 266 & 0.002 & 0.489 & 0.544 & -0.010 & -0.032 & 0.012 \\
decision-range & log & C vs B & less & 127 & -0.001 & 0.504 & 0.708 & -0.003 & -0.011 & 0.005 \\
decision-range & fos & C vs B & less & 127 & -0.001 & 0.504 & 0.622 & -0.006 & -0.016 & 0.004 \\
all & log & C vs A & less & 300 & -0.011 & 0.623 & <0.001 & -0.029 & -0.046 & -0.014 \\
\end{longtable}
\endgroup

\begingroup
\small
\setlength{\tabcolsep}{3pt}
\setlength{\LTcapwidth}{\textwidth}
\begin{longtable}{llllrrrrrrr}
\caption{Paired comparisons on test-stratified: Wilcoxon on d = |e1| - |e2| (smaller absolute errors) and bootstrap ΔRMSE. C$_0$, C$_0$-log, C$_0$-seq and C$_0$-logG are ablations: they train on the added points without the seed and carry no budget.}\label{tab:test-stratified__fos-rmse:comparisons} \\
\toprule
 &  &  &  & \multicolumn{4}{c}{Wilcoxon on d} & \multicolumn{3}{c}{bootstrap} \\
\cmidrule(lr){5-8} \cmidrule(lr){9-11}
subset & scale & comparison & alternative & n & median & P(d<0) & p & ΔRMSE & CI low & CI high \\
\midrule
\endfirsthead
\multicolumn{11}{l}{\textit{Table \thetable, continued}} \\
\toprule
 &  &  &  & \multicolumn{4}{c}{Wilcoxon on d} & \multicolumn{3}{c}{bootstrap} \\
\cmidrule(lr){5-8} \cmidrule(lr){9-11}
subset & scale & comparison & alternative & n & median & P(d<0) & p & ΔRMSE & CI low & CI high \\
\midrule
\endhead
\midrule
\multicolumn{11}{r}{\textit{continued on the next page}} \\
\endfoot
\bottomrule
\endlastfoot
\multicolumn{11}{l}{\textit{GBM}} \\
all & log & C vs B & less & 150 & -0.003 & 0.533 & 0.227 & -0.002 & -0.009 & 0.004 \\
all & fos & C vs B & less & 150 & -0.003 & 0.527 & 0.170 & -0.022 & -0.068 & 0.023 \\
selection-range & log & C vs B & less & 124 & -0.001 & 0.524 & 0.320 & -0.002 & -0.008 & 0.005 \\
selection-range & fos & C vs B & less & 124 & -0.001 & 0.516 & 0.238 & -0.003 & -0.028 & 0.023 \\
decision-range & log & C vs B & less & 36 & 0.002 & 0.444 & 0.528 & -0.005 & -0.017 & 0.007 \\
decision-range & fos & C vs B & less & 36 & 0.002 & 0.417 & 0.503 & -0.006 & -0.021 & 0.009 \\
all & log & C vs A & less & 150 & -0.013 & 0.667 & <0.001 & -0.027 & -0.040 & -0.017 \\
\end{longtable}
\endgroup

\end{document}